\documentclass[runningheads]{llncs}

\usepackage{graphicx}
\usepackage{amsmath}
\usepackage{amsfonts}
\usepackage{hyperref}
\usepackage{multirow}
\usepackage{rotating}
\usepackage{tikz}
\usepackage{float}
\usepackage{siunitx}
\usepackage[export]{adjustbox}
\usepackage{color}

\begin{document}
\title{Point Cloud Diffusion Models for Automatic Implant Generation}
\author{Paul Friedrich\inst{1} \and Julia Wolleb\inst{1} \and Florentin Bieder\inst{1} \and Florian M. Thieringer\inst{1,2}\and Philippe C. Cattin\inst{1}}
%
\authorrunning{P. Friedrich et al.}
\institute{Department of Biomedical Engineering, University of Basel, Allschwil, Switzerland \and
Department of Oral and Cranio-Maxillofacial Surgery, University Hospital Basel, Basel, Switzerland\\
\email{paul.friedrich@unibas.ch}}
\maketitle
\begin{abstract}
Advances in 3D printing of biocompatible materials make patient-specific implants increasingly popular. The design of these implants is, however, still a tedious and largely manual process. Existing approaches to automate implant generation are mainly based on \mbox{3D U-Net} architectures on downsampled or patch-wise data, which can result in a loss of detail or contextual information. Following the recent success of Diffusion Probabilistic Models, we propose a novel approach for implant generation based on a combination of 3D point cloud diffusion models and voxelization networks. Due to the stochastic sampling process in our diffusion model, we can propose an ensemble of different implants per defect, from which the physicians can choose the most suitable one. We evaluate our method on the SkullBreak and SkullFix datasets, generating high-quality implants and achieving competitive evaluation scores. The project page can be found at \url{https://pfriedri.github.io/pcdiff-implant-io}.
\end{abstract}
\section{Introduction}
\begin{figure}
    \includegraphics[width=\textwidth]{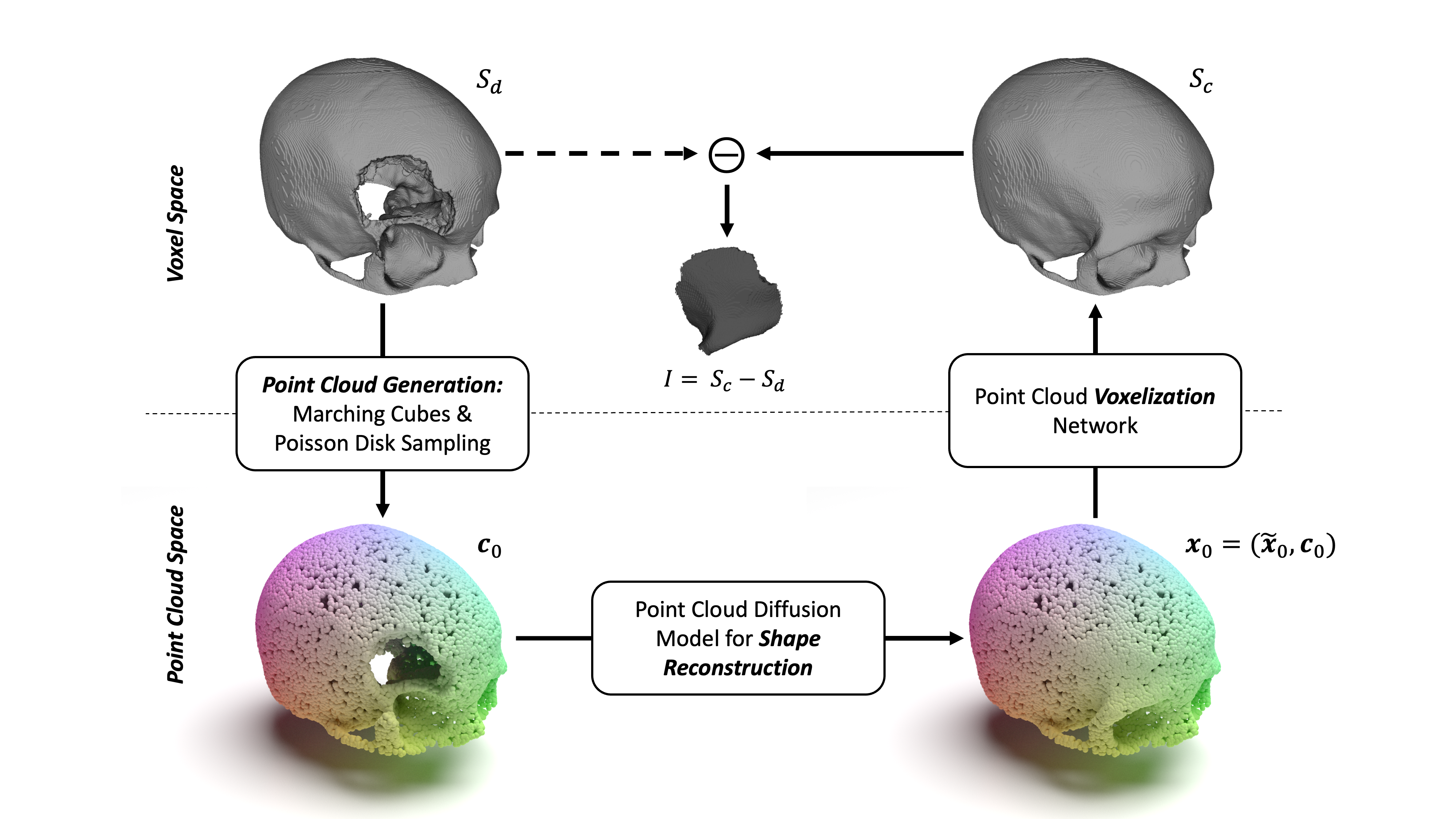}
    \caption{Proposed implant generation method with shape reconstruction in point cloud space. The implant geometry $I$ is defined as the Boolean subtraction between completed output $S_c$ and defective input $S_d$ in voxel space.} 
    \label{pipeline}
\end{figure}

The design of 3D-printed patient-specific implants, commonly used in cranioplasty and maxillofacial surgery, is a challenging and time-consuming task that is usually performed manually. To speed up the design process and enable point-of-care implant generation, approaches for automatically deriving suitable implant designs from medical images are needed. This paper presents a novel approach based on a Denoising Diffusion Probabilistic Model for 3D point clouds that reconstructs complete anatomical structures $S_c$ from segmented CT images of subjects showing bone defects $S_d$. An overview of the proposed method is shown in Figure \ref{pipeline}.
Since performing the anatomy reconstruction task in a high resolution voxel space would be memory inefficient and computationally expensive, we propose a method that builds upon a sparse surface point cloud representation of the input anatomy. This point cloud $\boldsymbol{c}_0$, which can be obtained from the defective segmentation mask $S_d$, serves as input to a Denoising Diffusion Probabilistic Model \cite{Zhou2021} that, conditioned on this input $\boldsymbol{c}_0$, reconstructs the complete anatomy $\boldsymbol{x}_0$ by generating missing points $\boldsymbol{\tilde{x}}_0$. The second network transforms the point cloud $\boldsymbol{x}_0$ back into voxel space using a Differentiable Poisson Solver \cite{Peng2021}. The final implant $I$ is generated by the Boolean subtraction of the completed and defective anatomical structure. We thereby ensure a good fit at the junction between implant and skull. Our main contributions are:
\begin{itemize}
    \item We employ 3D point cloud diffusion models for an automatic patient-specific implant generation task. The stochastic sampling process of diffusion models allows for the generation of multiple anatomically reasonable implant designs per subject, from which physicians can choose the most suitable one.
    \item We evaluate our method on the SkullBreak and SkullFix datasets, generating high-quality implants and achieving competitive evaluation scores.
\end{itemize}
\subsubsection{Related work}
Previous work on automatic implant generation methods mainly derived from the AutoImplant challenges \cite{Li2020Towards,Li2021Towards,Li2021Summary} at the 2020/21 MICCAI conferences. Most of the proposed methods were based on 2D slice-wise U-Nets \cite{Shi2020} and 3D U-Nets on downsampled \cite{Kodym2020,Matzkin2020,Wodzinski2020} or patch-wise data \cite{Jin2020,Li2021,Pathak2021}. Other approaches were based on Statistical Shape Models \cite{Yu2021}, Generative Adversarial Networks \cite{Pimentel2020} or Variational Autoencoders \cite{Wang2020}. This work is based on Diffusion Models \cite{Ho2020,Sohl2015}, which achieved good results in 2D reconstruction tasks like image inpainting \cite{Lugmayr2022,Saharia2021} and were also already applied to 3D generative tasks like point cloud generation \cite{Luo2021,Nichol2022,Zeng2022} and point cloud completion \cite{Lyu2021,Zhou2021}. For retrieving a dense voxel representation of a point cloud, many approaches rely on a combination of surface meshing \cite{Gropp2020,Hanocka2020,Kazhdan2013} and ray casting algorithms. Since surface meshing of unoriented point clouds with a non-uniform distribution is challenging and ray casting implies additional computational effort, we look at point cloud voxelization based on a Differentiable Poisson Solver \cite{Peng2021}.
\section{Methods}
As presented in Figure \ref{pipeline}, we propose a multi-step approach for generating an implant $I$ from a binary voxel representation $S_d$ of a defective anatomical structure.
\subsubsection{Point Cloud Generation}
Since the proposed method for shape reconstruction works in the point cloud space, we first need to derive a point cloud $\boldsymbol{c}_0\in \mathbb{R}^{N\times 3}$ from $S_d$. We therefore create a surface mesh of $S_d$ using Marching Cubes \cite{Lorensen1987}.  Then we sample $N$ points from this surface mesh using Poisson Disk Sampling \cite{Yuksel2015}. During training, we generate the ground truth point cloud $\boldsymbol{\tilde{x}}_0 \in \mathbb{R}^{M\times 3}$ by sampling $M$ points from the ground truth implant using the same approach.
\subsubsection{Diffusion Model for Shape Reconstruction} 
\begin{figure}
    \includegraphics[width=\textwidth]{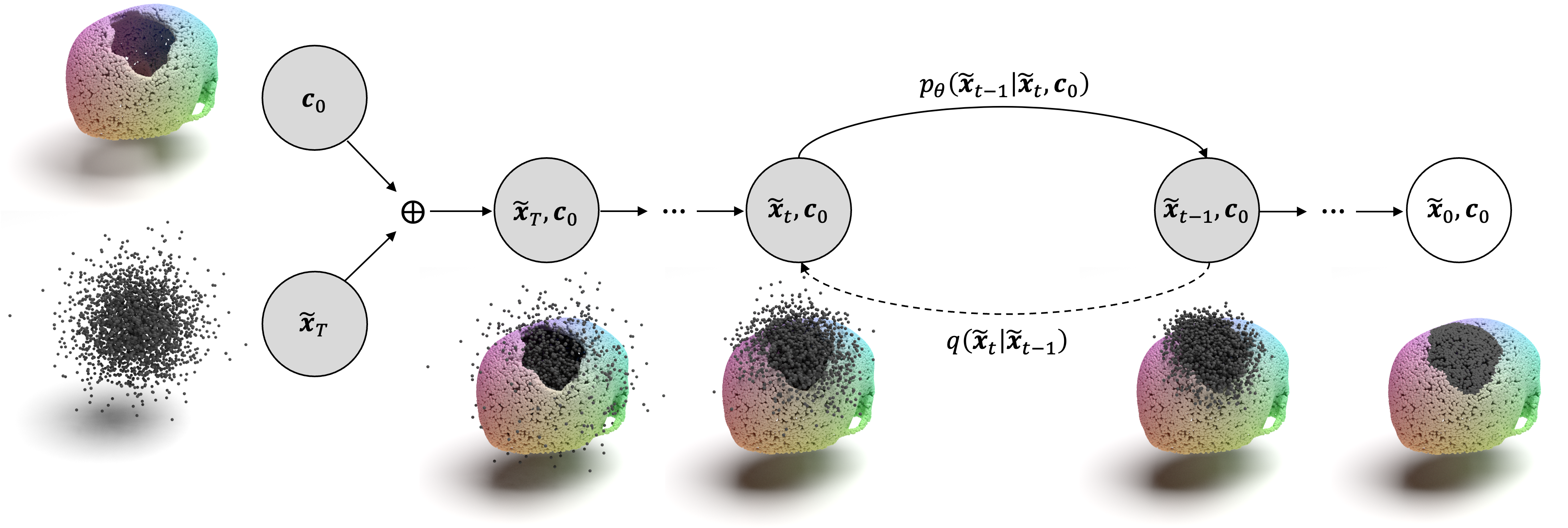}
    \caption{Conditional diffusion model for anatomy reconstruction in point cloud space. Points belonging to the defective anatomical structure $\boldsymbol{c}_0$ are shown in color and remain unchanged throughout the whole process. The forward and reverse diffusion processes therefore only affect the gray points $\boldsymbol{\tilde{x}}_{0:T}$ belonging to the implant.} 
    \label{diffusion_completion}
\end{figure}

Reconstructing the shape of an anatomical structure can be seen as a conditional generation process.  We train a diffusion model $\epsilon_\theta$ to reconstruct the point cloud $\boldsymbol{x}_0 = (\boldsymbol{\tilde{x}}_0,\boldsymbol{c}_0)$ that describes the complete anatomical structure $S_c$. The generation process is conditioned on the points $\boldsymbol{c}_0$ belonging to the known defective anatomical structure $S_d$. An overview is given in Figure \ref{diffusion_completion}. For describing the diffusion model, we follow the formulations in \cite{Zhou2021}. Starting from $\boldsymbol{x}_0$, we first define the forward diffusion process, that gradually adds small amounts of noise to $\boldsymbol{\tilde{x}}_0$, while keeping $\boldsymbol{c}_0$ unchanged and thus produces a series of point clouds $\{{\boldsymbol{x}_0=(\boldsymbol{\tilde{x}}_0,\boldsymbol{c}_0)}, {\boldsymbol{x}_1=(\boldsymbol{\tilde{x}}_1,\boldsymbol{c}_0)}, {...}, {\boldsymbol{x}_T=(\boldsymbol{\tilde{x}}_T,\boldsymbol{c}_0)\}}$. 
This \textit{conditional forward diffusion process} can be modeled as a Markov chain with a defined number of timesteps $T$ and transfers $\boldsymbol{\tilde{x}}_0$ into a noise distribution:
\begin{equation}
    q(\boldsymbol{\tilde{x}}_{0:T}) = q(\boldsymbol{\tilde{x}}_0)\prod_{t=1}^T q(\boldsymbol{\tilde{x}}_t|\boldsymbol{\tilde{x}}_{t-1}).
\end{equation}
Each transition is modeled as a parameterized Gaussian and follows a predefined variance schedule $\beta_1, ..., \beta_T$ that controls the diffusion rate of the process:
\begin{equation}
    q(\boldsymbol{\tilde{x}}_t|\boldsymbol{\tilde{x}}_{t-1}) := \mathcal{N}(\sqrt{1-\beta_t}\boldsymbol{\tilde{x}}_{t-1}, \beta_t\boldsymbol{I}).
\end{equation}
The goal of the diffusion model is to learn the \textit{reverse diffusion process} that is able to gradually remove noise from $\boldsymbol{\tilde{x}}_T \sim \mathcal{N}(0, \boldsymbol{I})$. This reverse process is also modeled as a Markov chain
\begin{equation}
    p_{\theta}(\boldsymbol{\tilde{x}}_{0:T}) = p(\boldsymbol{\tilde{x}}_T)\prod_{t=1}^T p_{\theta}(\boldsymbol{\tilde{x}}_{t-1}|\boldsymbol{\tilde{x}}_t, \boldsymbol{c}_0),
\end{equation}
with each transition being defined as a Gaussian, with the estimated mean $\mu_\theta$:
\begin{equation}
    p_{\theta}(\boldsymbol{\tilde{x}}_{t-1}|\boldsymbol{\tilde{x}}_t, \boldsymbol{c}_0) := \mathcal{N}(\mu_\theta(\boldsymbol{\tilde{x}}_t, \boldsymbol{c}_0, t), \beta_t\boldsymbol{I}).
\end{equation}
As derived in \cite{Zhou2021}, the network $\epsilon_\theta$ can be adapted to predict the noise $\epsilon_\theta(\boldsymbol{\tilde{x}}_t, \boldsymbol{c}_0, t)$ to be removed from a noisy point cloud $\boldsymbol{\tilde{x}}_t$. During training, we compute $\boldsymbol{\tilde{x}}_t$ at a random timestep $t \in \{1, ..., T\}$ and optimize a Mean Squared Error (MSE) loss
\begin{equation}
    \mathcal{L}_t = \| \epsilon - \epsilon_\theta(\boldsymbol{\tilde{x}}_t, \boldsymbol{c}_0, t) \|^2, \quad \text{where} \quad \boldsymbol{\tilde{x}}_t = \sqrt{\tilde{\alpha}_t}\boldsymbol{\tilde{x}}_0 + \sqrt{1-\tilde{\alpha}_t}\epsilon,
\end{equation} 
with $\epsilon \sim \mathcal{N}(0, \boldsymbol{I})$, $\alpha_t=1-\beta_t$, and $\tilde{\alpha}_t = \prod_{s=1}^t\alpha_s$. To perform shape reconstruction with the trained network, we start with a point cloud $\boldsymbol{x}_T = (\boldsymbol{\tilde{x}}_T, \boldsymbol{c}_0)$ with $\boldsymbol{\tilde{x}}_T \sim \mathcal{N}(0, \boldsymbol{I})$. This point cloud is then passed through the reverse diffusion process
\begin{equation}
    \boldsymbol{\tilde{x}}_{t-1} =\frac{1}{\sqrt{\alpha_t}}\biggl(\boldsymbol{\tilde{x}}_t-\frac{1-\alpha_t}{\sqrt{1-\tilde{\alpha}}_t}\epsilon_\theta(\boldsymbol{\tilde{x}}_t, \boldsymbol{c}_0, t)\biggr)+\sqrt{\beta_t}\boldsymbol{z},
\end{equation}
with $\boldsymbol{z}\sim \mathcal{N}(0, \boldsymbol{I})$, for $t=T,...,1$. While this reverse diffusion process gradually removes noise from $\boldsymbol{\tilde{x}}_T$, the points belonging to the defective anatomical structure $\boldsymbol{c}_0$ remain unchanged. As proposed in \cite{Zhou2021}, our network $\epsilon_\theta$ is based on a PointNet++ \cite{Qi2017} architecture with Point-Voxel Convolutions \cite{Liu2019}. Details on the network architecture can be found in the supplementary material.
\subsubsection{Voxelization}
\begin{figure}
    \includegraphics[width=\textwidth]{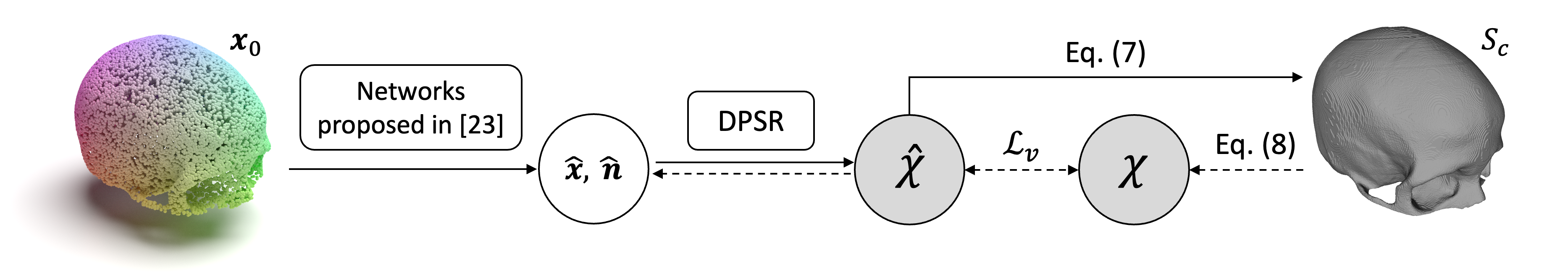}
    \caption{Voxelization pipeline based on a trainable network and a Differentiable Poisson Solver (DPSR). The trainable part learns to produce an upsampled point cloud $\boldsymbol{\hat{x}}$ with estimated normal vectors $\boldsymbol{\hat{n}}$. The DPSR transforms $(\boldsymbol{\hat{x}}, \boldsymbol{\hat{n}})$ into an indicator function $\hat{\chi}$ from which a voxel representation can be derived.} 
    \label{voxelization}
\end{figure}

To create an implant, the point cloud of the restored anatomy must be converted back to voxel space. We follow a learning-based pipeline proposed in \cite{Peng2021}. The pipeline shown in Figure \ref{voxelization} takes an unoriented point cloud $\boldsymbol{x}_0$ as input and learns to predict an upsampled, oriented point cloud $\boldsymbol{\hat{x}}$ with normal vectors $\boldsymbol{\hat{n}}$. From this upsampled point cloud, an indicator grid $\hat{\chi}$ can be produced using a Differentiable Poisson Solver (DPSR).
The complete voxel representation $S_c$ can then be obtained by evaluating the following equation for every voxel position $i$:
\begin{equation}
    S_c(i) = 
    \begin{cases}
        1, & \text{if } \hat{\chi}(i) \leq 0 \\
        0, & \text{if } \hat{\chi}(i) > 0 \\
    \end{cases}.
\end{equation}
During training, the ground truth indicator grid $\chi$ can be obtained directly from the ground truth voxel representation $S_{c}$ and, as described in \cite{Kazdhan2006}, is defined as:
\begin{equation}
    \chi(i) = 
    \begin{cases}
    0.5, & \text{if } S_{c}(i)=0 \\
    -0.5, & \text{if } S_{c}(i)=1
    \end{cases}.
\end{equation}
Due to the differentiability of the used Poisson solver, the networks can be trained with an MSE loss between the estimated and ground truth indicator grid:
\begin{equation}
    \mathcal{L}_v = \|\hat{\chi} - \chi \|^2.
\end{equation}
For further information on the used network architectures, we refer to \cite{Peng2021}.
\subsubsection{Implant Generation}
With the predicted complete anatomical structure $S_c$ and the defective input structure $S_d$, an implant geometry $I$ can be derived by the Boolean subtraction between $S_c$ and $S_d$:
\begin{equation}
    I = S_c-S_d.
\end{equation}
To further improve the implant quality and remove noise, we apply a median filter as well as binary opening to the generated implant.
\subsubsection{Ensembling}
As the proposed point cloud diffusion model features a stochastic generation process, we can sample multiple anatomically reasonable implant designs for each defect. This offers physicians the opportunity of selecting from various possible implants and allows the determination of a mean implant from a previously generated ensemble of $n$ different implants. As presented in \cite{Wolleb2021}, the ensembling strategy can also be used to create voxel-wise variance over an ensemble. These variance maps highlight areas with high differences between multiple anatomically reasonable implants.
\section{Experiments}
We evaluated our method on the publicly available parts of the SkullBreak and SkullFix datasets. For the point cloud diffusion model we chose a total number of $\num{30720}$ points $(N=\num{27648}, M=\num{3072})$, set $T=\num{1000}$, followed a linear variance schedule between $\beta_0 = 10^{-4}$ and $\beta_T = 0.02$, used the Adam optimizer with a learning rate of $2\times10^{-4}$, a batch size of 8, and trained the network for $\num{15000}$ epochs. This took about \SI{20}{\day}/\SI{4}{\day} for the SkullBreak/SkullFix dataset. For training the voxelization network, we used the Adam optimizer with a learning rate of $5\times10^{-4}$, a batch size of 2 and trained the networks for 1300/500 epochs on the SkullBreak/SkullFix dataset. This took about \SI{72}{\hour}/\SI{5}{\hour}. All experiments were performed on an NVIDIA A100 GPU using PyTorch as the framework.
\subsubsection{SkullBreak/SkullFix}
Both datasets, SkullBreak and SkullFix \cite{Kodym2021}, contain binary segmentation masks of head CT images with artificially created skull defects. While the SkullFix dataset mainly features rectangular defect patterns with additional craniotomy drill holes, SkullBreak offers more diverse defect patterns. The SkullFix dataset was resampled to an isotropic voxel size of \SI{0.45}{\milli\metre}, zero padded to a volume size of $512 \times 512 \times 512$, and split into a training set with 75 and a test set with 25 volumes. The SkullBreak dataset already has an isotropic voxel size of \SI{0.4}{\milli\metre} and a volume size of $512 \times 512 \times 512$.  We split the SkullBreak dataset into a training set with 430 and a test set with 140 volumes. All point clouds sampled from these datasets were normalized to a range between $[-3,3]$ in all spatial dimensions. The SkullBreak and SkullFix datasets were both adapted from the publicly available head CT dataset CQ500, which is licensed under a CC-BY-NC-SA 4.0 and End User License Agreement (EULA). The SkullBreak and SkullFix datasets were adapted and published under the same licenses.
\section{Results and Discussion}
\begin{figure}[H]
\centering
\resizebox{\textwidth}{!}{
	\begin{tikzpicture}
	    \node[] at (0, 0)       {\adjincludegraphics[width=0.16\textwidth, trim={.15\width, .5\width, .15\width, 0}, clip]{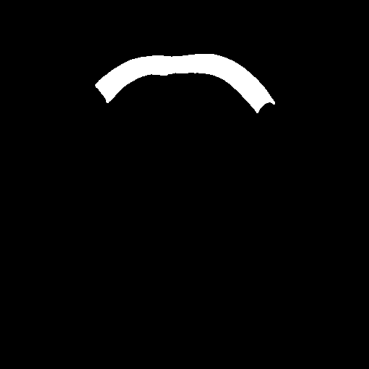}};
    	\node[] at (0, 1)       {\scriptsize Implant 1};
    	\node[] at (2, 0)       {\adjincludegraphics[width=0.16\textwidth, trim={.15\width, .5\width, .15\width, 0}, clip]{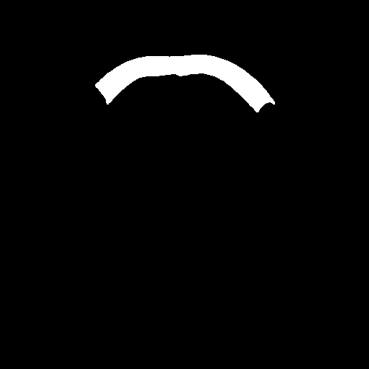}};
    	\node[] at (2, 1)       {\scriptsize Implant 2};
    	\node[] at (3.5, 0)     {$...$};
    	\node[] at (5, 0)       {\adjincludegraphics[width=0.16\textwidth, trim={.15\width, .5\width, .15\width, 0}, clip]{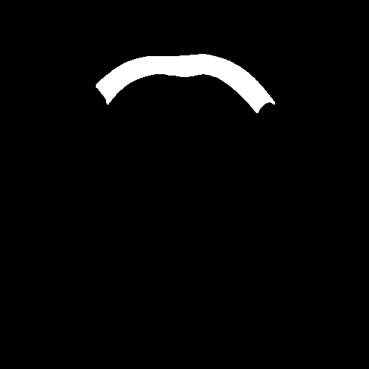}};
    	\node[] at (5, 1)       {\scriptsize Implant 5};
    	\node[] at (7, 0)       {\adjincludegraphics[width=0.16\textwidth, trim={.15\width, .5\width, .15\width, 0}, clip]{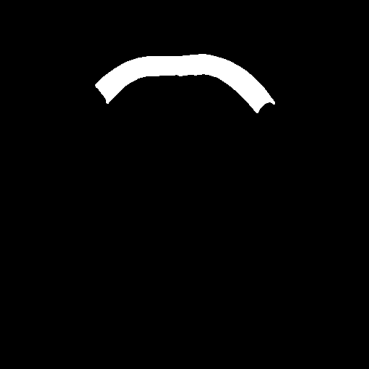}};
    	\node[] at (7, 1)       {\scriptsize Mean};
    	\node[] at (9, 0)       {\adjincludegraphics[width=0.16\textwidth, trim={.15\width, .5\width, .15\width, 0}, clip]{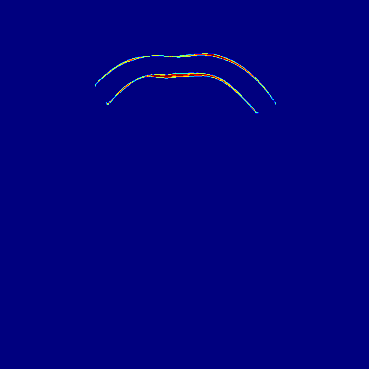}};
    	\node[] at (9, 1)       {\scriptsize Variance Map};
	\end{tikzpicture}}
\caption{Different implants, mean implant and variance map for a single skull defect.}
\label{Ensembling}
\end{figure}
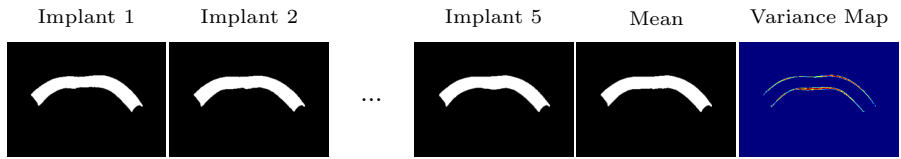

For evaluating our approach, we compared it to three methods from AutoImplant 2021 challenge: the winning 3D U-Net based approach \cite{Wodzinski2020}, a 3D U-Net based approach with sparse convolutions \cite{Kroviakov2020} and a slice-wise 2D U-Net approach \cite{Yang2021}. We also evaluated the mean implant produced by the proposed ensembling strategy ($n=5$). The implant generation time ranges from $\sim$\SI{1000}{\second} for SkullBreak ($n=1$) to $\sim$\SI{1200}{\second} for SkullFix ($n=5$), with the diffusion model requiring most of this time. In Table \ref{scores}, the Dice score (DSC), the \SI{10}{\milli\metre} boundary DSC (bDSC), as well as the 95 percentile Hausdorff Distance (HD95) are presented as mean values over the respective test sets. 
\begin{table}[ht]
\caption{Mean evaluation scores on SkullBreak and SkullFix test sets.}
\begin{center}
    \begin{tabular}{l|c c c|c c c }
        \multirow{2}{*}{\textbf{Model}} & \multicolumn{3}{c|}{\textbf{SkullBreak}} & \multicolumn{3}{c}{\textbf{SkullFix}}\\
        & DSC $\uparrow$ & bDSC $\uparrow$ & HD95 $\downarrow$ & DSC $\uparrow$ & bDSC $\uparrow$ & HD95 $\downarrow$ \\
        \hline
        3D U-Net \cite{Wodzinski2020}           & 0.87 & 0.91 & 2.32 & 0.91 & 0.95 & 1.79\\
        3D U-Net (sparse) \cite{Kroviakov2020}  & 0.71 & 0.80 & 4.60 & 0.81 & 0.87 & 3.04\\
        2D U-Net \cite{Yang2021}                & 0.87 & 0.89 & 2.13 & 0.89 & 0.92 & 1.98\\
        \textbf{Ours}                           & 0.86 & 0.88 & 2.51 & 0.90 & 0.92 & 1.73\\
        \textbf{Ours (n=5)}                     & 0.87 & 0.89 & 2.45 & 0.90 & 0.93 & 1.69
    \end{tabular}
    \label{scores}
    \end{center}
\end{table}
Qualitative results of the different implant generation methods are shown in Figure \ref{qualiresults} and Figure \ref{SkullBreak}, as well as in the supplementary material. Implementation detail for the comparing methods, more detailed runtime information, as well as the used code can be found at \url{https://github.com/pfriedri/pcdiff-implant}. 
\begin{figure}[H]
\centering
\resizebox{0.79\textwidth}{!}{
	\begin{tikzpicture}
    	\node[] at (0, 0)       {\includegraphics[height=0.2\textwidth]{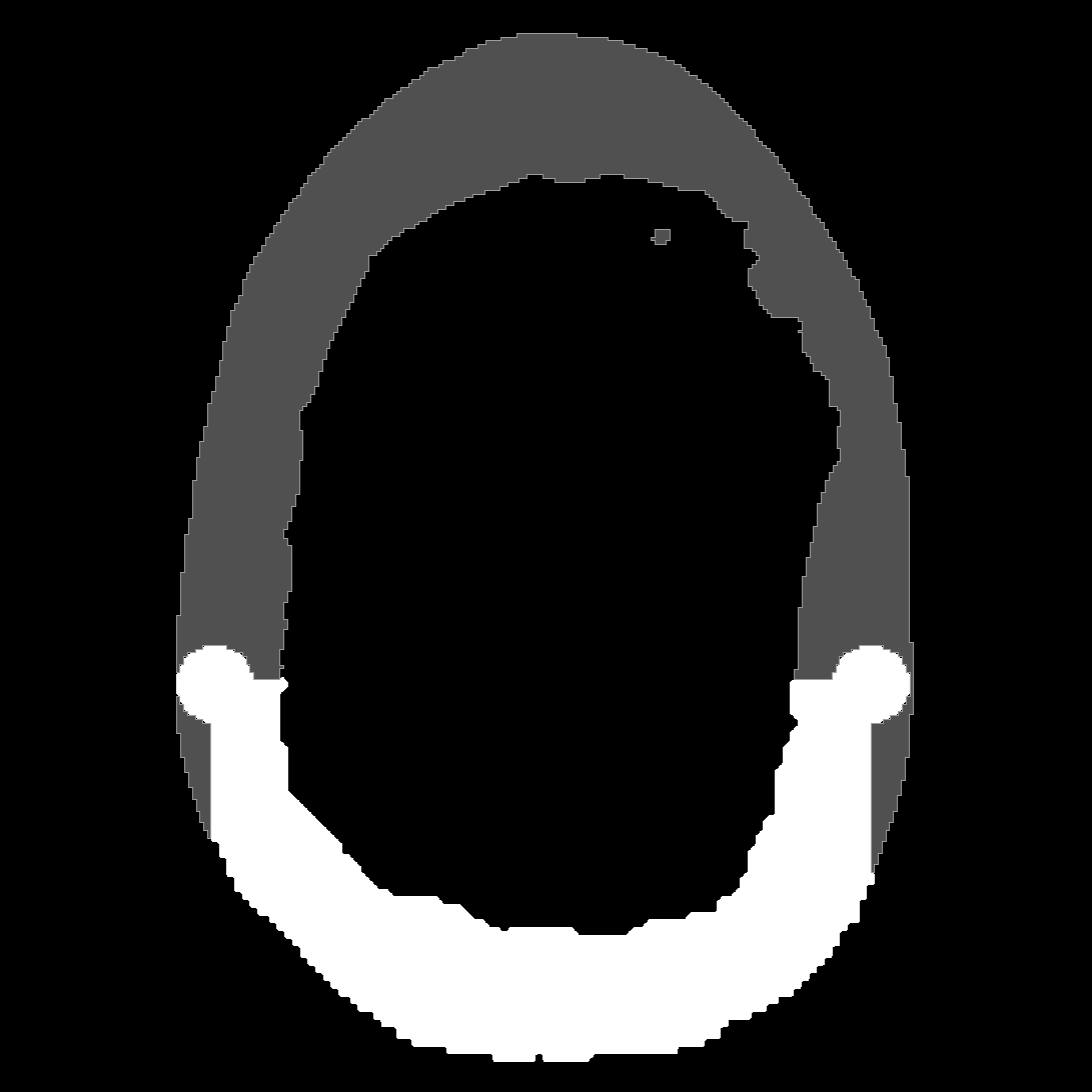}};
    	\node[] at (0, 1.5)     {\scriptsize Axial};
    	\node[] at (2.5, 0)     {\includegraphics[height=0.2\textwidth]{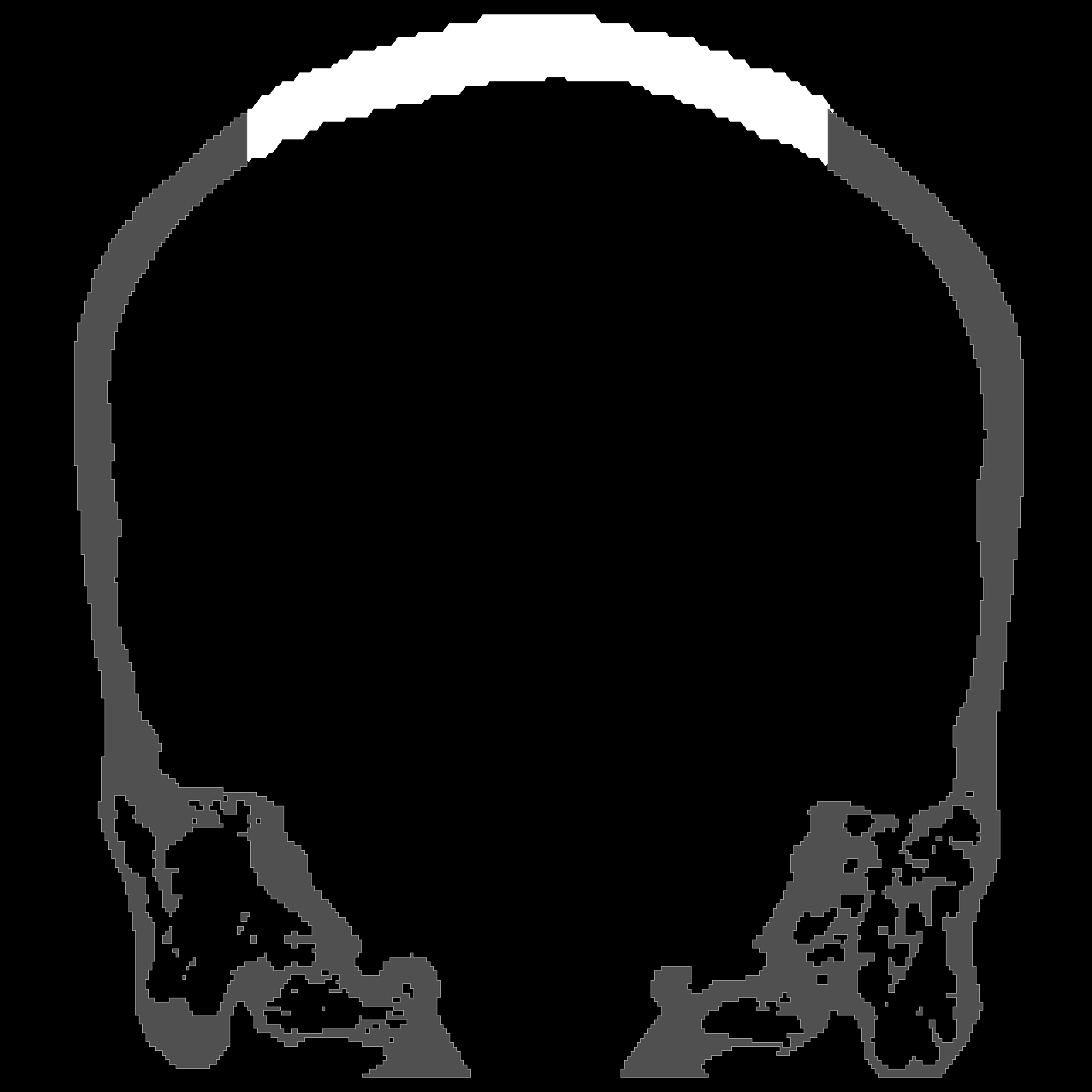}};
    	\node[] at (2.5, 1.5)   {\scriptsize Coronal};
    	\node[] at (5, 0)       {\includegraphics[height=0.2\textwidth]{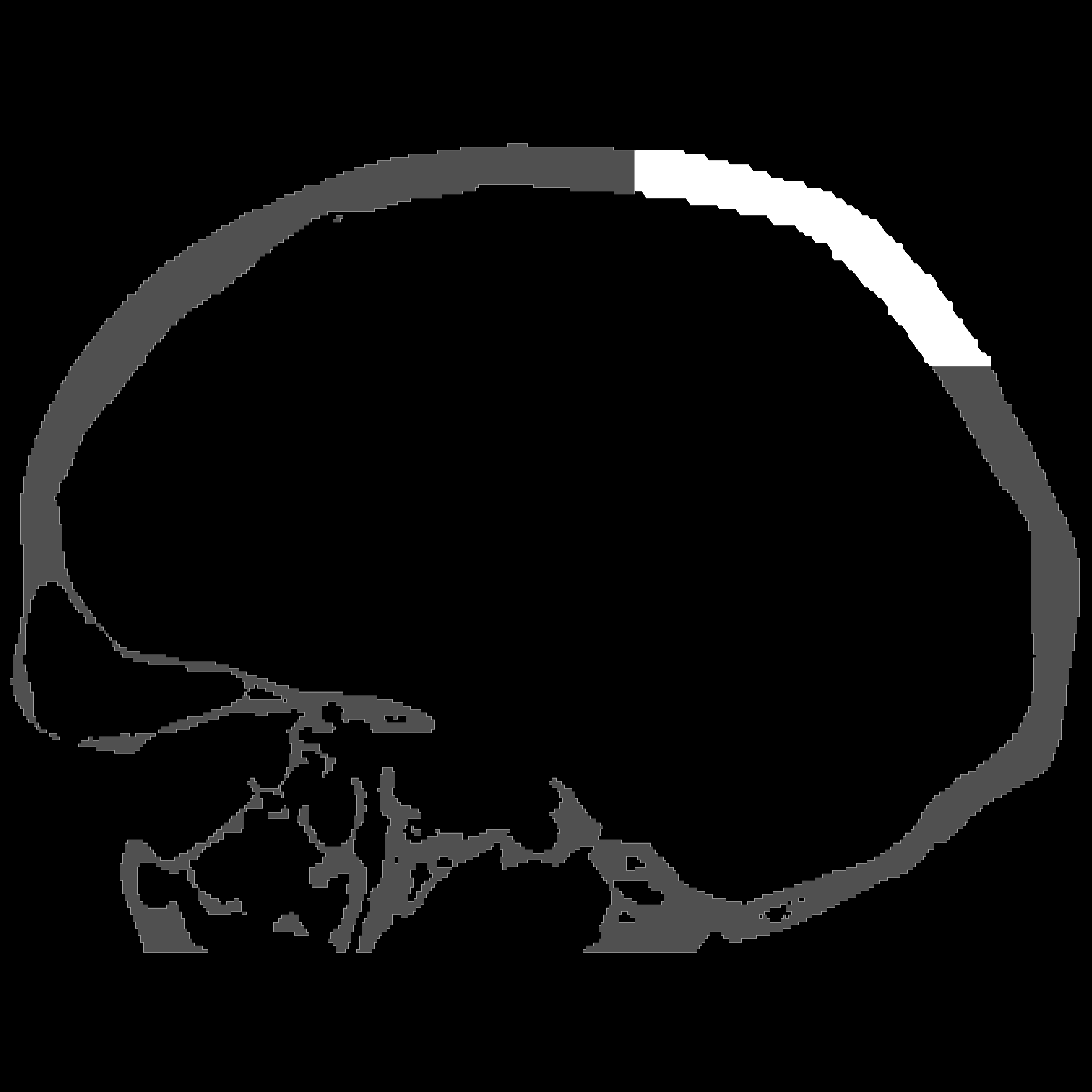}};
    	\node[] at (5, 1.5)     {\scriptsize Sagittal};
    	\node[] at (7.5, 0)     {\includegraphics[height=0.2\textwidth]{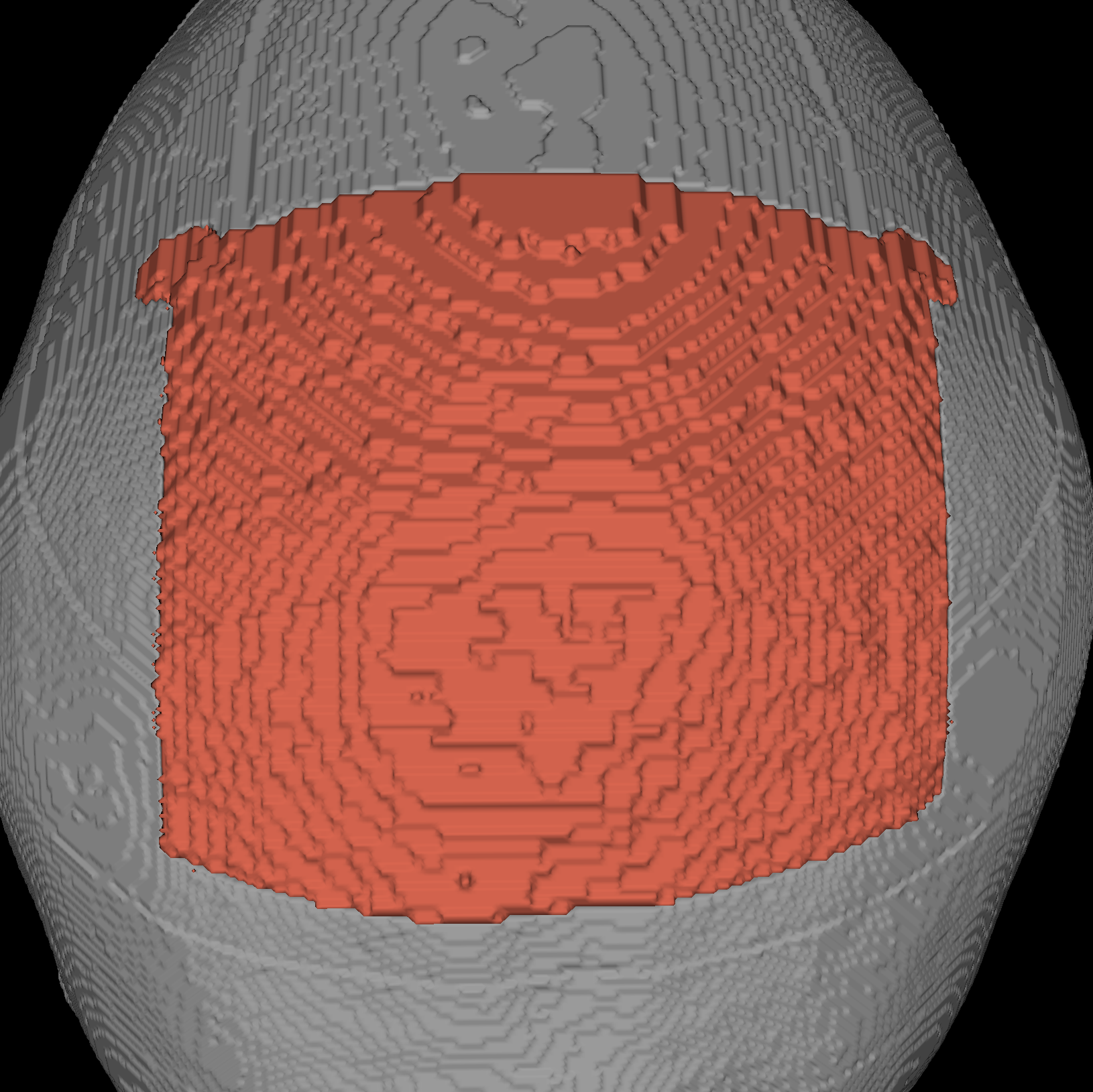}};
        \node[] at (7.5, 1.5)   {\scriptsize Overview};
        \node[] at (0, -2.5)    {\includegraphics[height=0.2\textwidth]{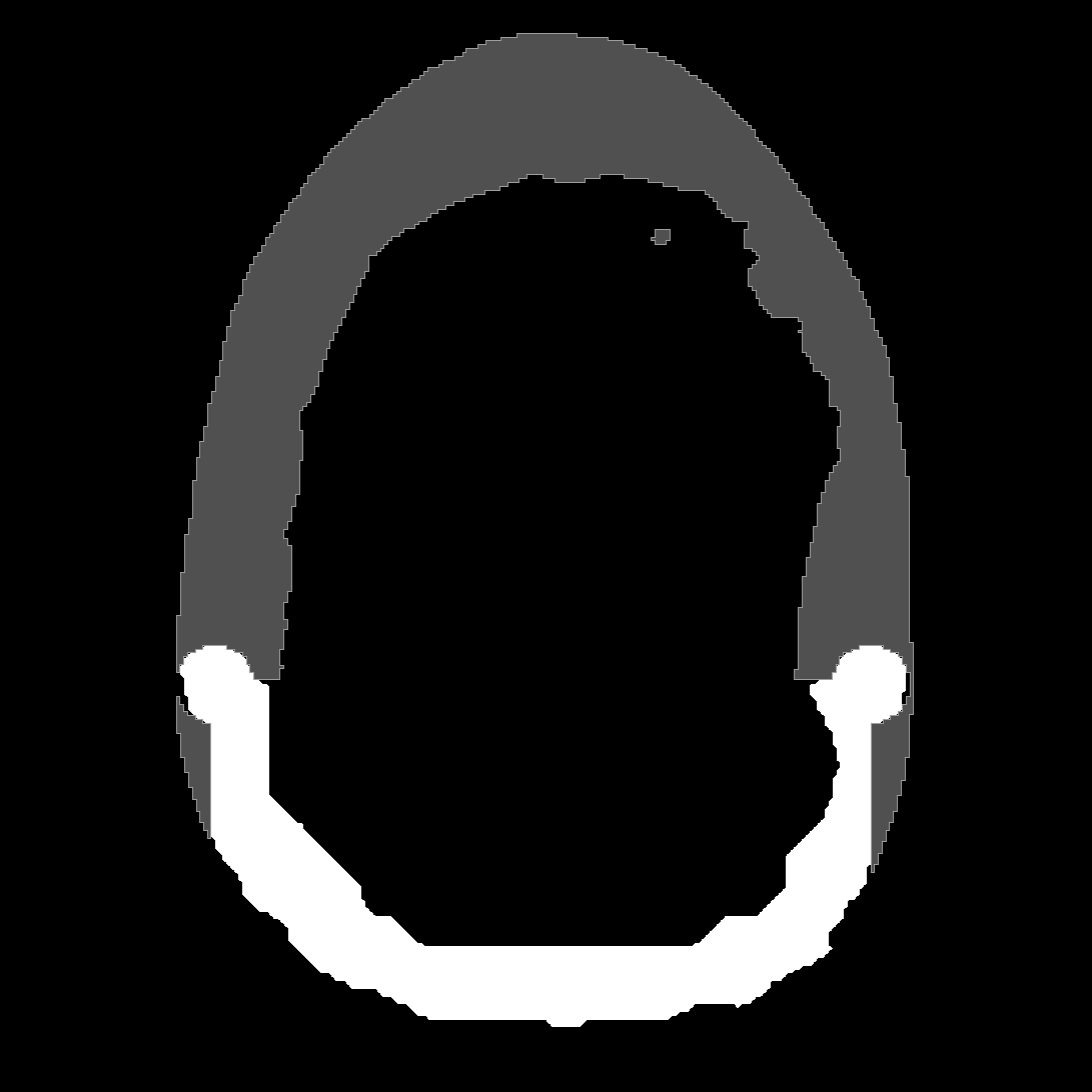}};
        \node[] at (2.5, -2.5)  {\includegraphics[height=0.2\textwidth]{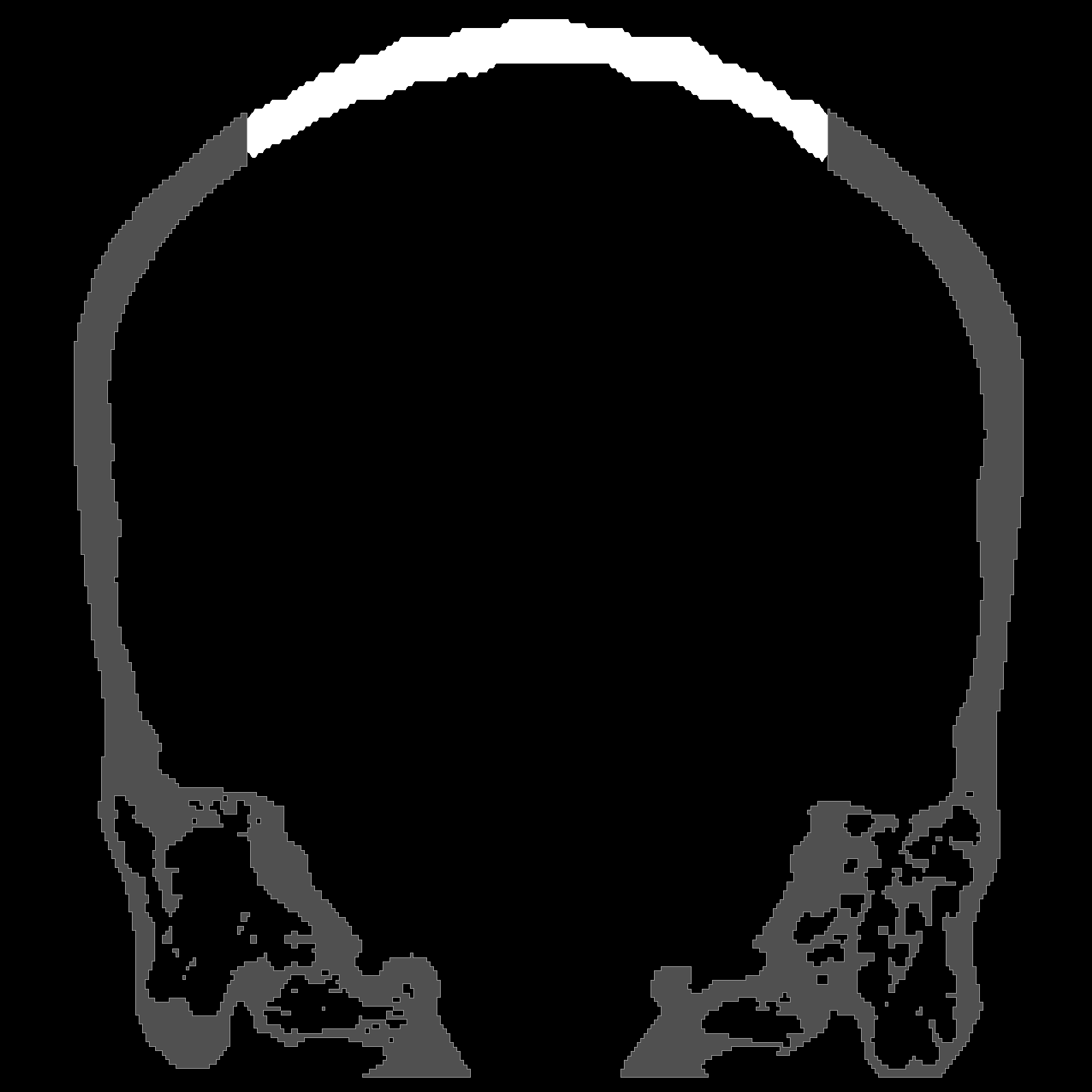}};
        \node[] at (5, -2.5)    {\includegraphics[height=0.2\textwidth]{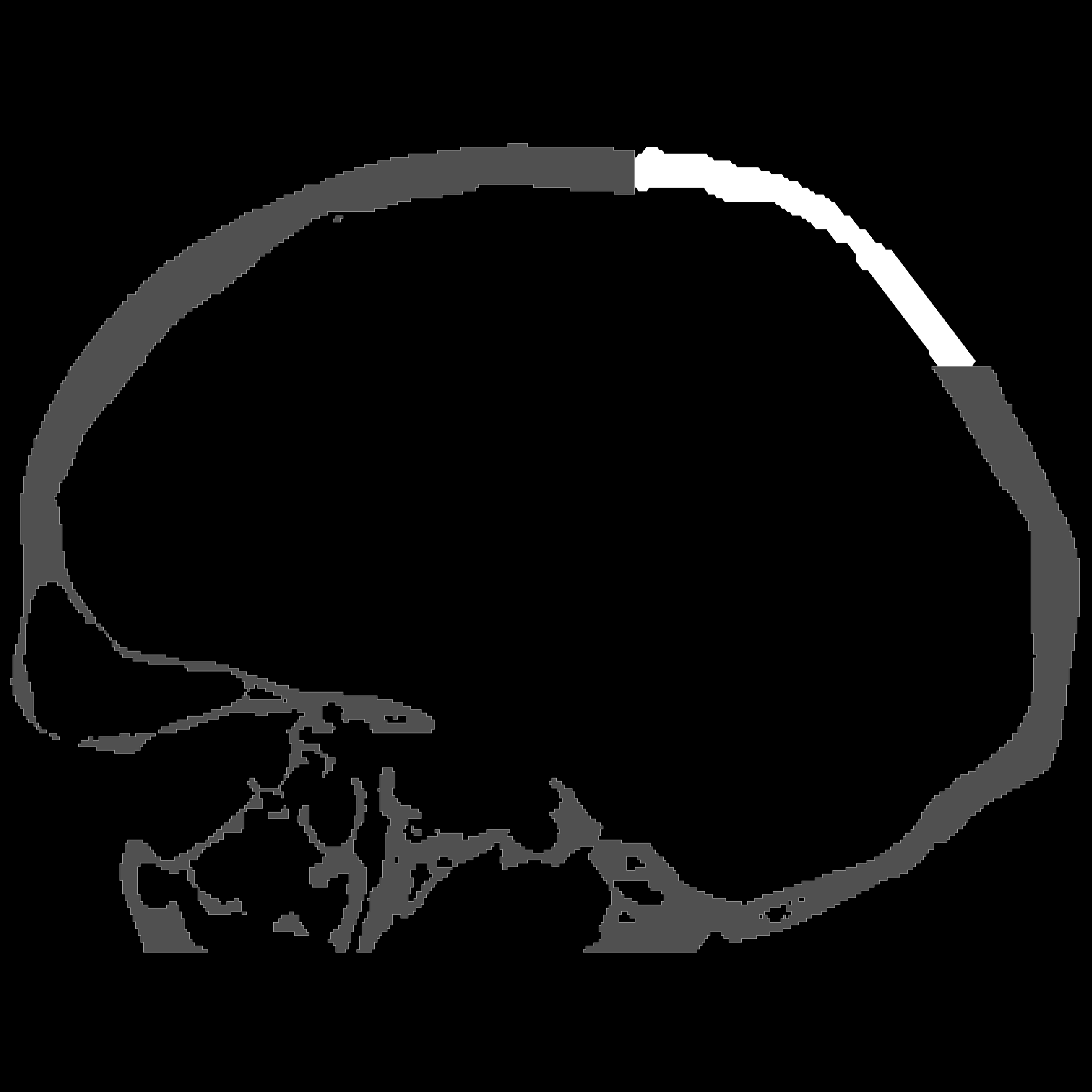}};
        \node[] at (7.5, -2.5)  {\includegraphics[height=0.2\textwidth]{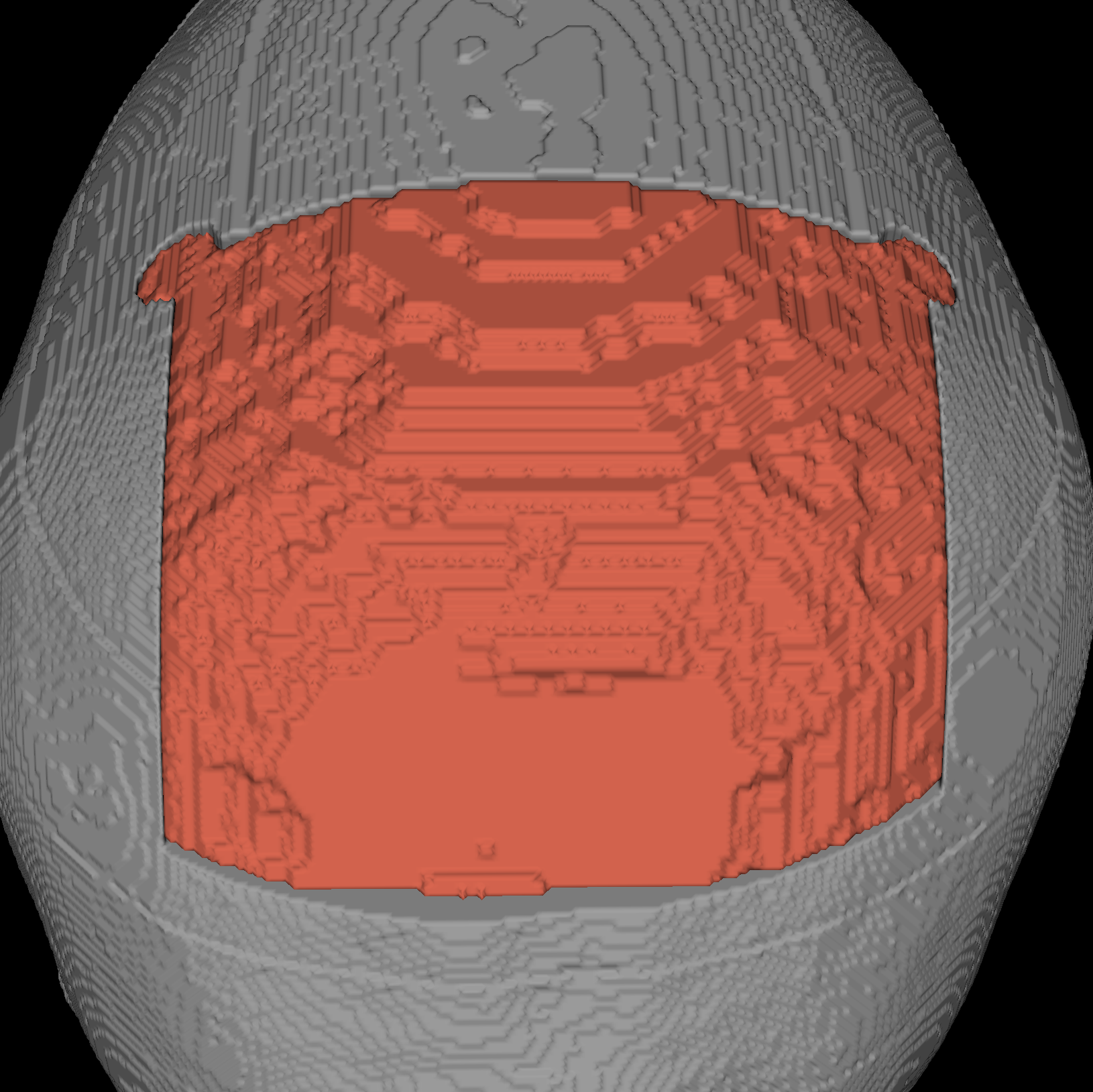}};
        \node[] at (0, -5)      {\includegraphics[height=0.2\textwidth]{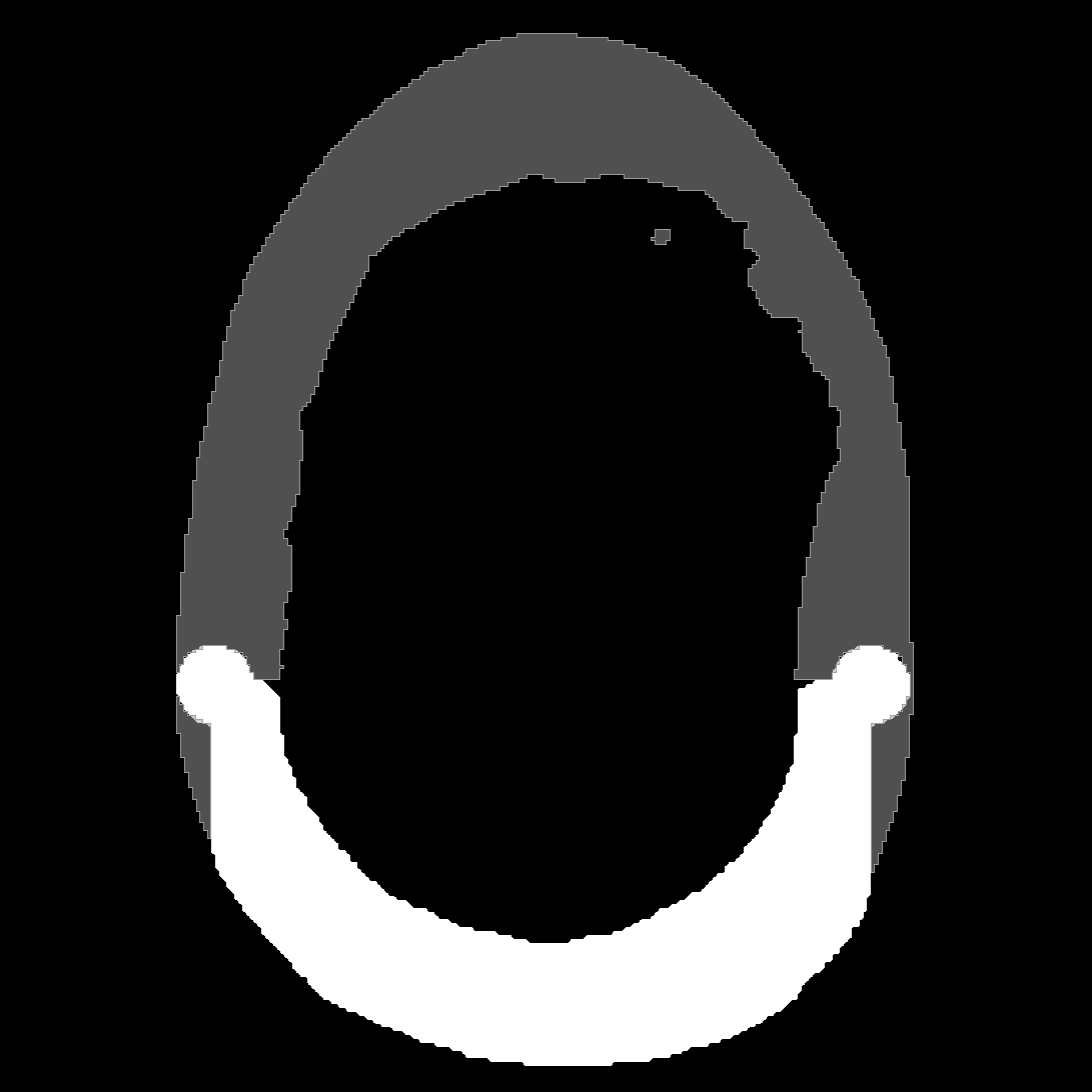}};
        \node[] at (2.5, -5)    {\includegraphics[height=0.2\textwidth]{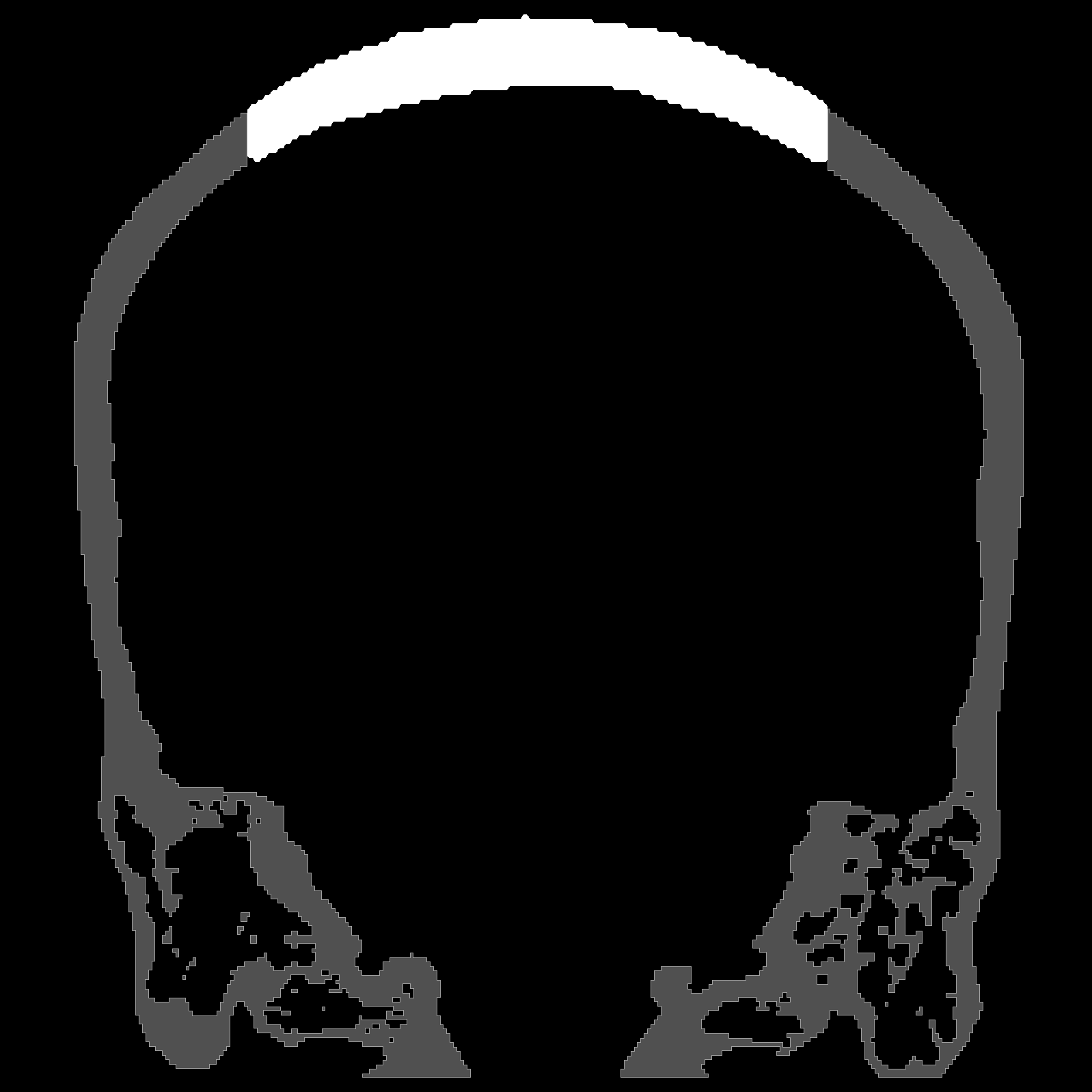}};
        \node[] at (5, -5)      {\includegraphics[height=0.2\textwidth]{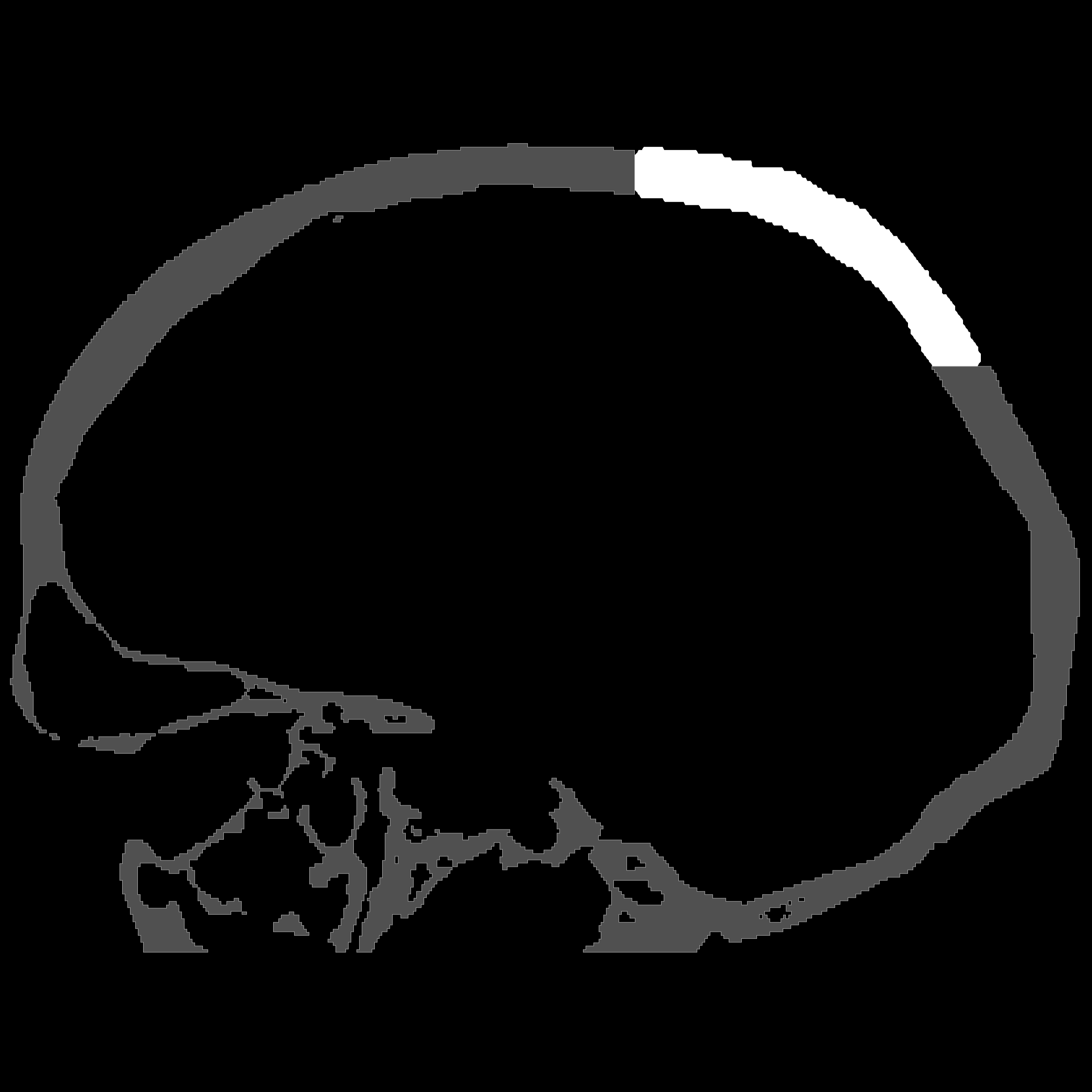}};
        \node[] at (7.5, -5)    {\includegraphics[height=0.2\textwidth]{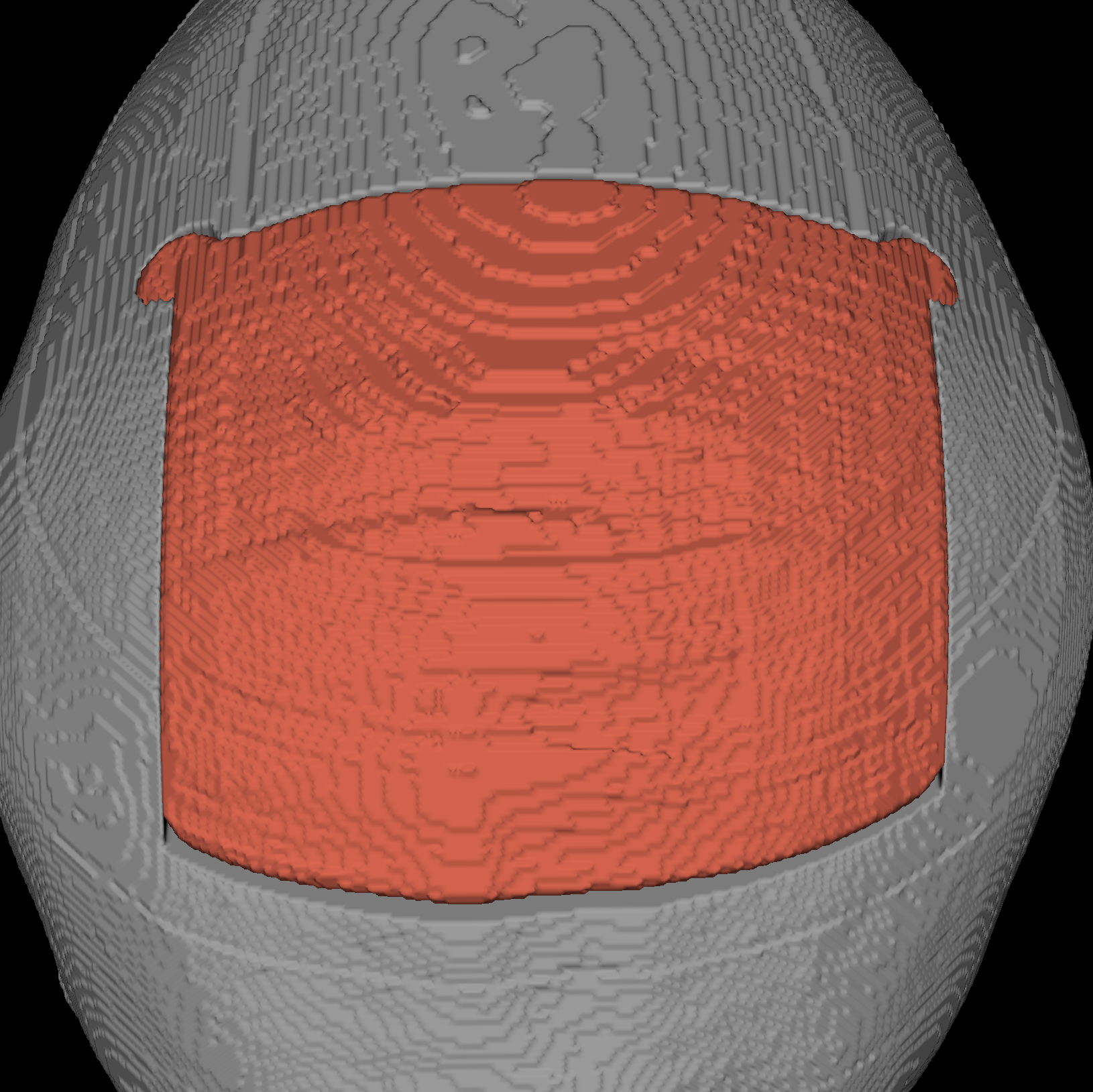}};
        \node[] at (0, -7.5)    {\includegraphics[height=0.2\textwidth]{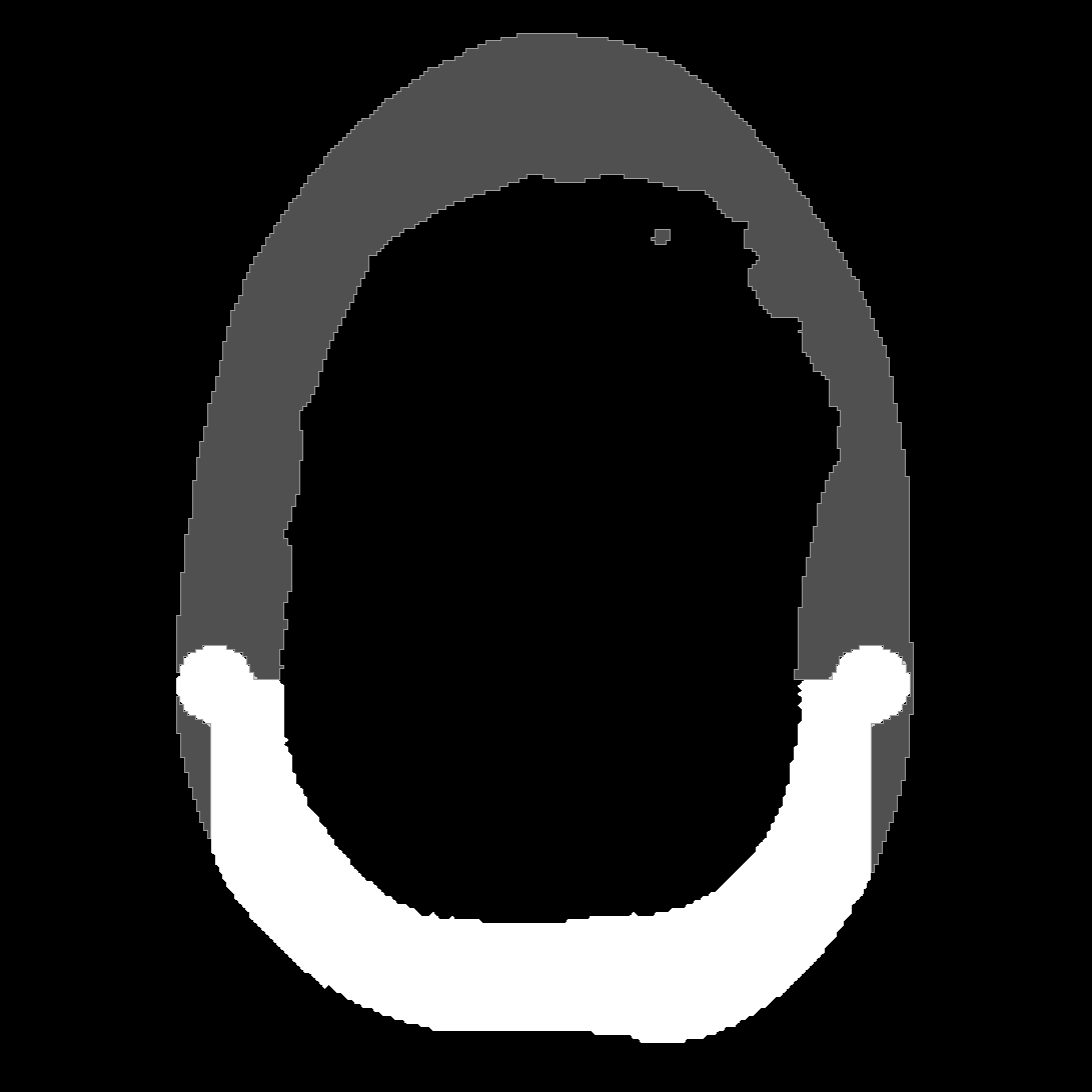}};
        \node[] at (2.5, -7.5)  {\includegraphics[height=0.2\textwidth]{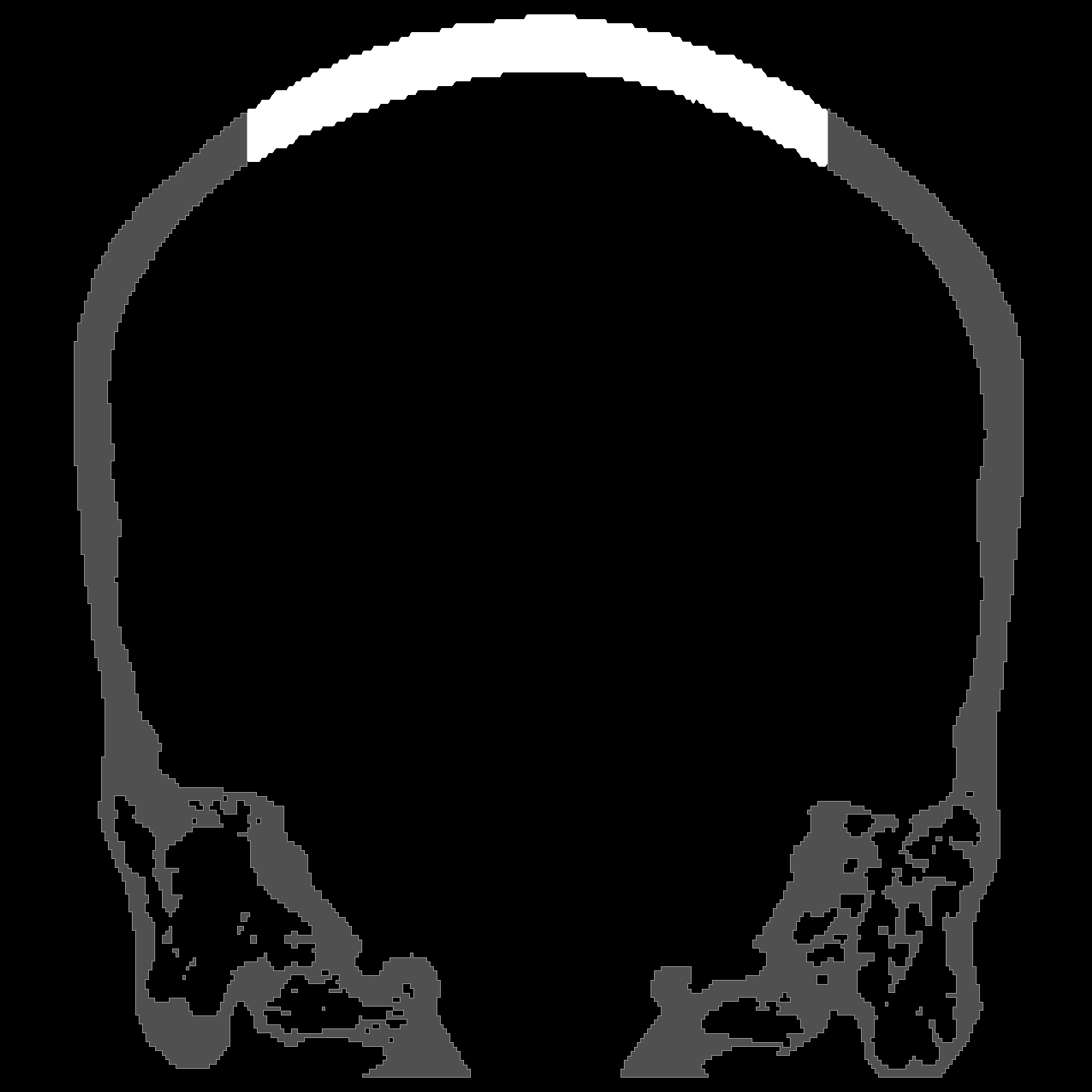}};
        \node[] at (5, -7.5)    {\includegraphics[height=0.2\textwidth]{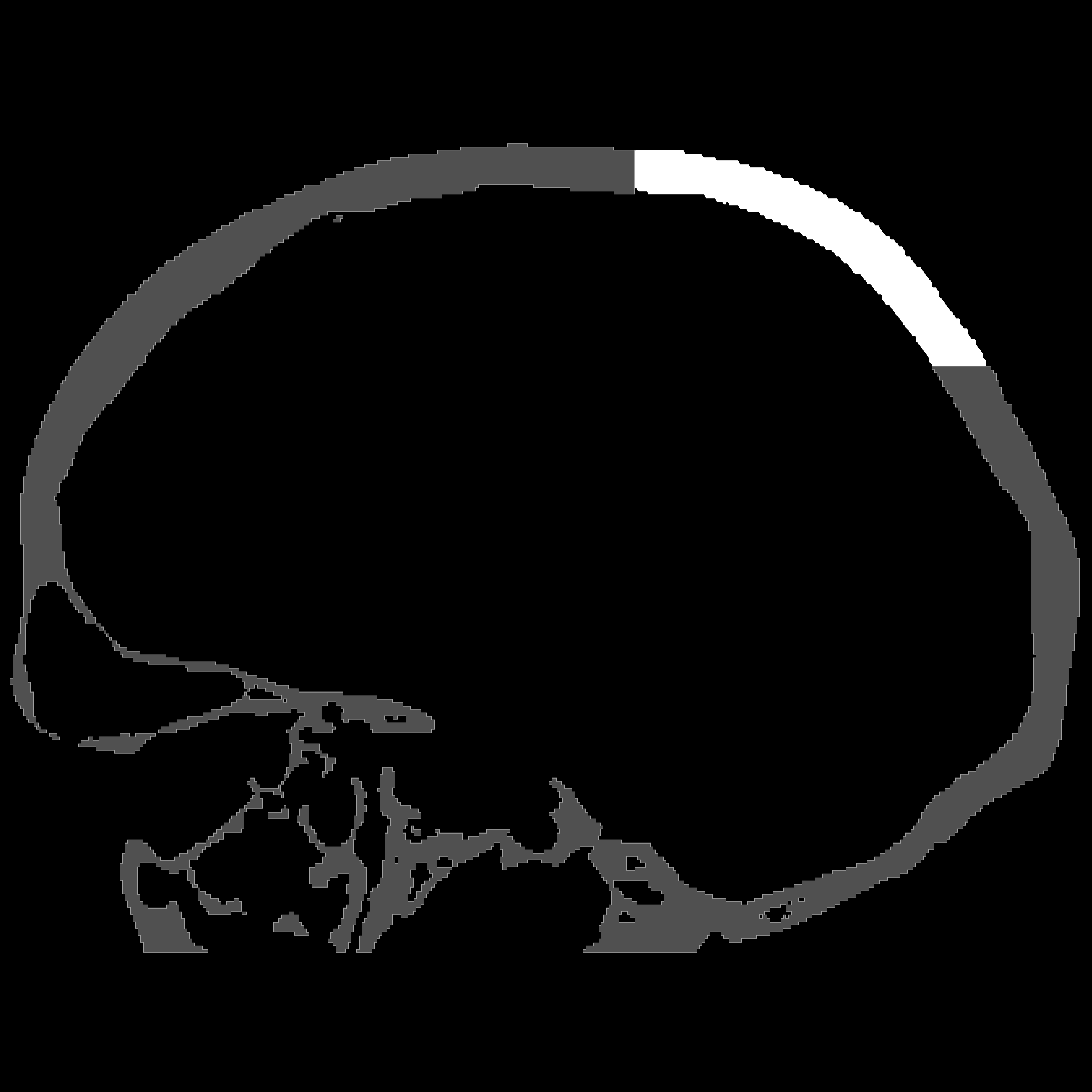}};
        \node[] at (7.5, -7.5)  {\includegraphics[height=0.2\textwidth]{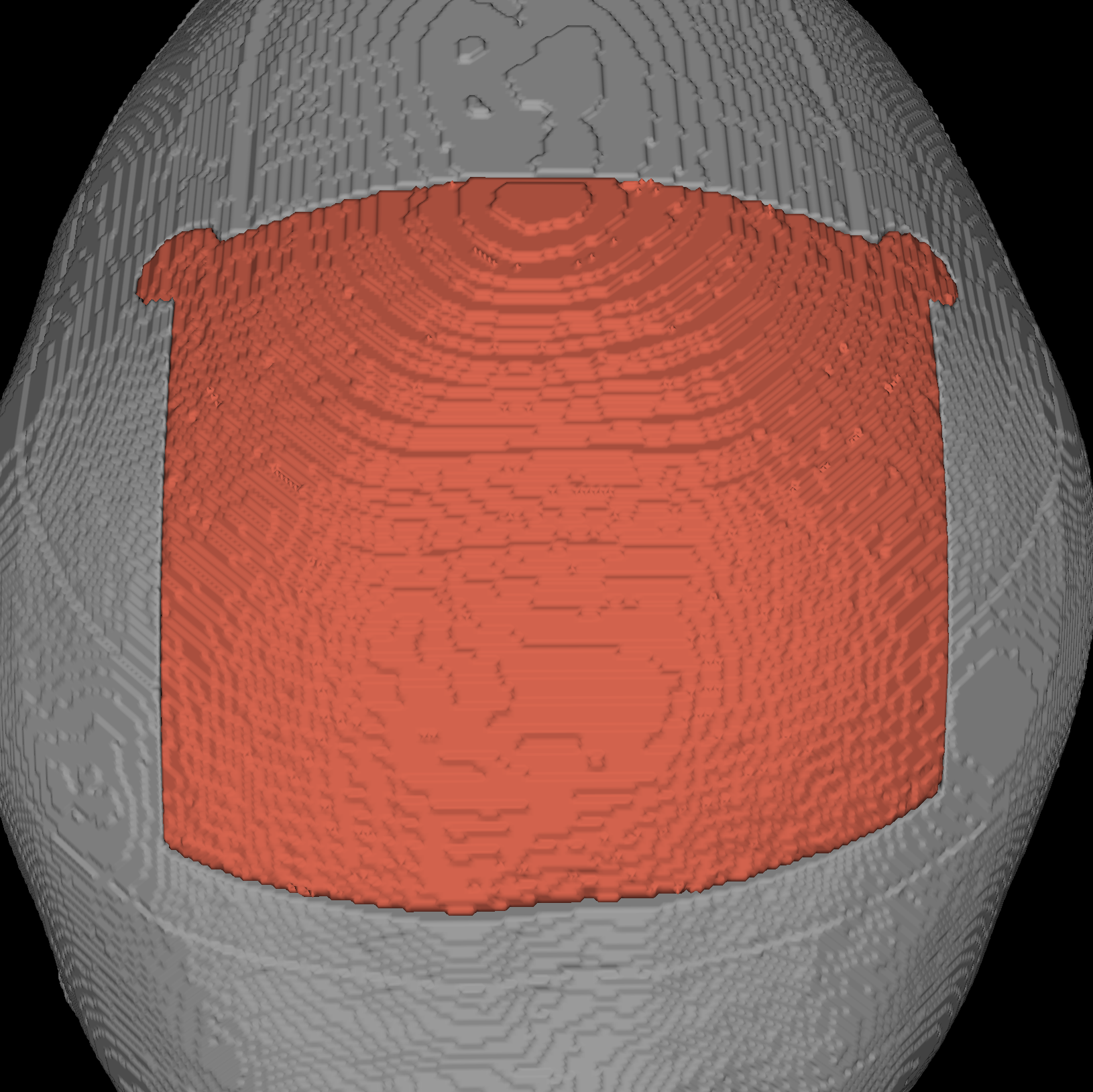}};
        \node[] at (0, -10)     {\includegraphics[height=0.2\textwidth]{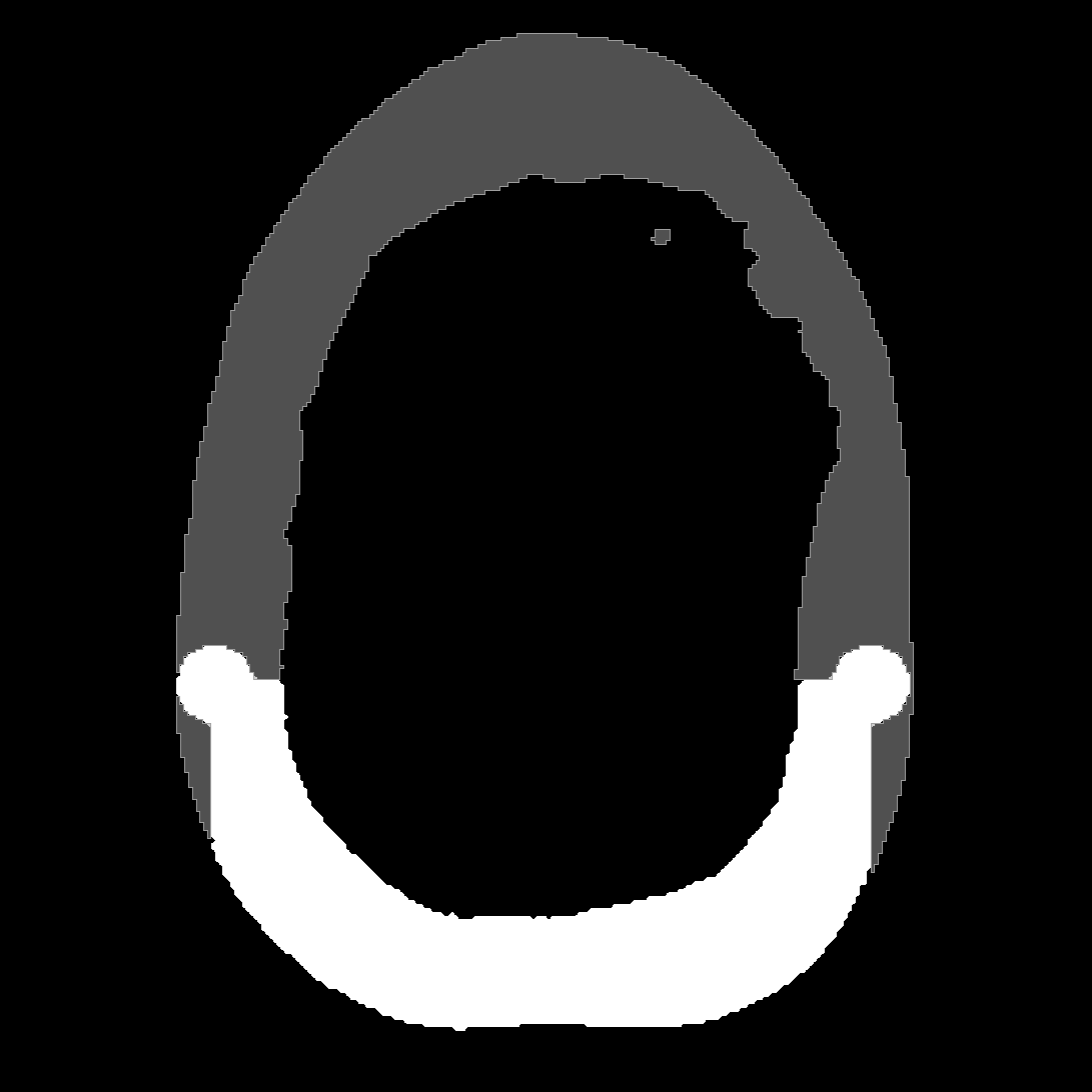}};
        \node[] at (2.5, -10)   {\includegraphics[height=0.2\textwidth]{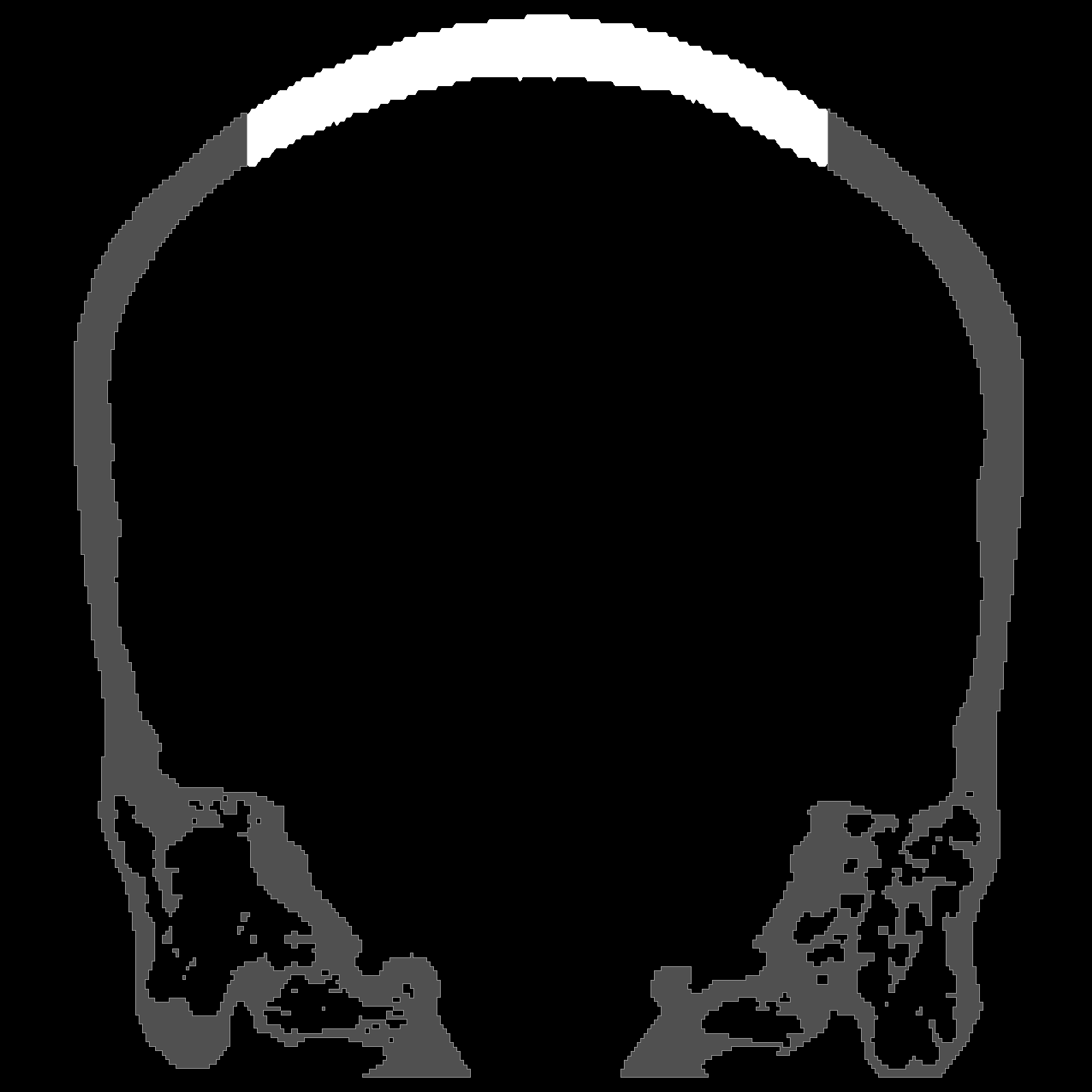}};
        \node[] at (5, -10)     {\includegraphics[height=0.2\textwidth]{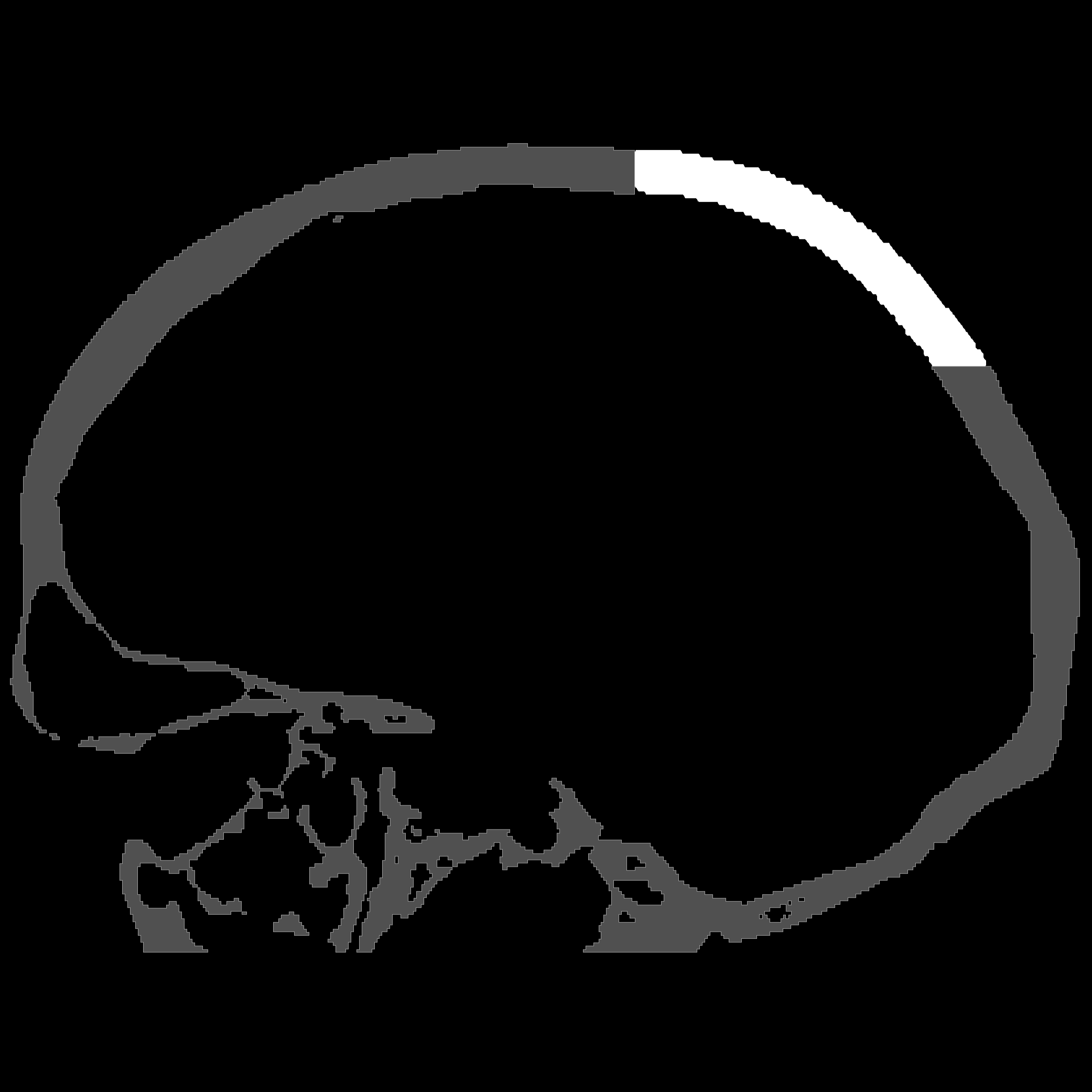}};
        \node[] at (7.5, -10)   {\includegraphics[height=0.2\textwidth]{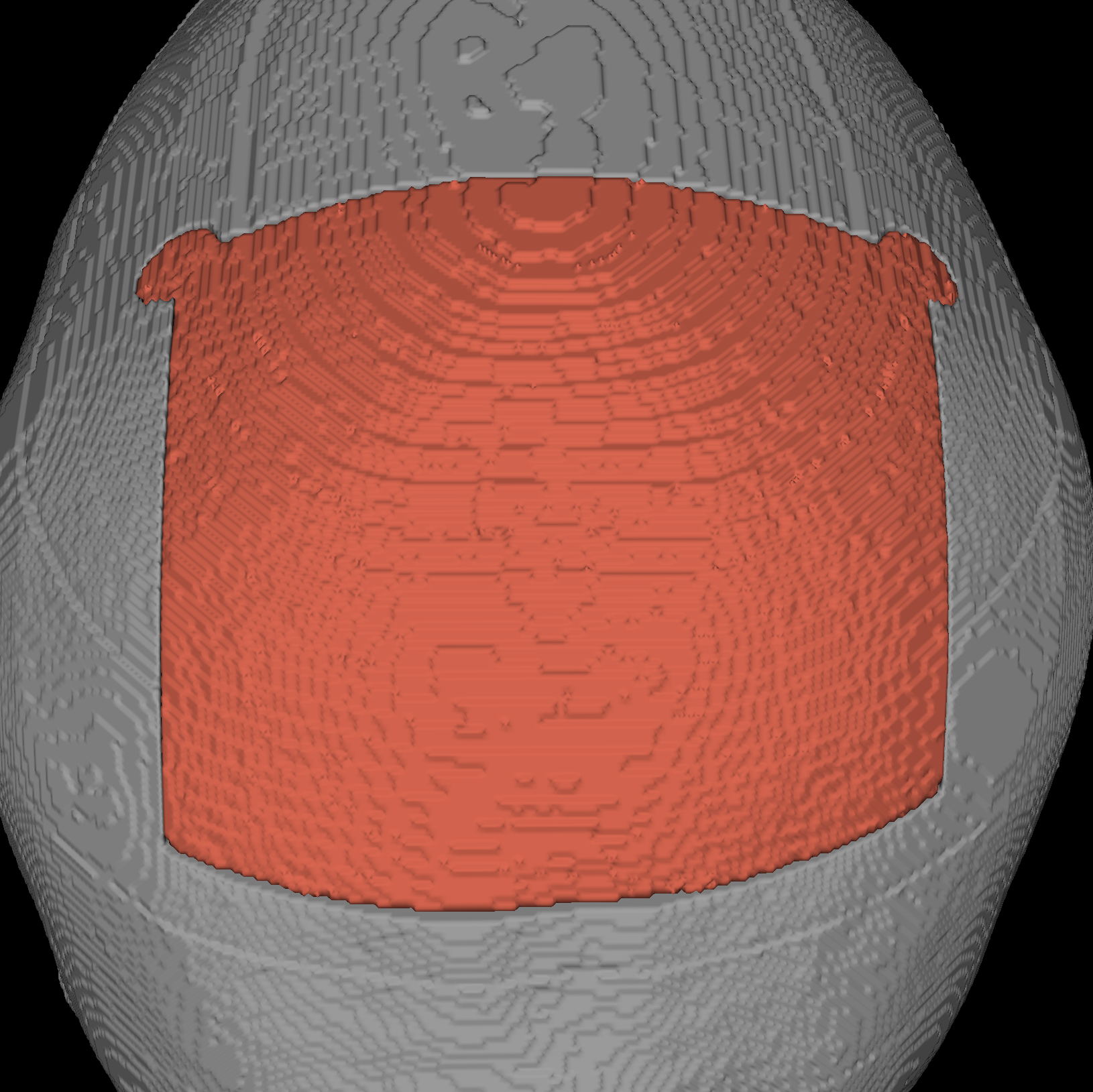}};
        \node[] at (0, -12.5)   {\includegraphics[height=0.2\textwidth]{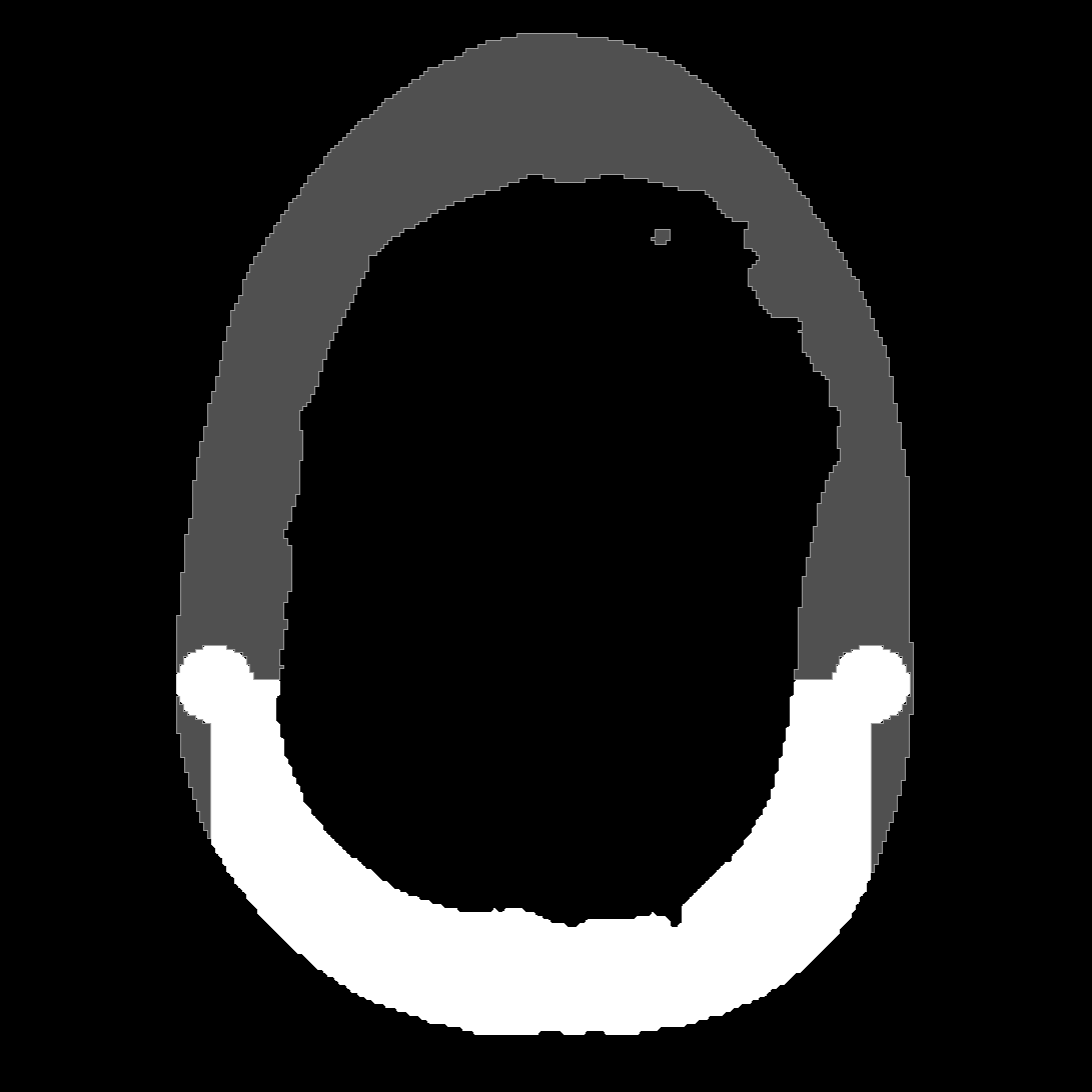}};
        \node[] at (2.5, -12.5) {\includegraphics[height=0.2\textwidth]{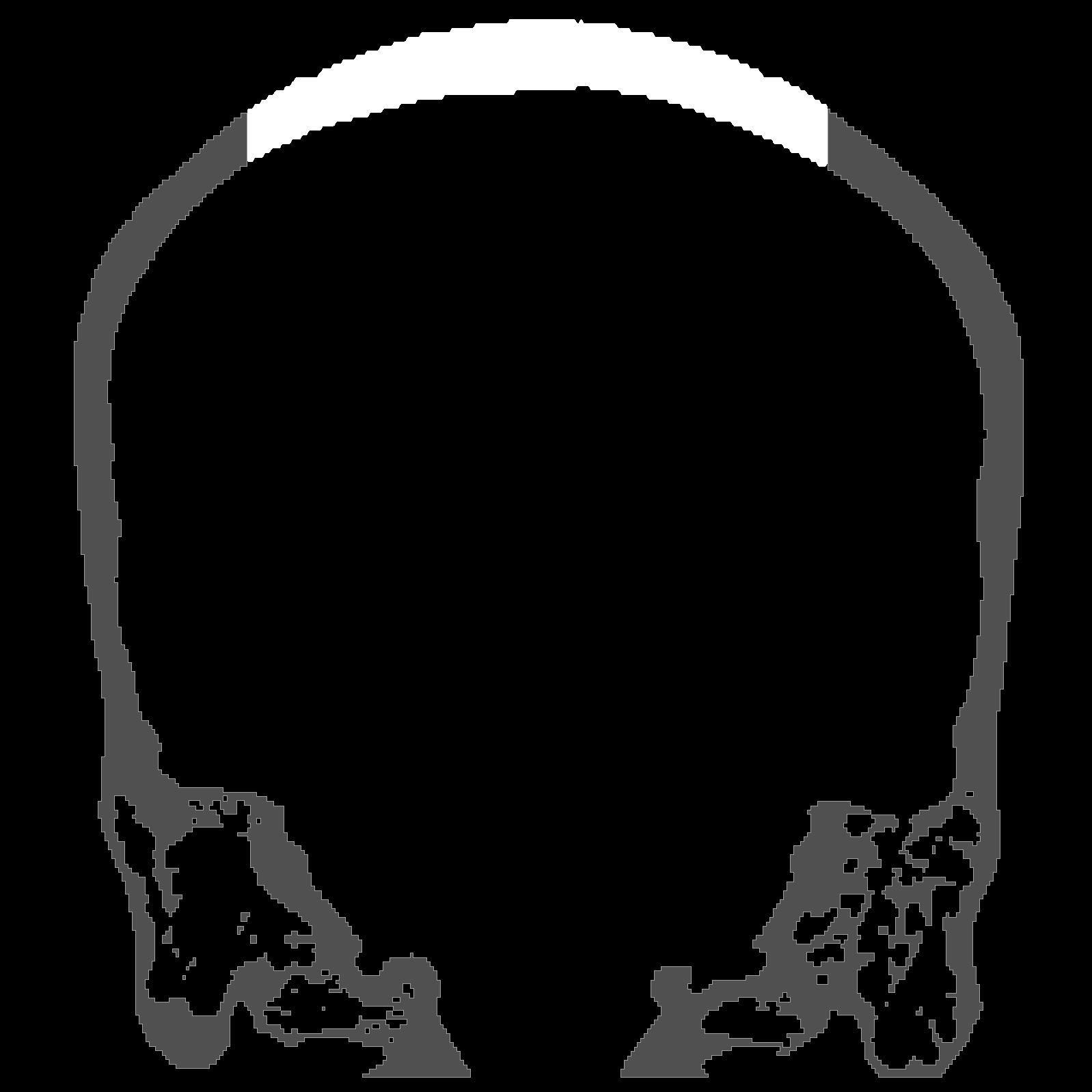}};
        \node[] at (5, -12.5)   {\includegraphics[height=0.2\textwidth]{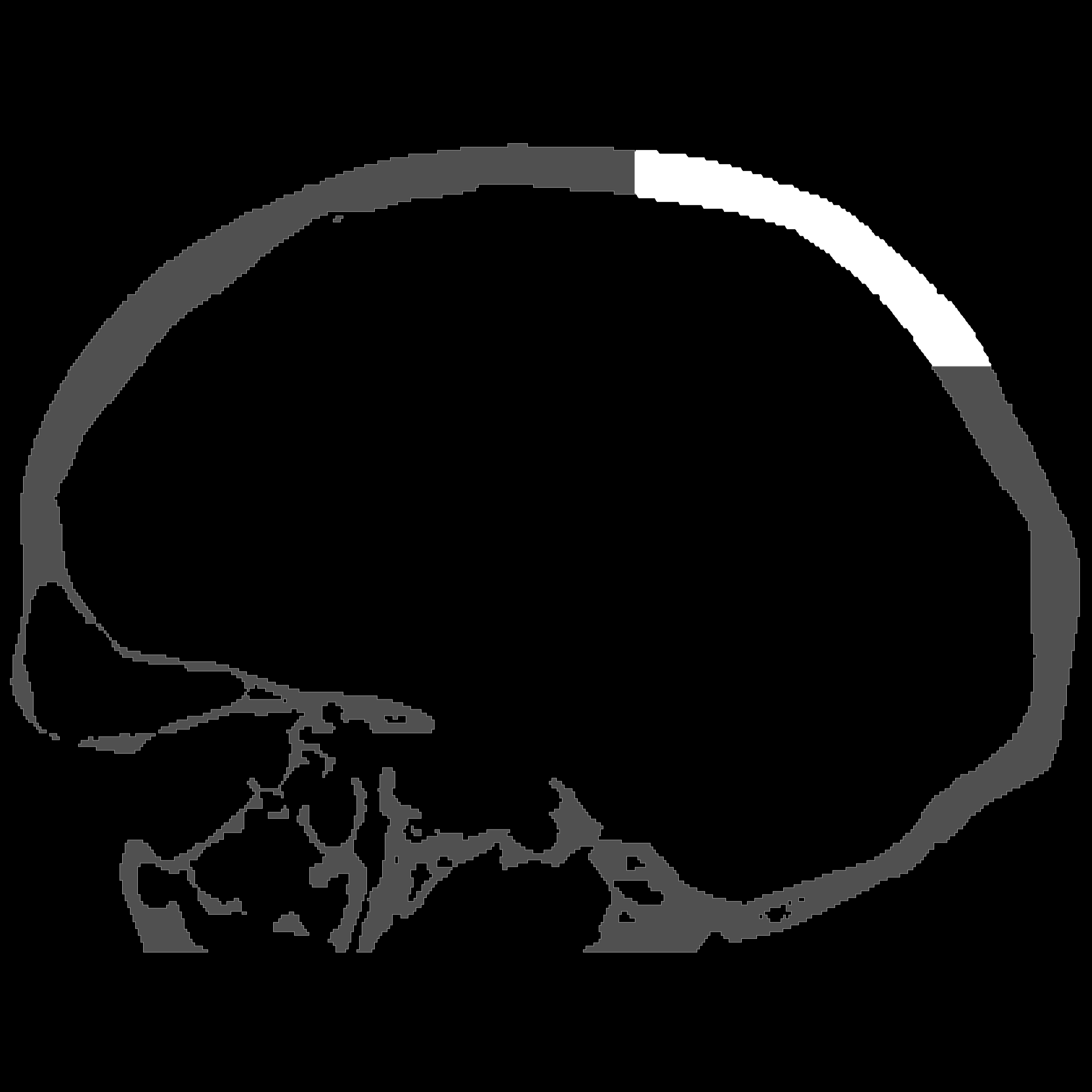}};
        \node[] at (7.5, -12.5) {\includegraphics[height=0.2\textwidth]{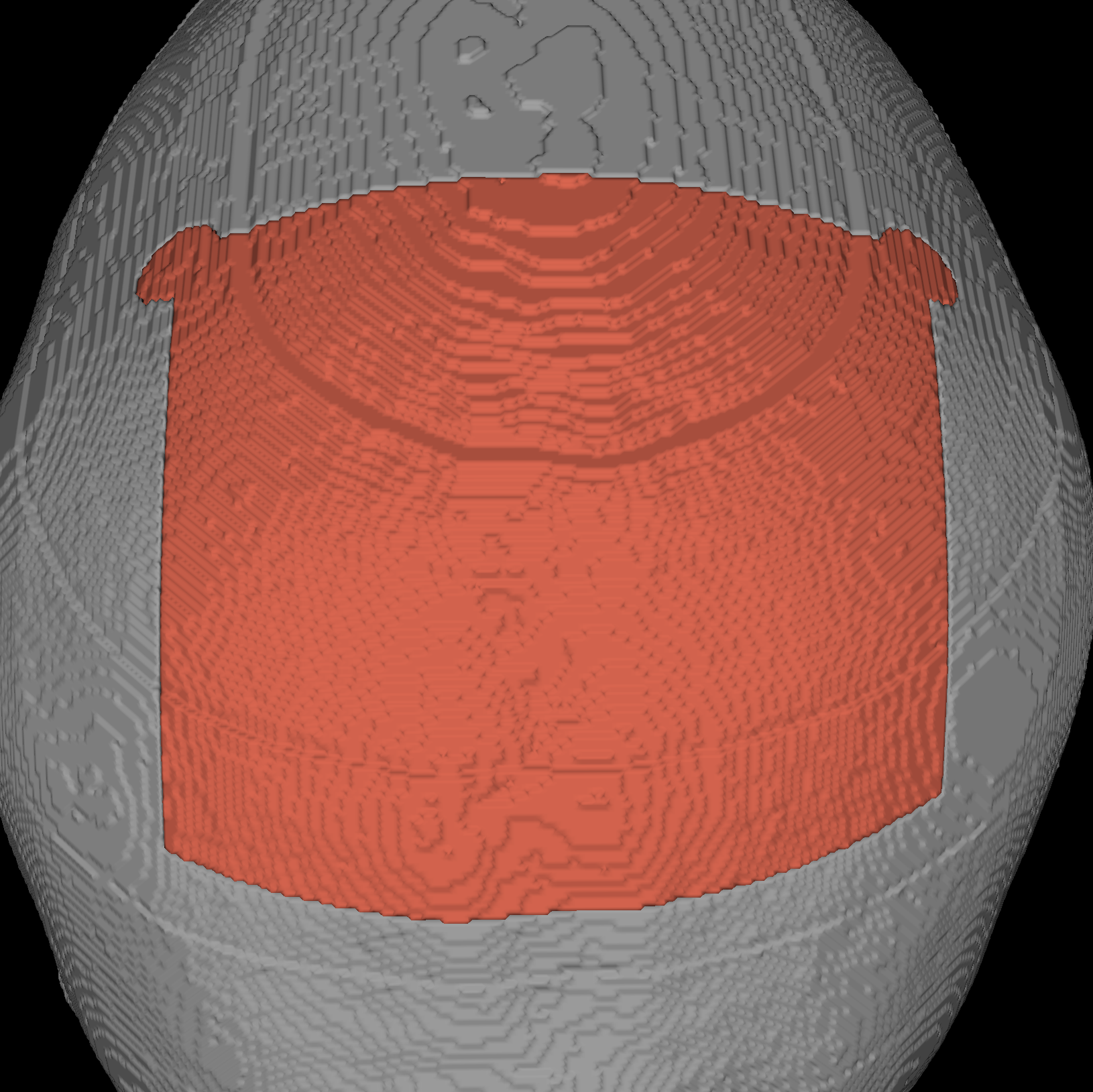}};
    	\node[rotate=90] at (-1.5, 0)       {\scriptsize 3D U-Net};
    	\node[rotate=90] at (-1.5, -2.5)    {\scriptsize 3D U-Net (sparse)};
    	\node[rotate=90] at (-1.5,-5)       {\scriptsize 2D U-Net};
    	\node[rotate=90] at (-1.5,-7.5)     {\scriptsize \textbf{Ours}};
    	\node[rotate=90] at (-1.5,-10)      {\scriptsize \textbf{Ours (n=5)}};
    	\node[rotate=90] at (-1.5,-12.5)    {\scriptsize Ground Truth};
	\end{tikzpicture}}
\caption{Predicted implants for a skull defect from the SkullFix dataset.}
\label{qualiresults}
\end{figure}
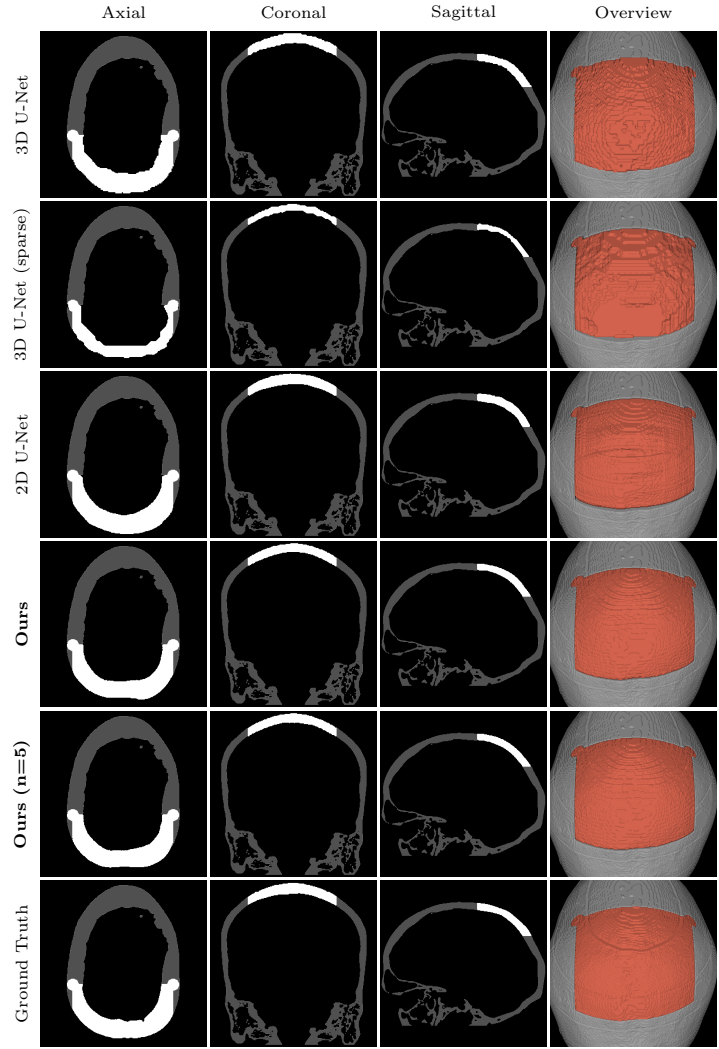
\noindent We outperformed the sparse 3D U-Net, while achieving comparable results to the challenge winning 3D U-Net and the 2D U-Net based approach. By visually comparing the results in Figure \ref{qualiresults}, our method produces significantly smoother surfaces that are more similar to the ground truth implant. In Figure \ref{SkullBreak}, we show that our method reliably reconstructs defects of various sizes, as well as complicated geometric structures. For large defects, however, we achieve lower evaluation scores, which can be explained with multiple anatomically reasonable implant solutions. This was also reported in \cite{Wodzinski2020} and \cite{Yu2021}.
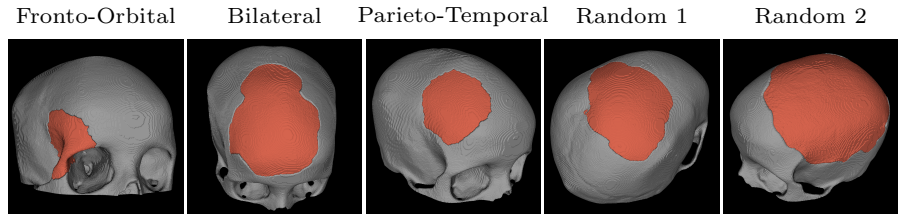
\begin{figure}[H]
\centering
\resizebox{\textwidth}{!}{
	\begin{tikzpicture}
	    \node[] at (0, 0)     {\includegraphics[height=0.16\textwidth]{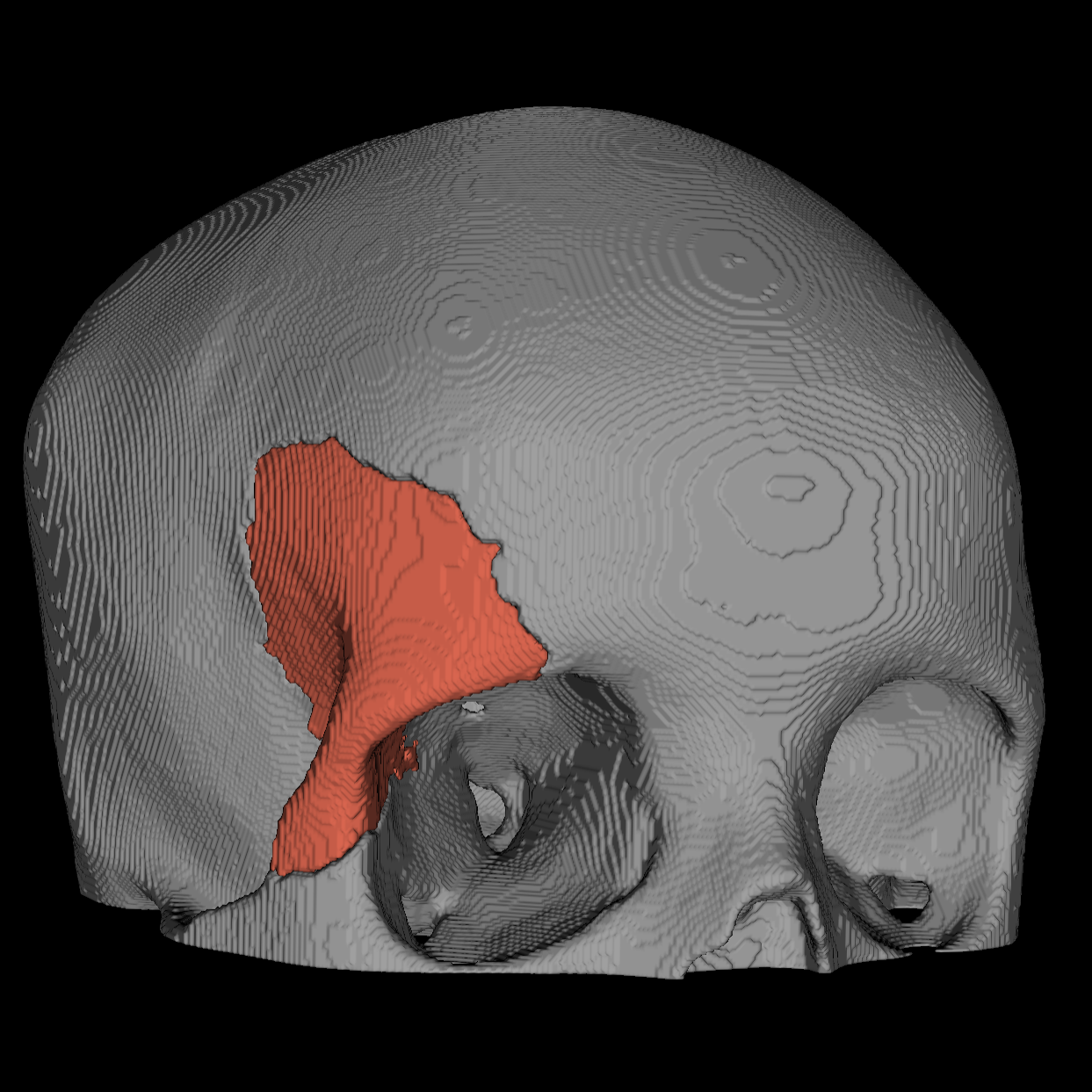}};
    	\node[] at (0, 1.25)  {\scriptsize Fronto-Orbital};
    	\node[] at (6, 0)     {\includegraphics[height=0.16\textwidth]{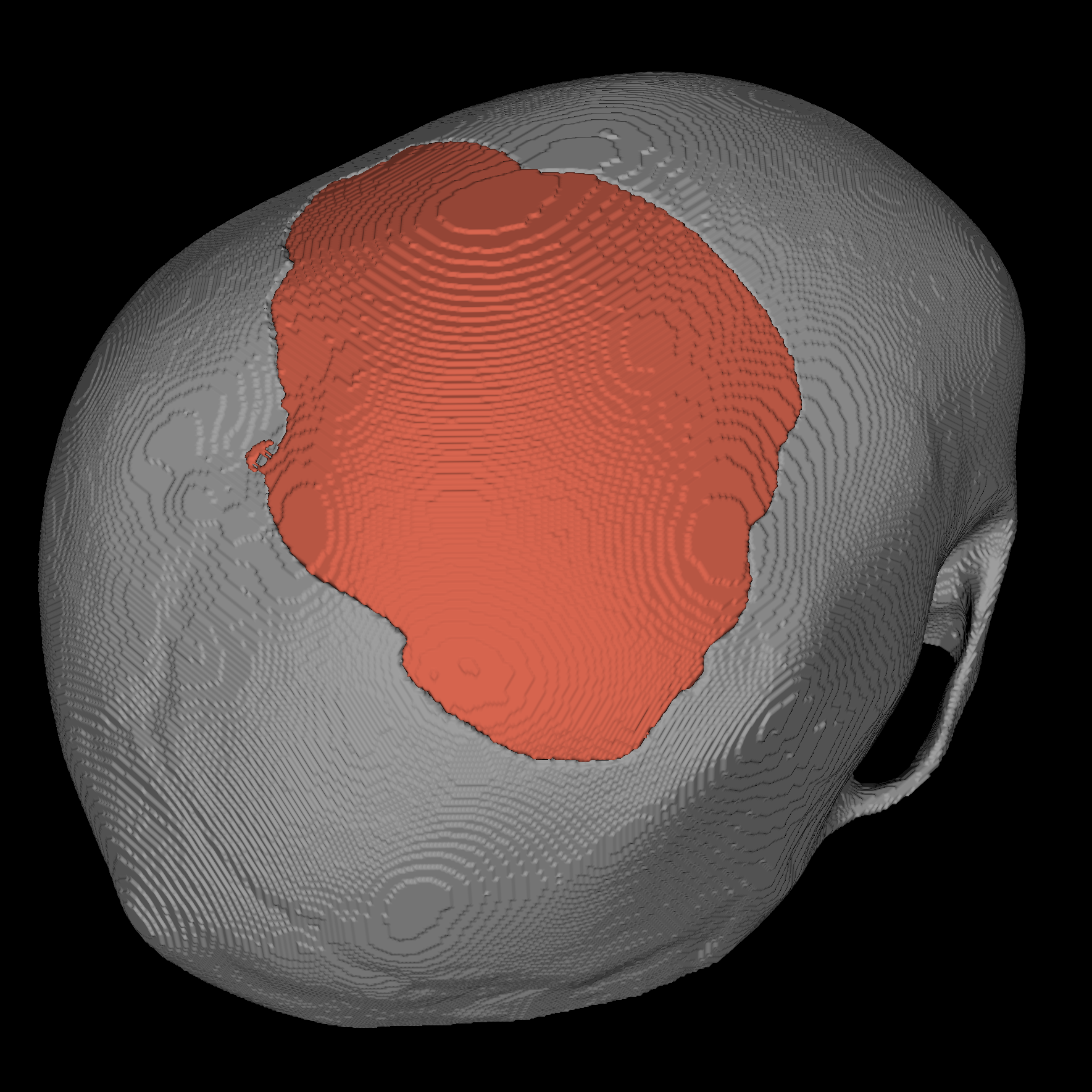}};
    	\node[] at (6, 1.25)  {\scriptsize Random 1};
    	\node[] at (4, 0)     {\includegraphics[height=0.16\textwidth]{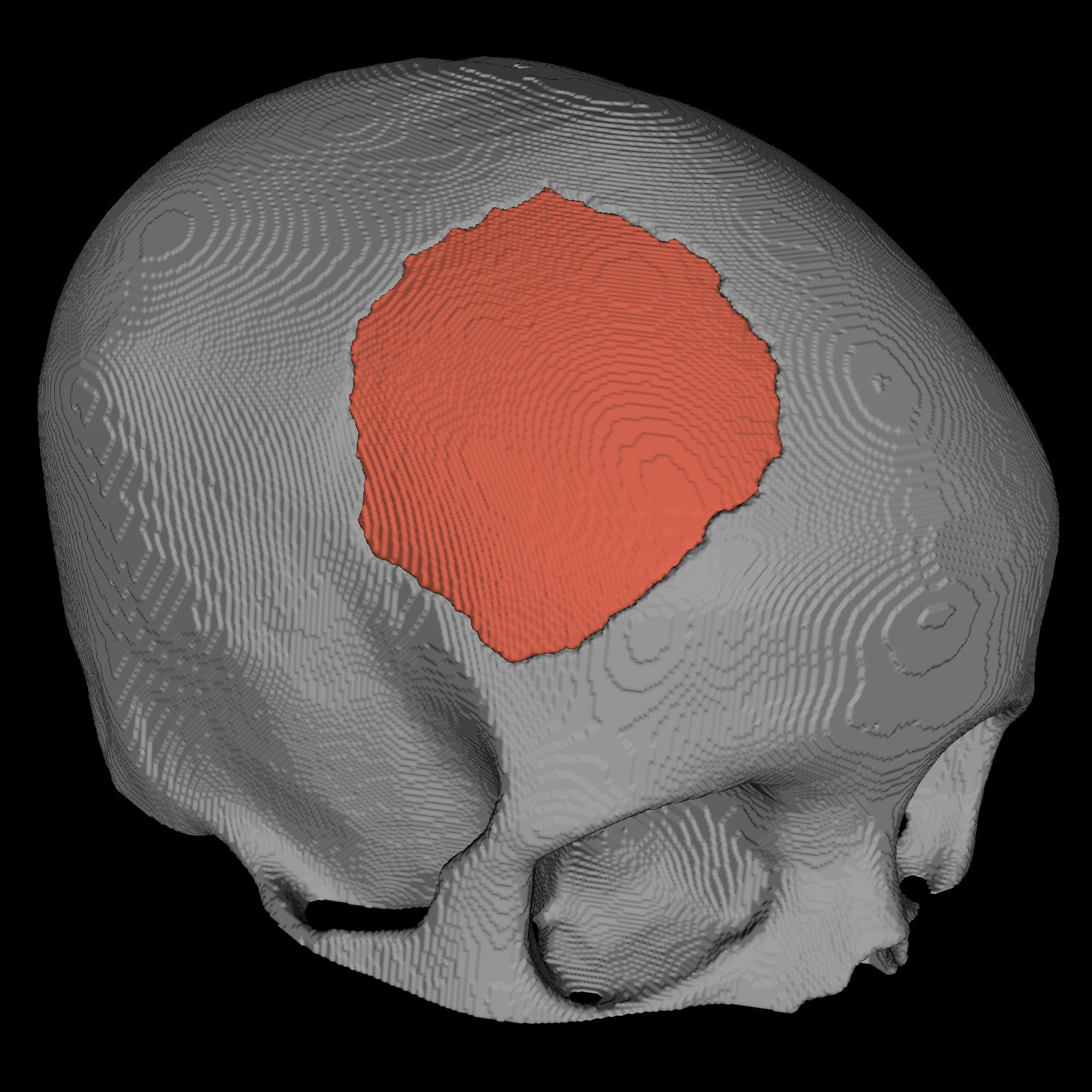}};
    	\node[] at (4, 1.23)  {\scriptsize Parieto-Temporal};
    	\node[] at (2, 0)     {\includegraphics[height=0.16\textwidth]{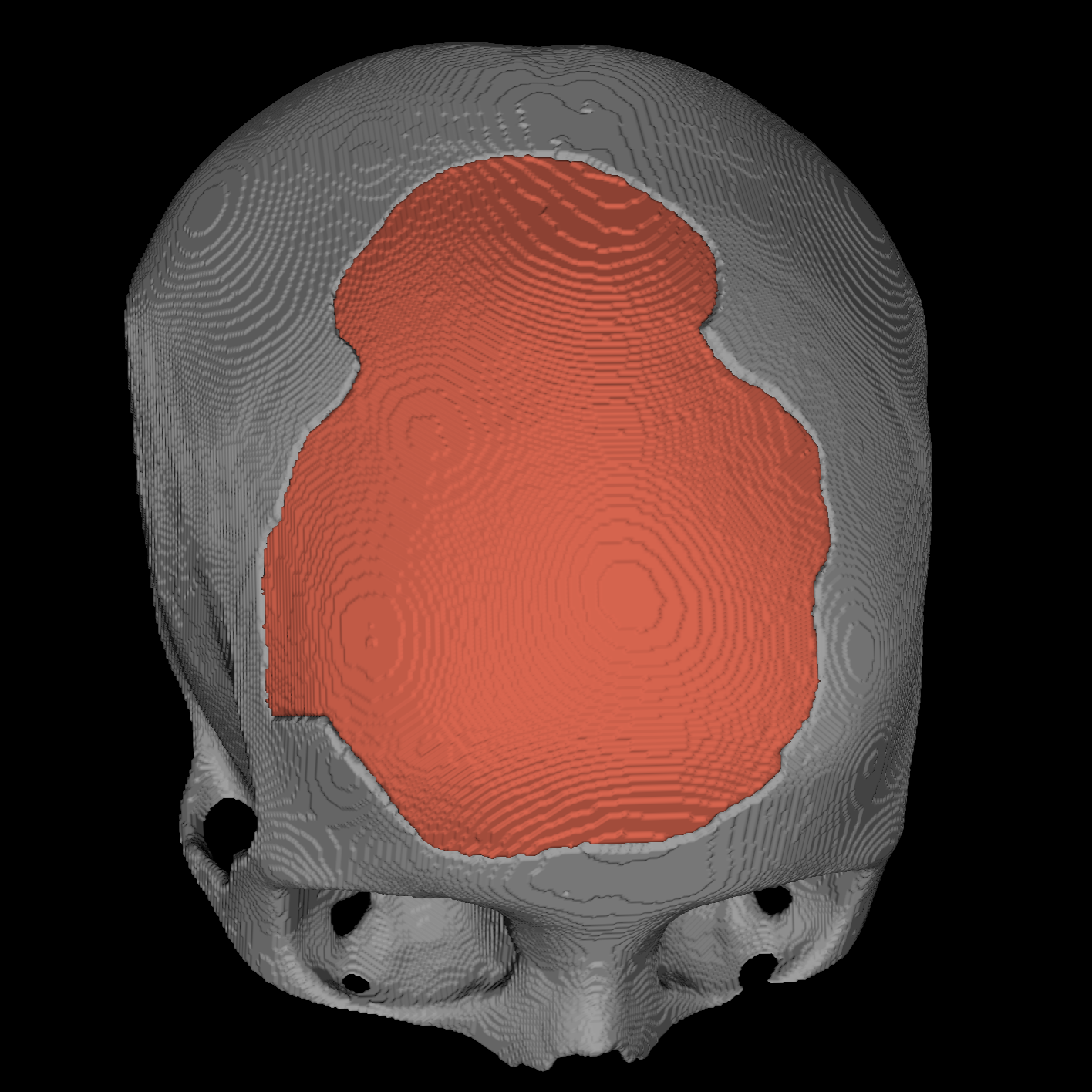}};
    	\node[] at (2, 1.25)  {\scriptsize Bilateral};
    	\node[] at (8, 0)     {\includegraphics[height=0.16\textwidth]{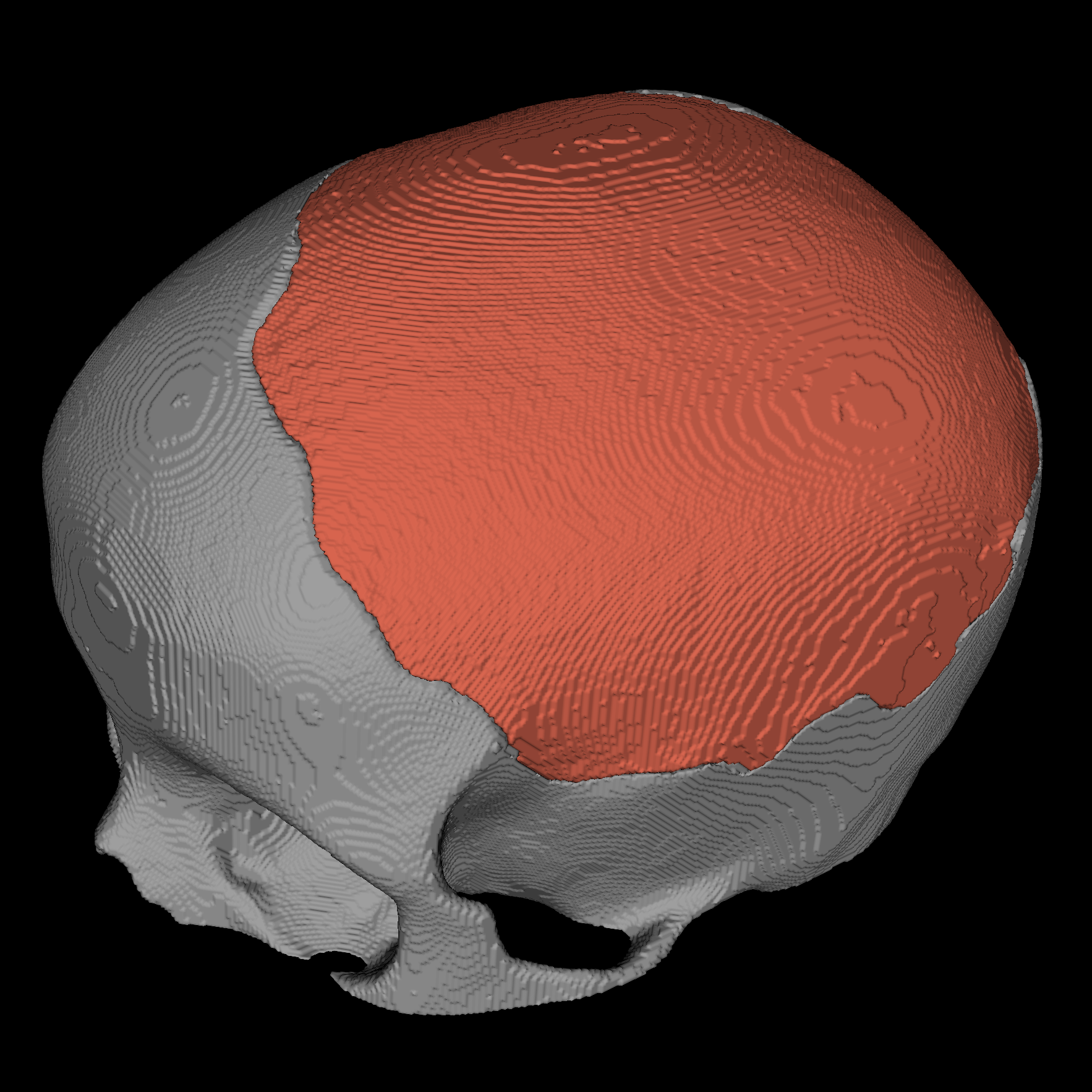}};
    	\node[] at (8, 1.25)  {\scriptsize Random 2};
	\end{tikzpicture}}
\caption{Results of our method for the five defect classes of the SkullBreak dataset.}
\label{SkullBreak}
\end{figure}
\noindent In Figure \ref{Ensembling}, we present the proposed ensembling strategy for an exemplary defect. Apart from generating multiple implants, we can compute the mean implant and the variance map. To the best of our knowledge, we are the first to combine such an approach with an automatic implant generation method. Not only do we offer a choice of implants, but we can also provide physicians with information about implant areas with more anatomical variation.
\section{Conclusion}
We present a novel approach for automatic implant generation based on a combination of a point cloud diffusion model and a voxelization network. Due to the sparse point cloud representation of the anatomical structure, the proposed approach can directly handle high resolution input images without losing context information. We achieve competitive evaluation scores, while producing smoother, more realistic surfaces. Furthermore, our method is capable of producing different implants per defect, accounting for the anatomical variation seen in the training population. Thereby, we can propose several solutions to the physicians, from which they can choose the most suitable one. For future work, we plan to speed up the sampling process by using different sampling schemes, as proposed in \cite{Karras2022,Song2020}.
\subsubsection{Acknowledgments} This work was financially supported by the Werner Siemens Foundation through the MIRACLE II project.
\bibliographystyle{splncs04.bst}
\bibliography{bibliography.bib}

\begin{thebibliography}{10}
\providecommand{\url}[1]{\texttt{#1}}
\providecommand{\urlprefix}{URL }
\providecommand{\doi}[1]{https://doi.org/#1}

\bibitem{Gropp2020}
Gropp, A., Yariv, L., Haim, N., Atzmon, M., Lipman, Y.: Implicit geometric
  regularization for learning shapes. In: Proceedings of Machine Learning and
  Systems. pp. 3569--3579 (2020)

\bibitem{Hanocka2020}
Hanocka, R., Metzer, G., Giryes, R., Cohen-Or, D.: Point2mesh: A self-prior for
  deformable meshes. ACM Transactions on Graphics  \textbf{39} (2020)

\bibitem{Ho2020}
Ho, J., Jain, A., Abbeel, P.: Denoising diffusion probabilistic models. In:
  Advances in Neural Information Processing Systems (2020)

\bibitem{Jin2020}
Jin, Y., Li, J., Egger, J.: High-resolution cranial implant prediction via
  patch-wise training. In: Towards the Automatization of Cranial Implant Design
  in Cranioplasty. pp. 94--103 (2020)

\bibitem{Karras2022}
Karras, T., Aittala, M., Aila, T., Laine, S.: Elucidating the design space of
  diffusion-based generative models. In: Advances in Neural Information
  Processing Systems (2022)

\bibitem{Kazdhan2006}
Kazhdan, M., Bolitho, M., Hoppe, H.: Poisson surface reconstruction. In:
  Eurographics Symposium on Geometry Processing (2006)

\bibitem{Kazhdan2013}
Kazhdan, M., Hoppe, H.: Screened poisson surface reconstruction. ACM
  Transactions on Graphics  \textbf{32},  1--13 (2013)

\bibitem{Kodym2021}
Kodym, O., et~al.: Skullbreak / skullfix – dataset for automatic cranial
  implant design and a benchmark for volumetric shape learning tasks. Data in
  Brief  \textbf{35},  106902 (2021)

\bibitem{Kodym2020}
Kodym, O., Španěl, M., Herout, A.: Cranial defect reconstruction using
  cascaded cnn with alignment. In: Towards the Automatization of Cranial
  Implant Design in Cranioplasty. pp. 56--64 (2020)

\bibitem{Kroviakov2020}
Kroviakov, A., Li, J., Egger, J.: Sparse convolutional neural network for skull
  reconstruction. In: Towards the Automatization of Cranial Implant Design in
  Cranioplasty II. pp. 80--94 (2021)

\bibitem{Li2020Towards}
Li, J., Egger, J. (eds.): Towards the Automatization of Cranial Implant Design
  in Cranioplasty, vol. 12439. Springer International Publishing (2020)

\bibitem{Li2021Towards}
Li, J., Egger, J. (eds.): Towards the Automatization of Cranial Implant Design
  in Cranioplasty II, vol. 13123. Springer International Publishing (2021)

\bibitem{Li2021Summary}
Li, J., et~al.: Autoimplant 2020-first miccai challenge on automatic cranial
  implant design. IEEE Transactions on Medical Imaging  \textbf{40}(9),
  2329--2342 (2021)

\bibitem{Li2021}
Li, J., et~al.: Automatic skull defect restoration and cranial implant
  generation for cranioplasty. Medical Image Analysis  \textbf{73},  102171
  (2021)

\bibitem{Liu2019}
Liu, Z., Tang, H., Lin, Y., Han, S.: Point-voxel cnn for efficient 3d deep
  learning. In: Advances in Neural Information Processing Systems (2019)

\bibitem{Lorensen1987}
Lorensen, W.E., Cline, H.E.: Marching cubes: A high resolution 3d surface
  construction algorithm. ACM SIGGRAPH Computer Graphics  \textbf{21},
  163--169 (1987)

\bibitem{Lugmayr2022}
Lugmayr, A., Danelljan, M., Romero, A., Yu, F., Timofte, R., Gool, L.V.:
  Repaint: Inpainting using denoising diffusion probabilistic models. In:
  Proceedings of the IEEE/CVF Conference on Computer Vision and Pattern
  Recognition (2022)

\bibitem{Luo2021}
Luo, S., Hu, W.: Diffusion probabilistic models for 3d point cloud generation.
  In: Proceedings of the IEEE/CVF Conference on Computer Vision and Pattern
  Recognition (2021)

\bibitem{Lyu2021}
Lyu, Z., Kong, Z., Xu, X., Pan, L., Lin, D.: A conditional point
  diffusion-refinement paradigm for 3d point cloud completion. In:
  International Conference on Learning Representations (2022)

\bibitem{Matzkin2020}
Matzkin, F., Newcombe, V., Glocker, B., Ferrante, E.: Cranial implant design
  via virtual craniectomy with shape priors. In: Towards the Automatization of
  Cranial Implant Design in Cranioplasty. pp. 37--46 (2020)

\bibitem{Nichol2022}
Nichol, A., Jun, H., Dhariwal, P., Mishkin, P., Chen, M.: Point-e: A system for
  generating 3d point clouds from complex prompts. arXiv preprint
  arXiv:2212.08751  (2022)

\bibitem{Pathak2021}
Pathak, S., Sindhura, C., Gorthi, R.K.S.S., Kiran, D.V., Gorthi, S.: Cranial
  implant design using v-net based region of interest reconstruction. In:
  Towards the Automatization of Cranial Implant Design in Cranioplasty II. pp.
  116--128 (2021)

\bibitem{Peng2021}
Peng, S., Jiang, C.M., Liao, Y., Niemeyer, M., Pollefeys, M., Geiger, A.: Shape
  as points: A differentiable poisson solver. In: Advances in Neural
  Information Processing Systems (2021)

\bibitem{Pimentel2020}
Pimentel, P., et~al.: Automated virtual reconstruction of large skull defects
  using statistical shape models and generative adversarial networks. In:
  Towards the Automatization of Cranial Implant Design in Cranioplasty. pp.
  16--27 (2020)

\bibitem{Qi2017}
Qi, C.R., Yi, L., Su, H., Guibas, L.J.: Pointnet++: Deep hierarchical feature
  learning on point sets in a metric space. In: Advances in Neural Information
  Processing Systems (2017)

\bibitem{Saharia2021}
Saharia, C., Chan, W., Chang, H., Lee, C.A., Ho, J., Salimans, T., Fleet, D.J.,
  Norouzi, M.: Palette: Image-to-image diffusion models. In: ACM SIGGRAPH 2022
  Conference Proceedings (2021)

\bibitem{Shi2020}
Shi, H., Chen, X.: Cranial implant design through multiaxial slice inpainting
  using deep learning. In: Towards the Automatization of Cranial Implant Design
  in Cranioplasty. pp. 28--36 (2020)

\bibitem{Sohl2015}
Sohl-Dickstein, J., Weiss, E.A., Maheswaranathan, N., Ganguli, S.: Deep
  unsupervised learning using nonequilibrium thermodynamics. In: Proceedings of
  the 32nd International Conference on Machine Learning (2015)

\bibitem{Song2020}
Song, J., Meng, C., Ermon, S.: Denoising diffusion implicit models. In:
  International Conference on Learning Representations (2020)

\bibitem{Wang2020}
Wang, B., et~al.: Cranial implant design using a deep learning method with
  anatomical regularization. In: Towards the Automatization of Cranial Implant
  Design in Cranioplasty. pp. 85--93 (2020)

\bibitem{Wodzinski2020}
Wodzinski, M., Daniol, M., Hemmerling, D.: Improving the automatic cranial
  implant design in cranioplasty by linking different datasets. In: Towards the
  Automatization of Cranial Implant Design in Cranioplasty II. pp. 29--44
  (2021)

\bibitem{Wolleb2021}
Wolleb, J., Sandkühler, R., Bieder, F., Valmaggia, P., Cattin, P.C.: Diffusion
  models for implicit image segmentation ensembles. In: Proceedings of Machine
  Learning Research. pp. 1--13 (2022)

\bibitem{Yang2021}
Yang, B., Fang, K., Li, X.: Cranial implant prediction by learning an ensemble
  of slice-based skull completion networks. In: Towards the Automatization of
  Cranial Implant Design in Cranioplasty II. pp. 95--104 (2021)

\bibitem{Yu2021}
Yu, L., Li, J., Egger, J.: Pca-skull: 3d skull shape modelling using principal
  component analysis. In: Towards the Automatization of Cranial Implant Design
  in Cranioplasty II. pp. 105--115 (2021)

\bibitem{Yuksel2015}
Yuksel, C.: Sample elimination for generating poisson disk sample sets.
  Computer Graphics Forum  \textbf{34},  25--32 (2015)

\bibitem{Zeng2022}
Zeng, X., et~al.: Lion: Latent point diffusion models for 3d shape generation.
  In: Advances in Neural Information Processing Systems (2022)

\bibitem{Zhou2021}
Zhou, L., Du, Y., Wu, J.: 3d shape generation and completion through
  point-voxel diffusion. In: Proceedings of the IEEE/CVF International
  Conference on Computer Vision. pp. 5826--5835 (2021)

\end{thebibliography}
\end{document}


\title{Supplementary Material}
\author{}
\institute{}
\maketitle
\section{Additional Results on the SkullBreak Dataset}
\begin{figure}[h]
\centering
\resizebox{\textwidth}{!}{
	\begin{tikzpicture}
	    \node[] at (0, 0)       {\includegraphics[height=0.16\textwidth]{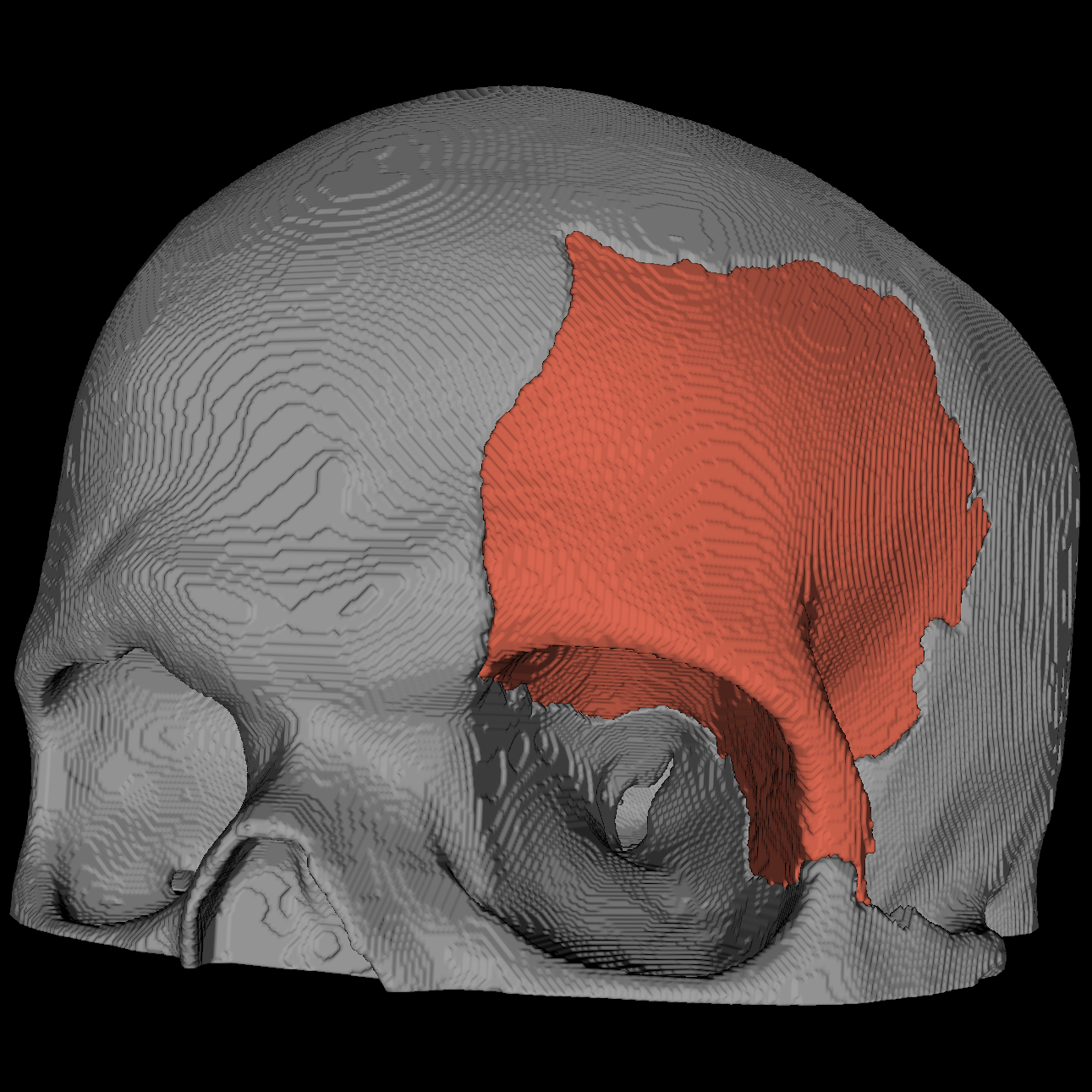}};
    	\node[] at (0, 1.25)    {\scriptsize Fronto-Orbital};
    	\node[] at (2, 0)       {\includegraphics[height=0.16\textwidth]{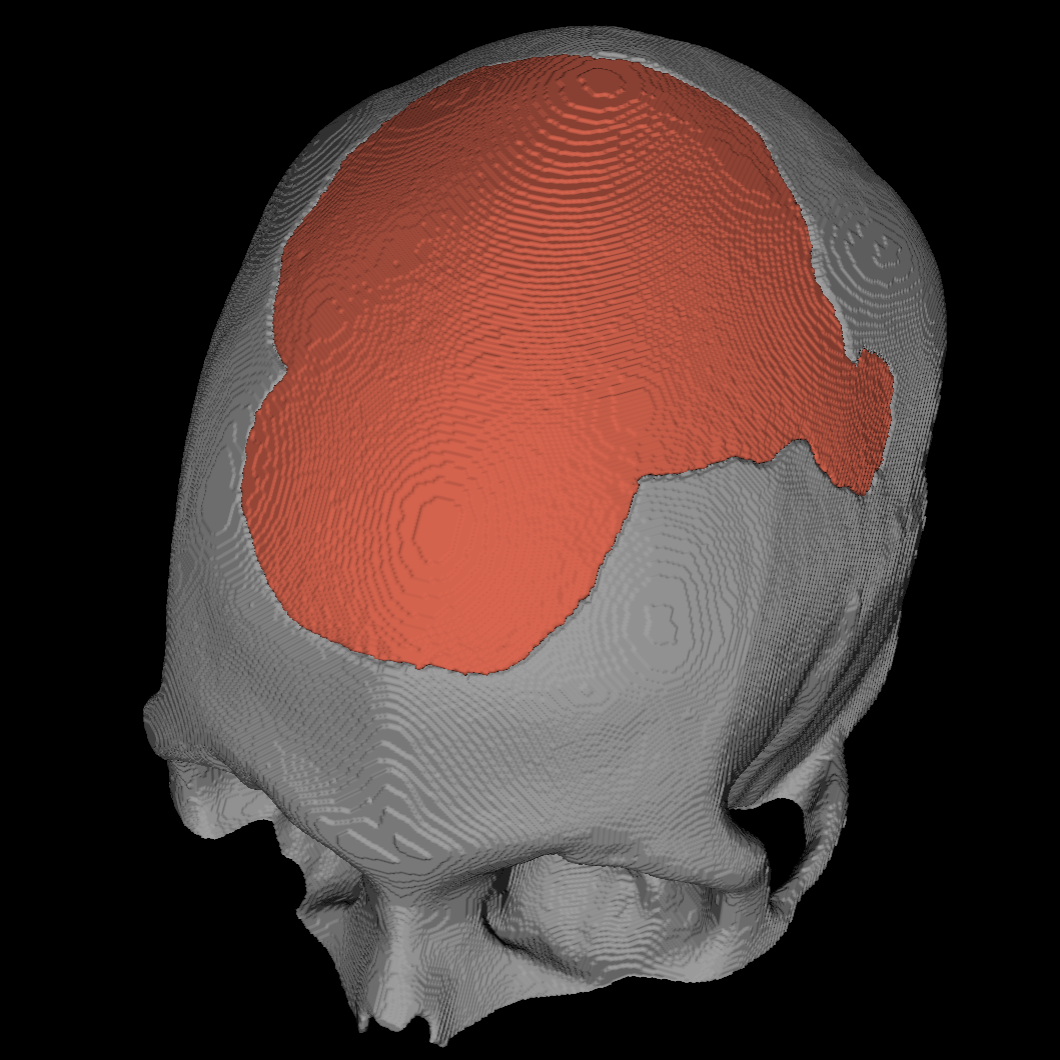}};
    	\node[] at (2, 1.25)    {\scriptsize Bilateral};
    	\node[] at (4, 0)       {\includegraphics[height=0.16\textwidth]{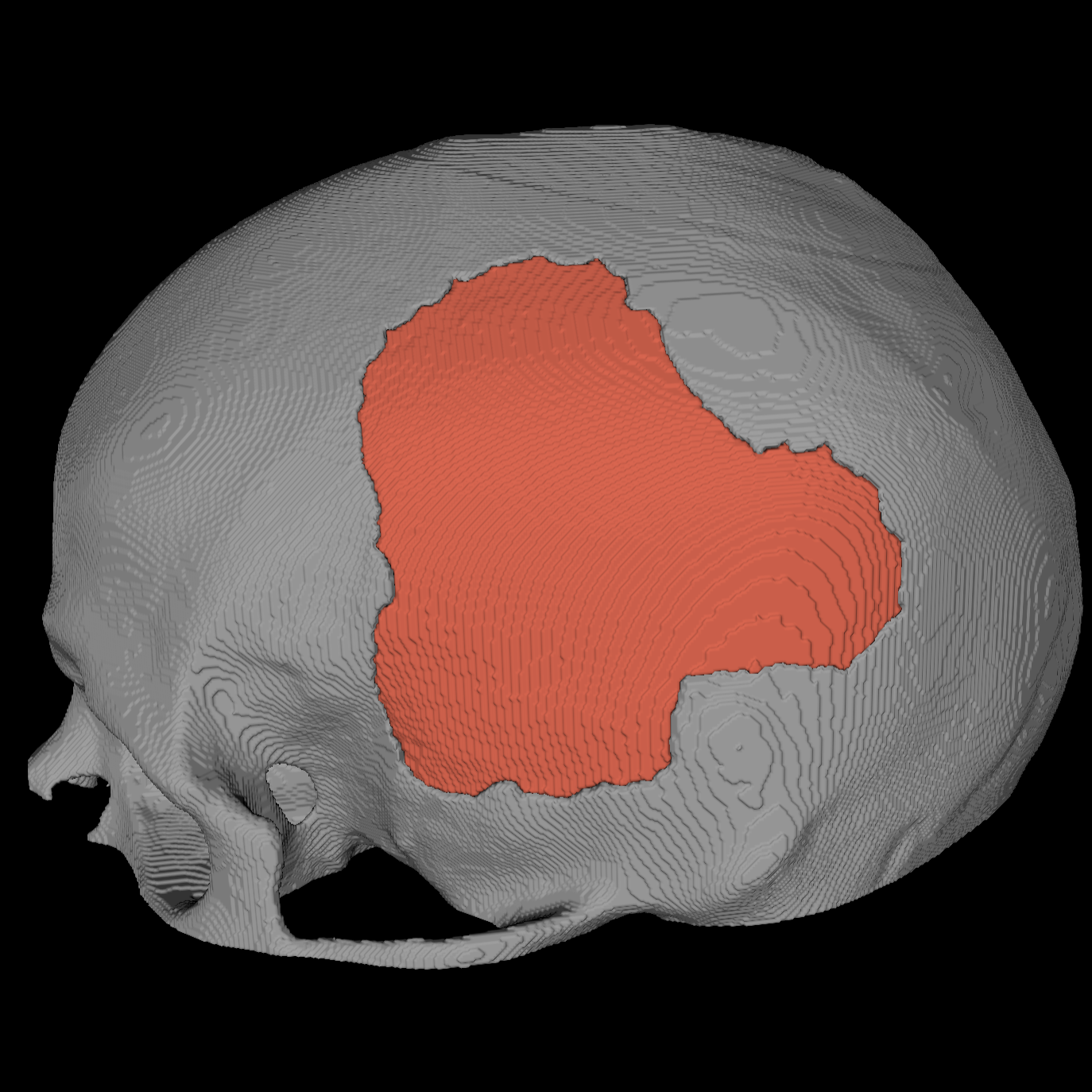}};
    	\node[] at (4, 1.23)    {\scriptsize Parieto-Temporal};
    	\node[] at (6, 0)       {\includegraphics[height=0.16\textwidth]{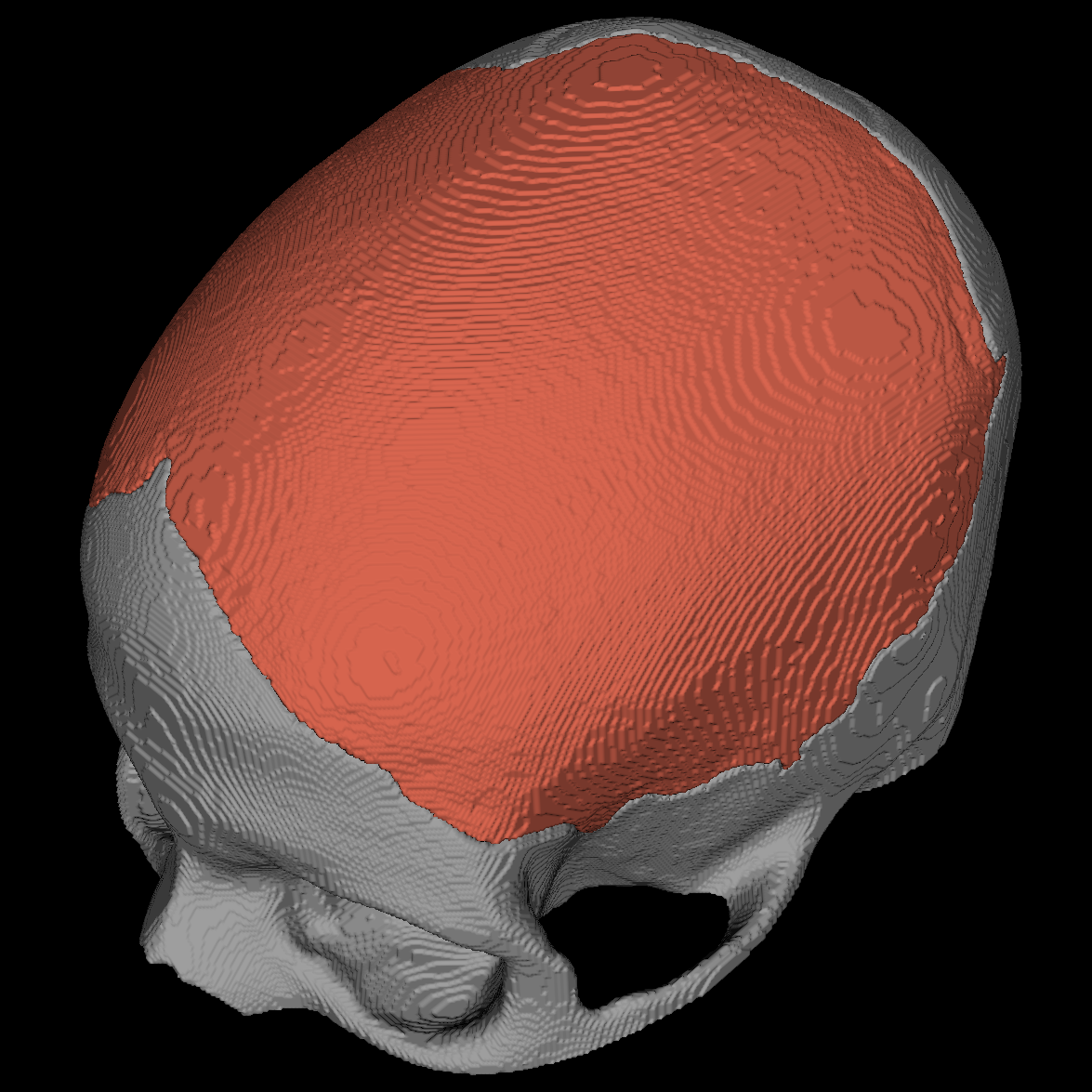}};
    	\node[] at (6, 1.25)    {\scriptsize Random 1};
    	\node[] at (8, 0)       {\includegraphics[height=0.16\textwidth]{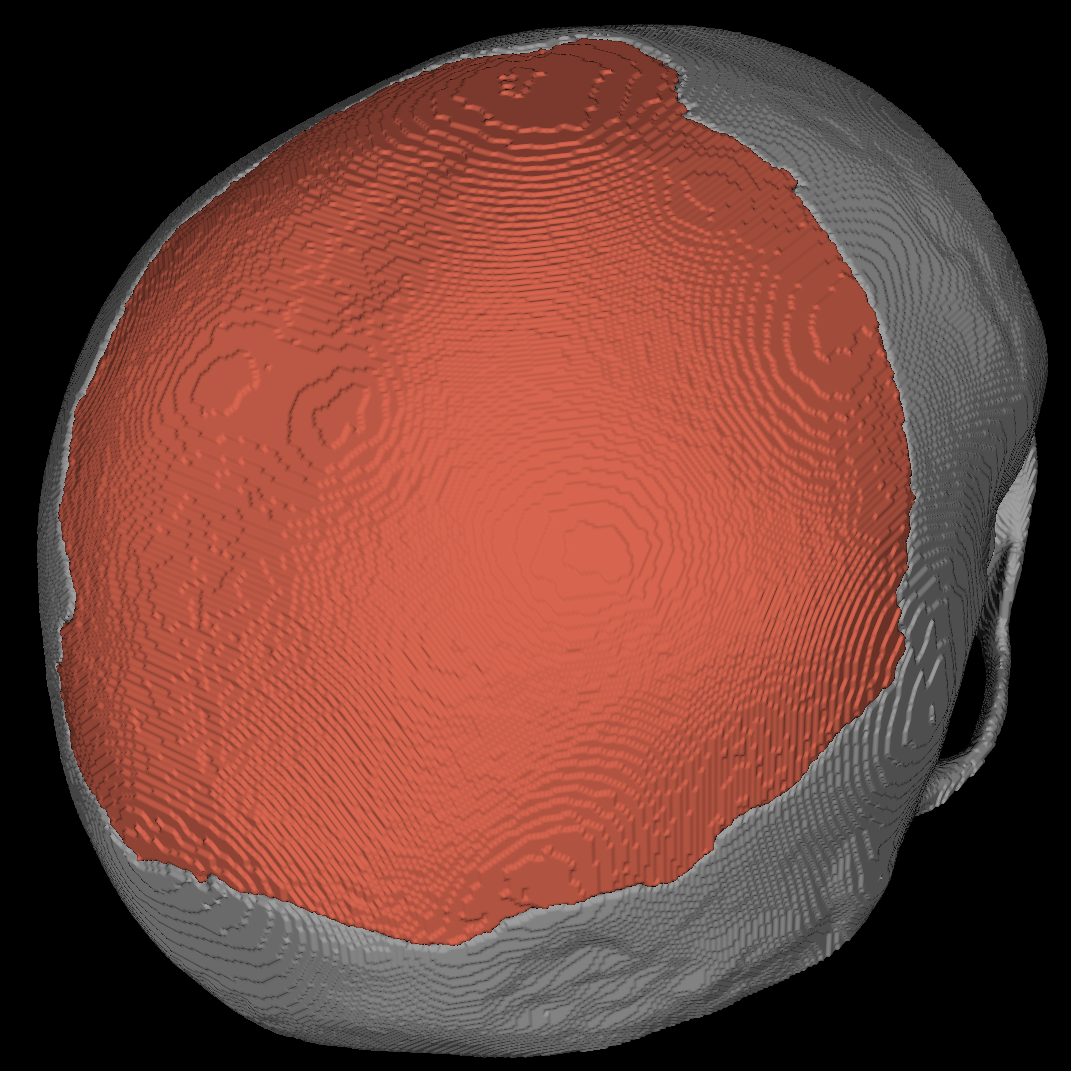}};
    	\node[] at (8, 1.25)    {\scriptsize Random 2};
    	\node[] at (0, -2)      {\includegraphics[height=0.16\textwidth]{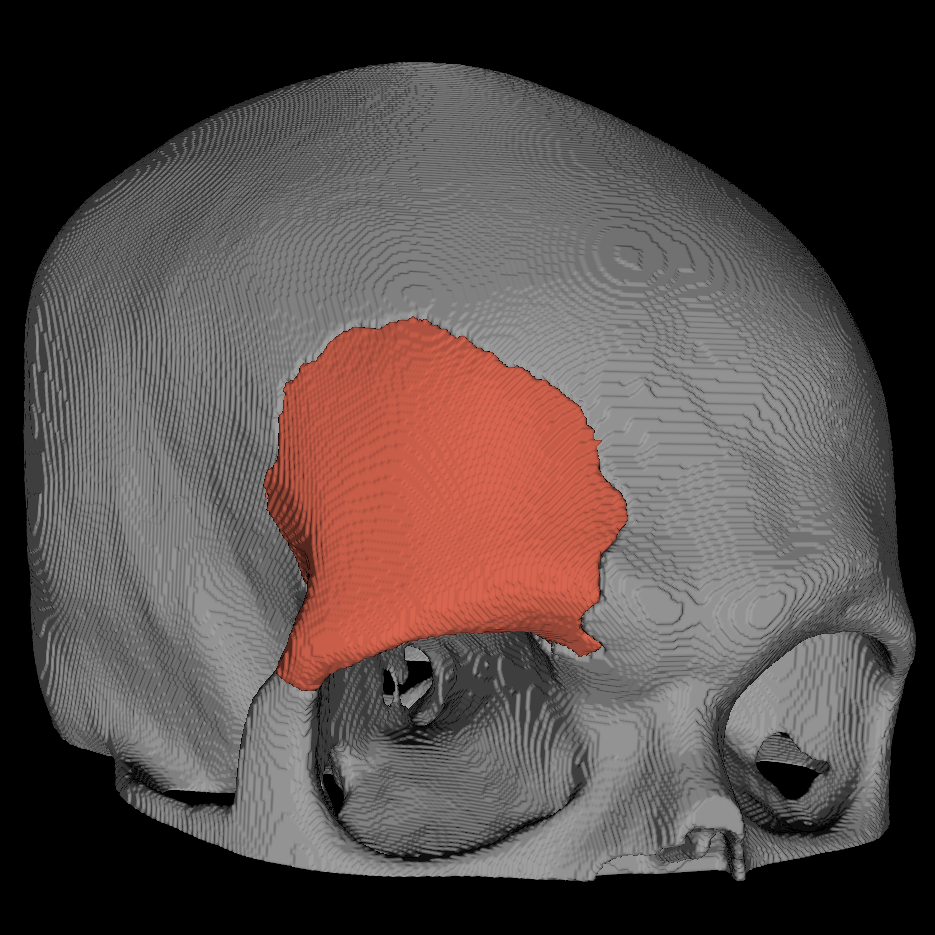}};
    	\node[] at (2, -2)      {\includegraphics[height=0.16\textwidth]{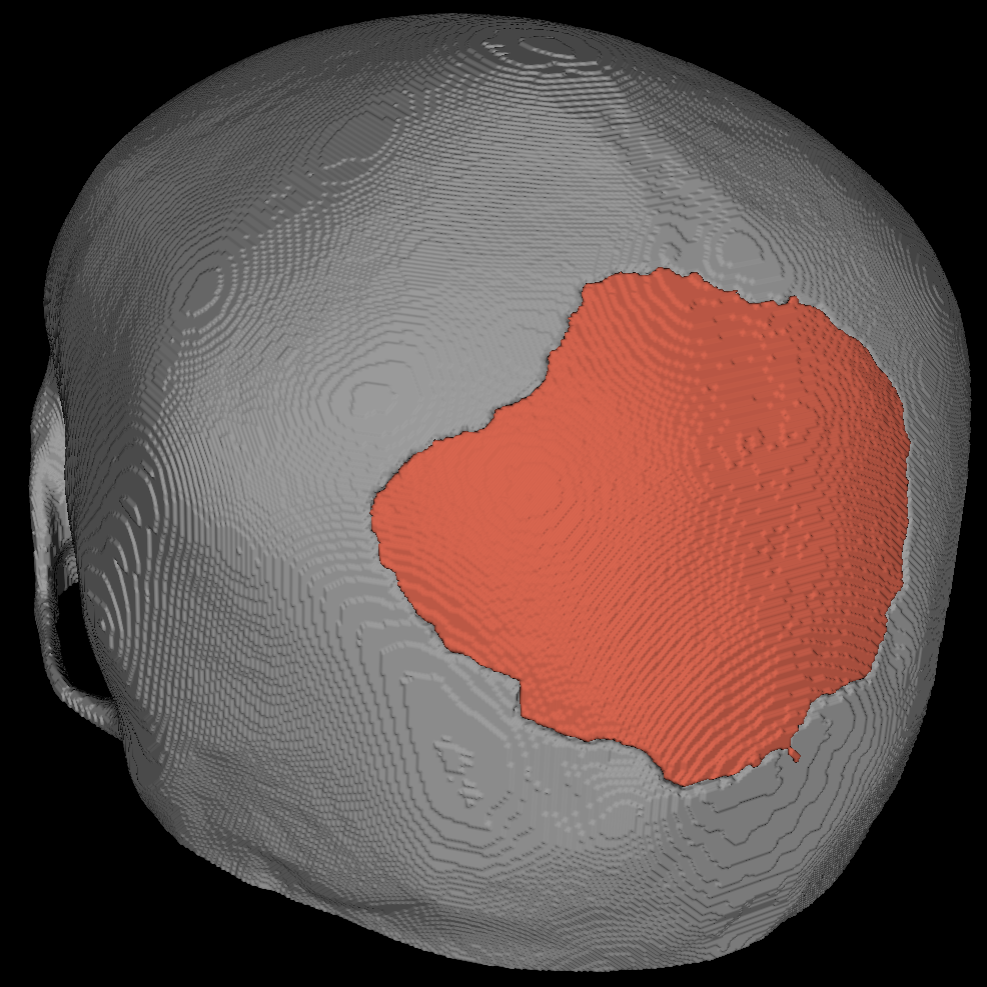}};
    	\node[] at (4, -2)      {\includegraphics[height=0.16\textwidth]{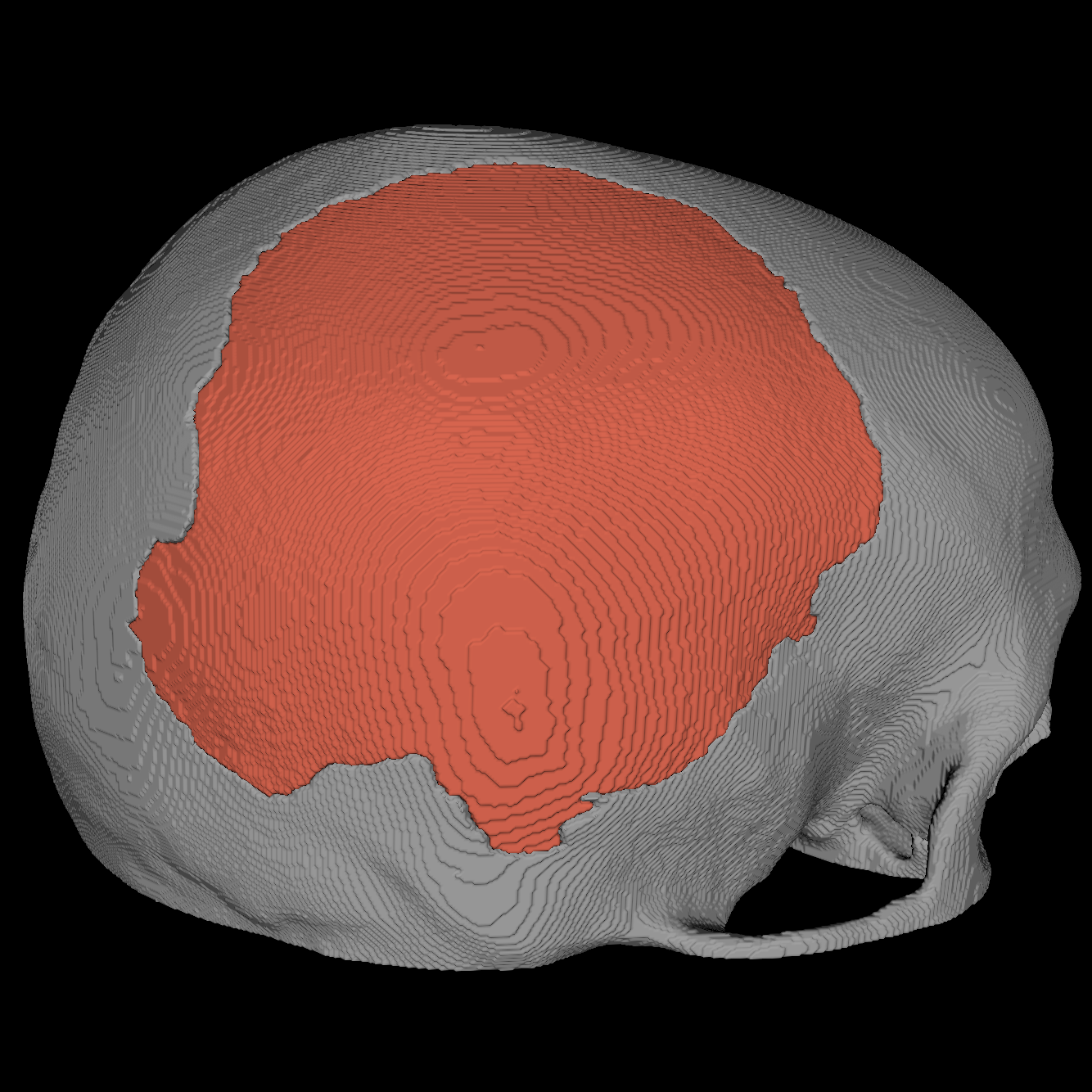}};
    	\node[] at (6, -2)      {\includegraphics[height=0.16\textwidth]{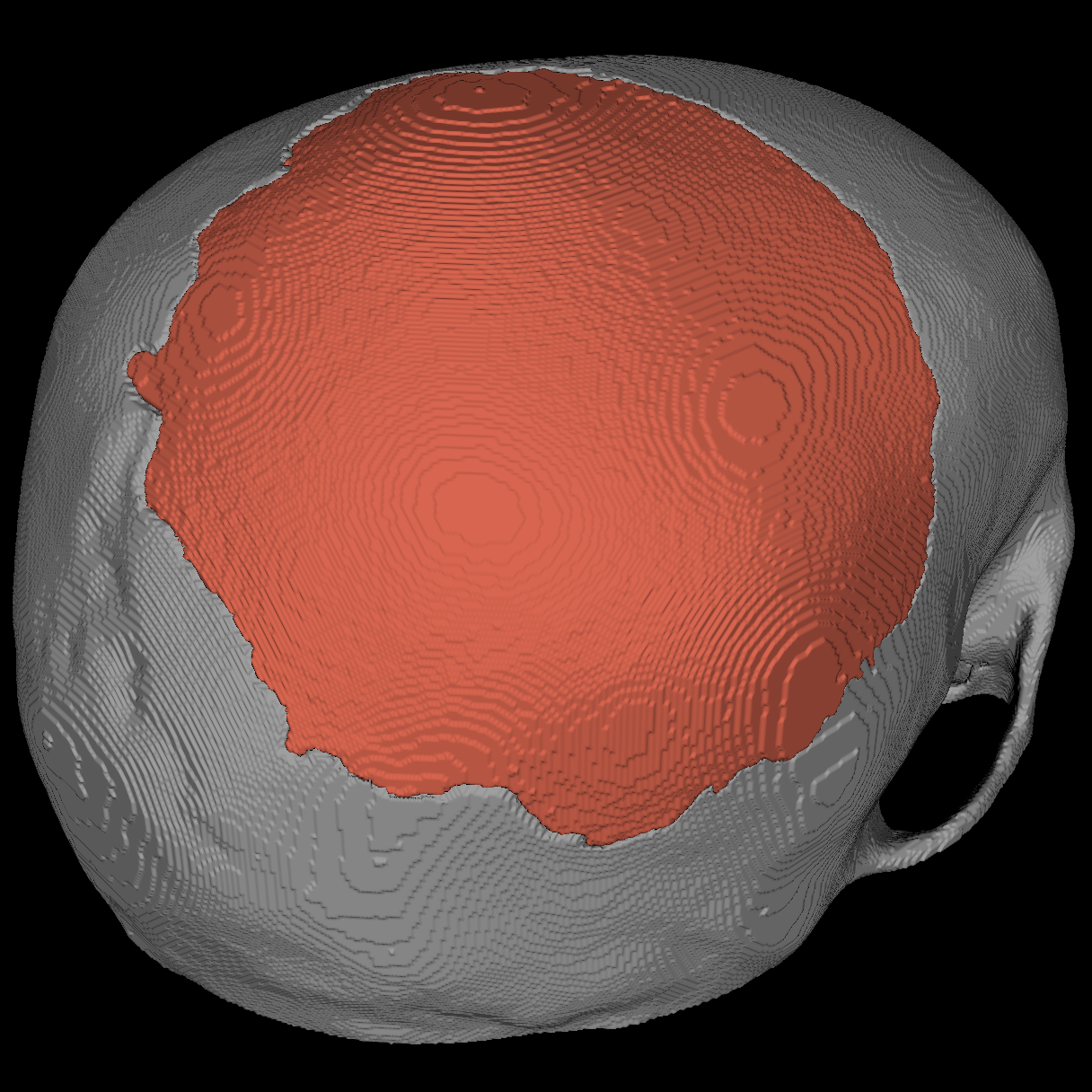}};
    	\node[] at (8, -2)      {\includegraphics[height=0.16\textwidth]{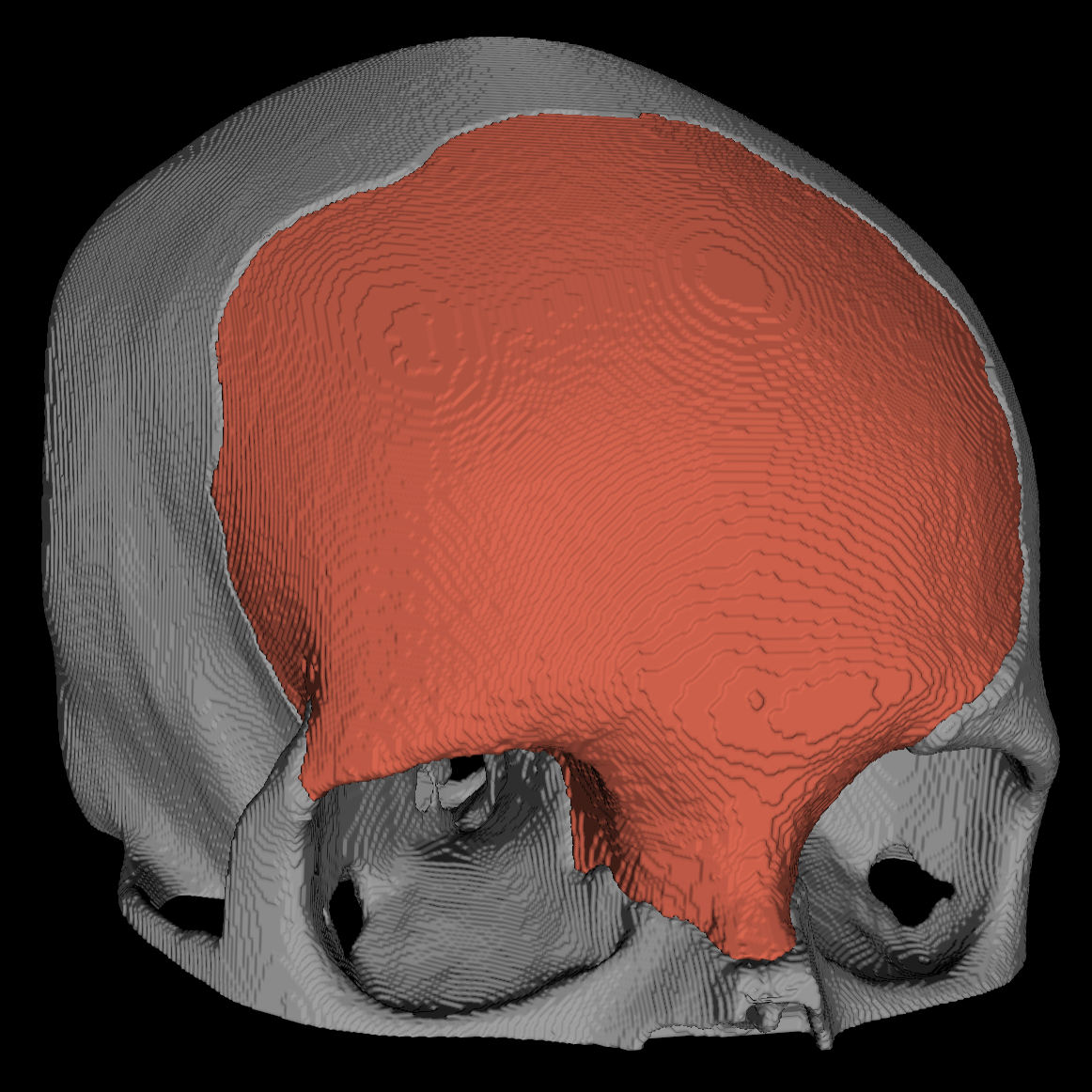}};
	\end{tikzpicture}}
\caption{Additional results of our method for the five different defect classes of the SkullBreak dataset.}
\label{SkullBreak}
\end{figure}
\section{Additional Results on the SkullFix Dataset}
\begin{figure}[h]
\centering
\resizebox{\textwidth}{!}{
	\begin{tikzpicture}
	    \node[] at (0, 0)   {\includegraphics[height=0.16\textwidth]{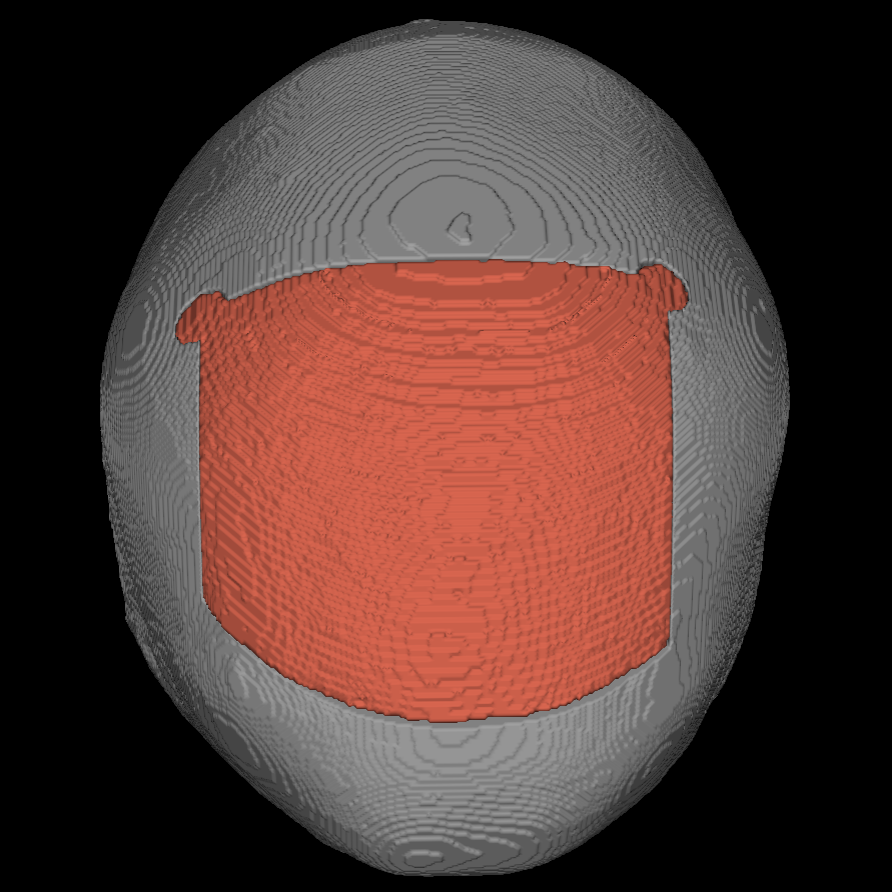}};
    	\node[] at (2, 0)   {\includegraphics[height=0.16\textwidth]{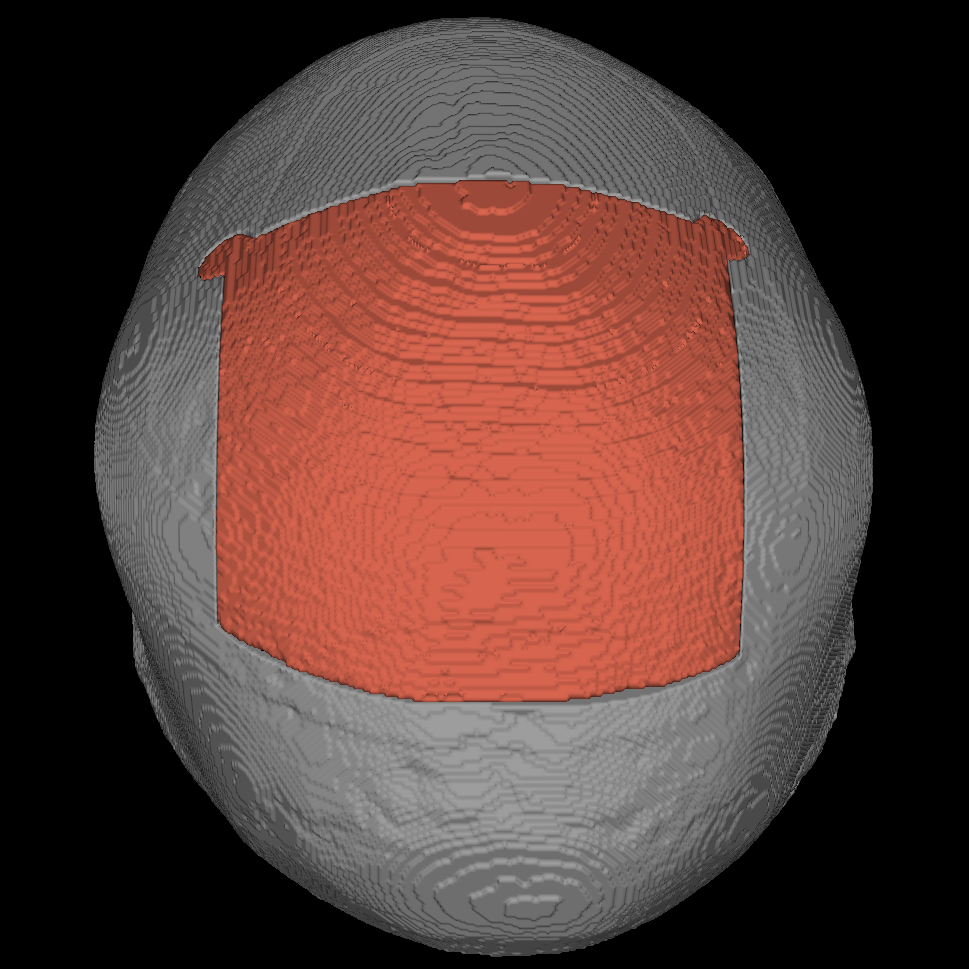}};
    	\node[] at (4, 0)   {\includegraphics[height=0.16\textwidth]{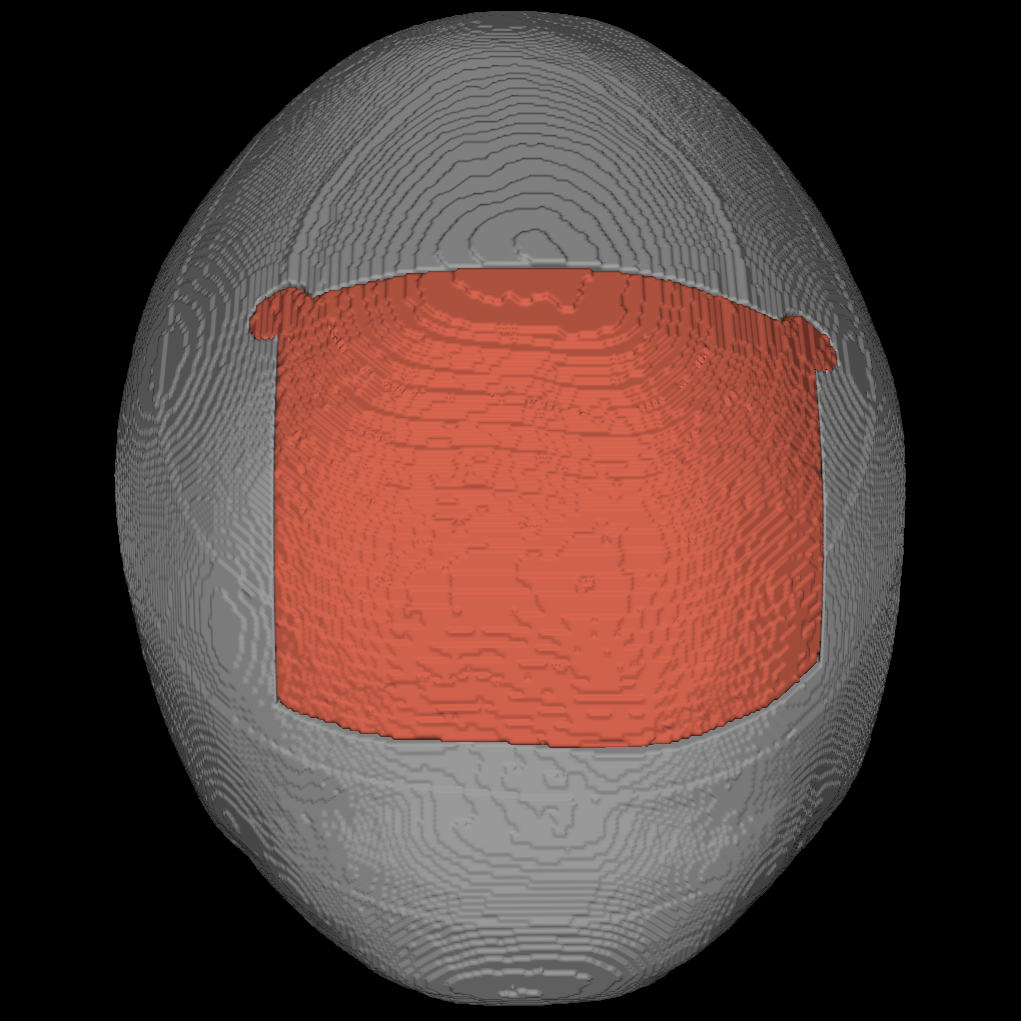}};
    	\node[] at (6, 0)   {\includegraphics[height=0.16\textwidth]{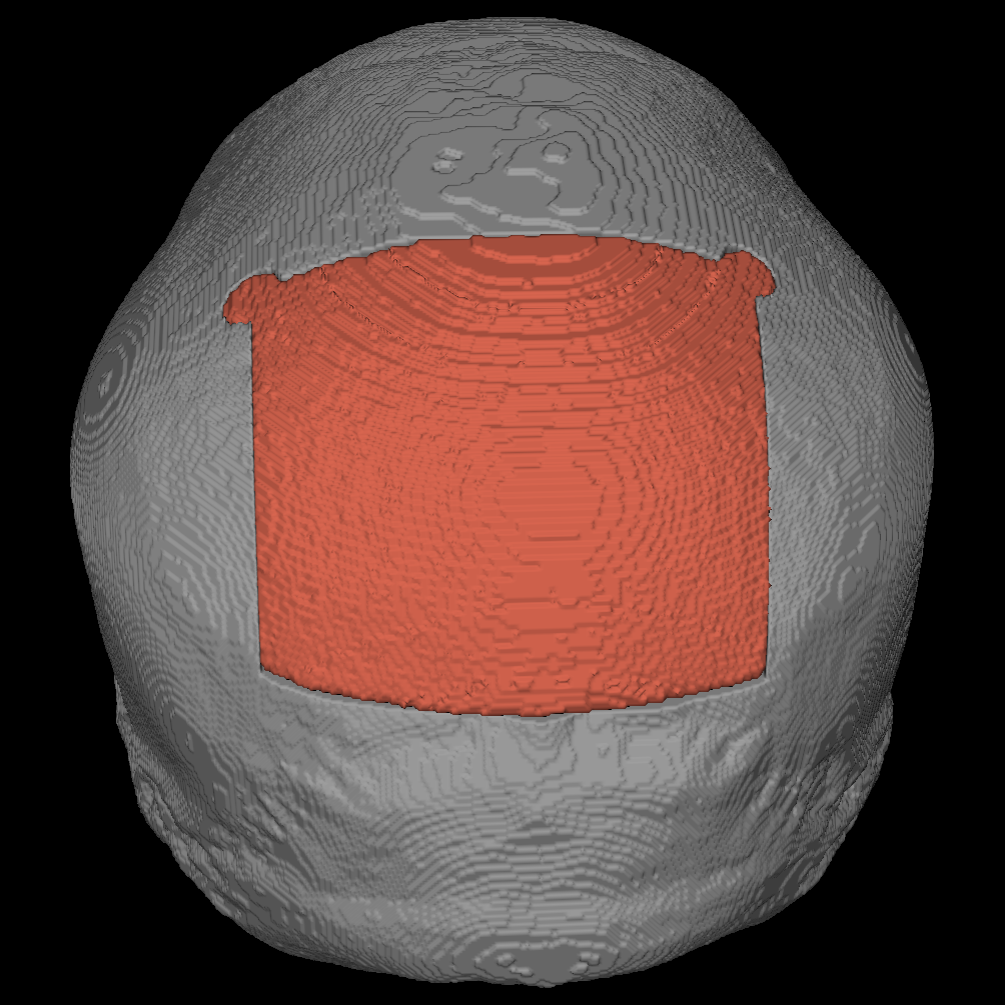}};
    	\node[] at (8, 0)   {\includegraphics[height=0.16\textwidth]{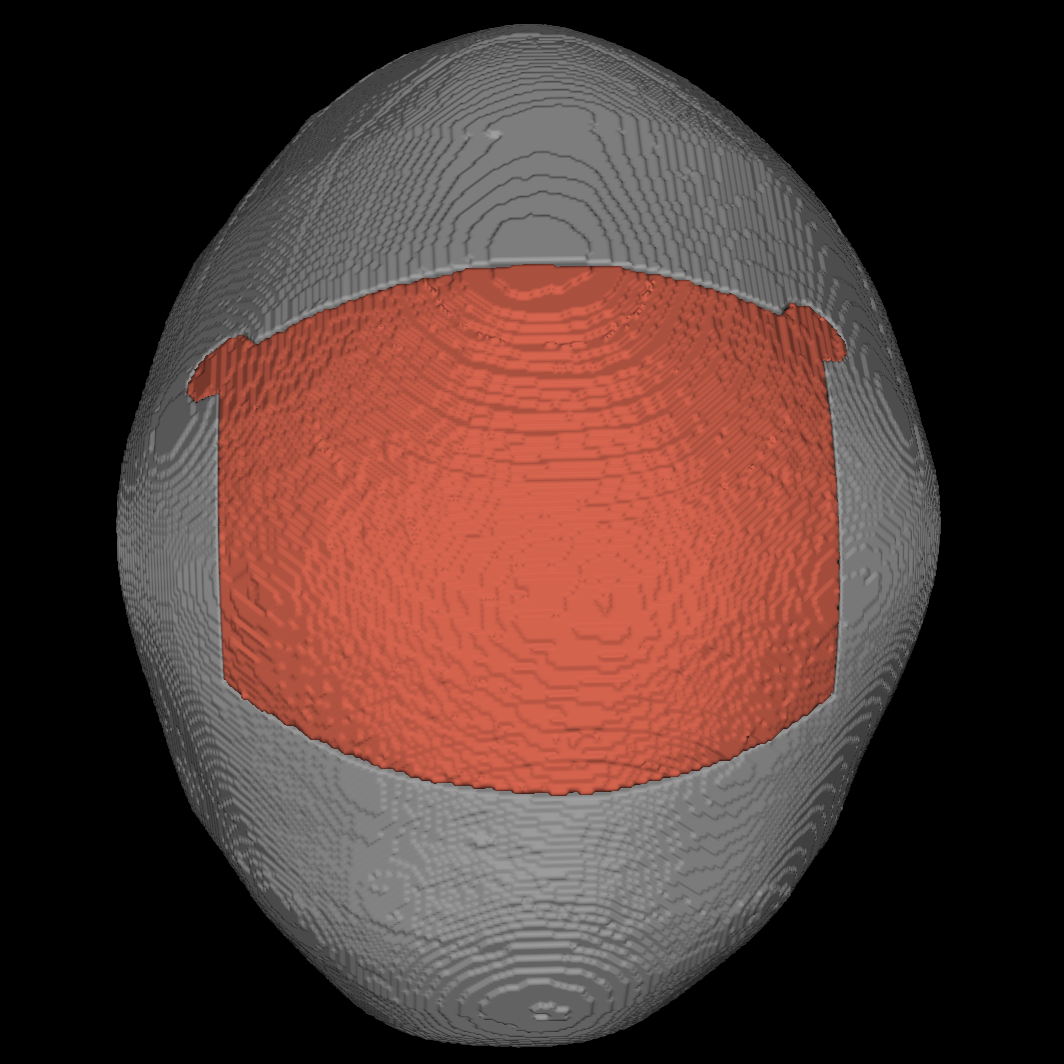}};
    	\node[] at (0, -2)  {\includegraphics[height=0.16\textwidth]{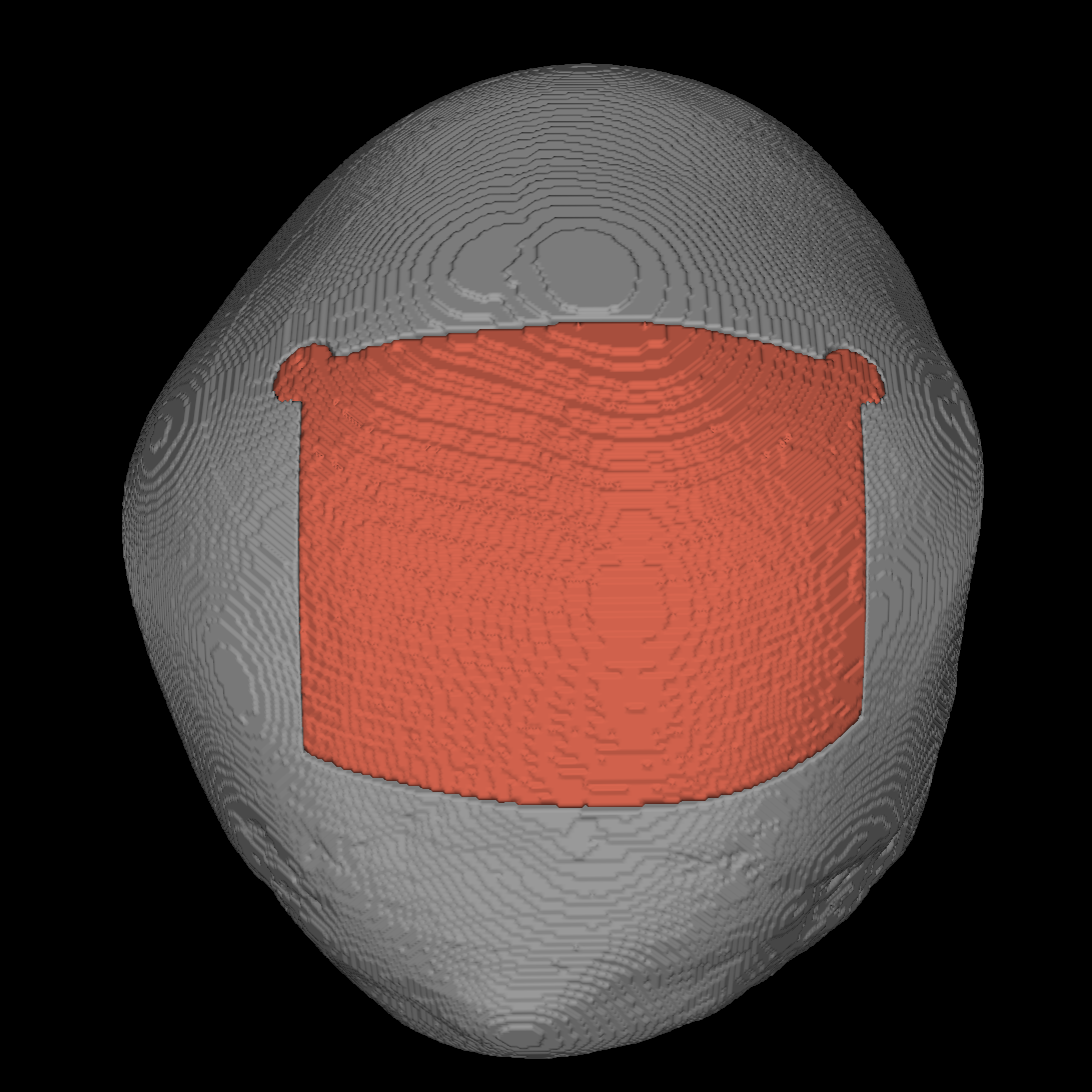}};
    	\node[] at (2, -2)  {\includegraphics[height=0.16\textwidth]{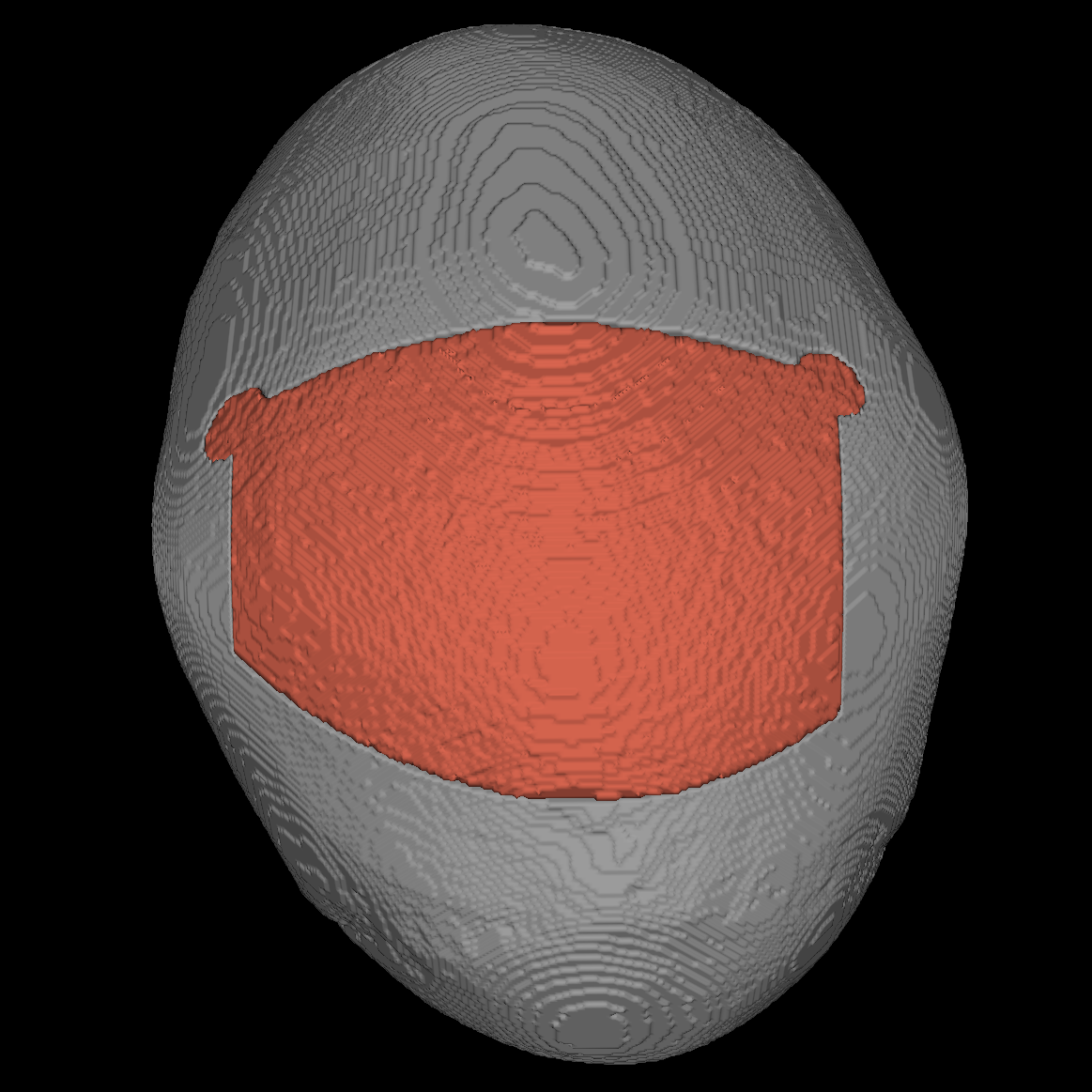}};
    	\node[] at (4, -2)  {\includegraphics[height=0.16\textwidth]{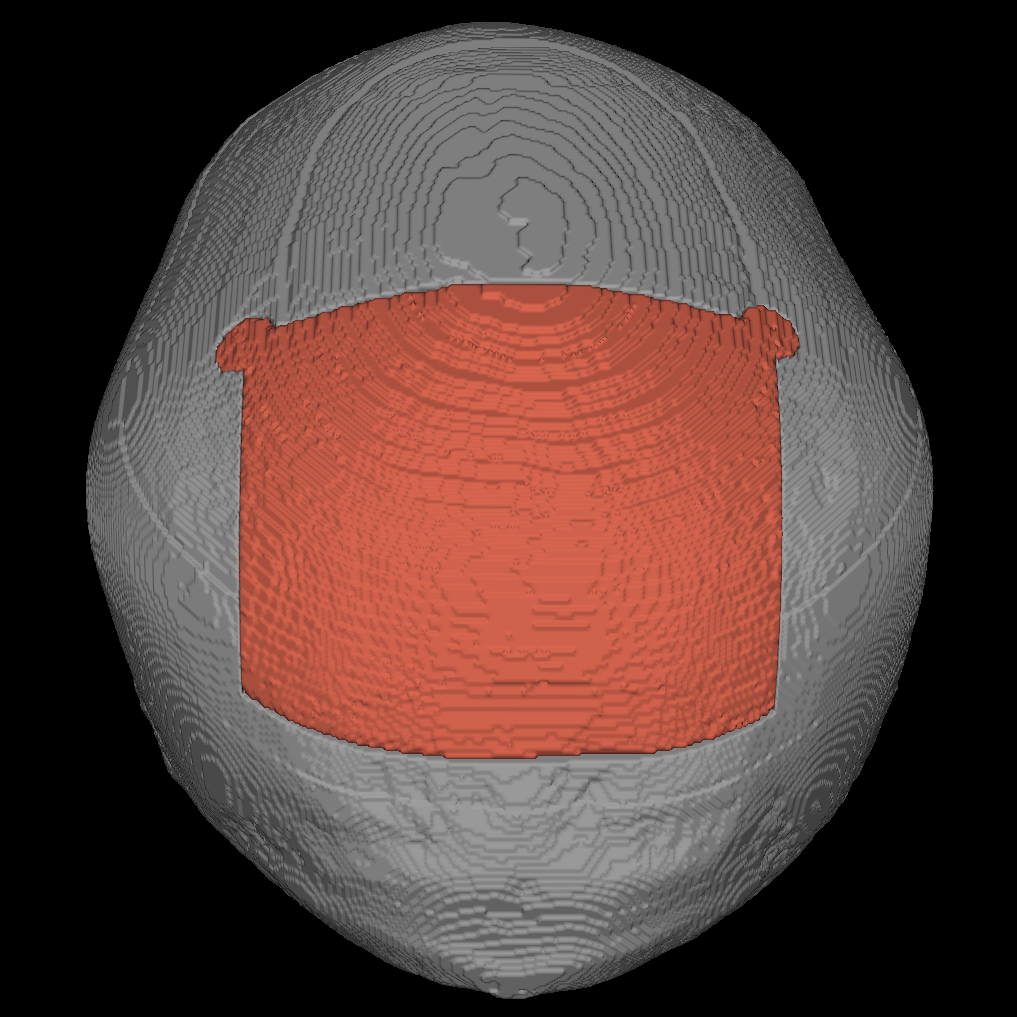}};
    	\node[] at (6, -2)  {\includegraphics[height=0.16\textwidth]{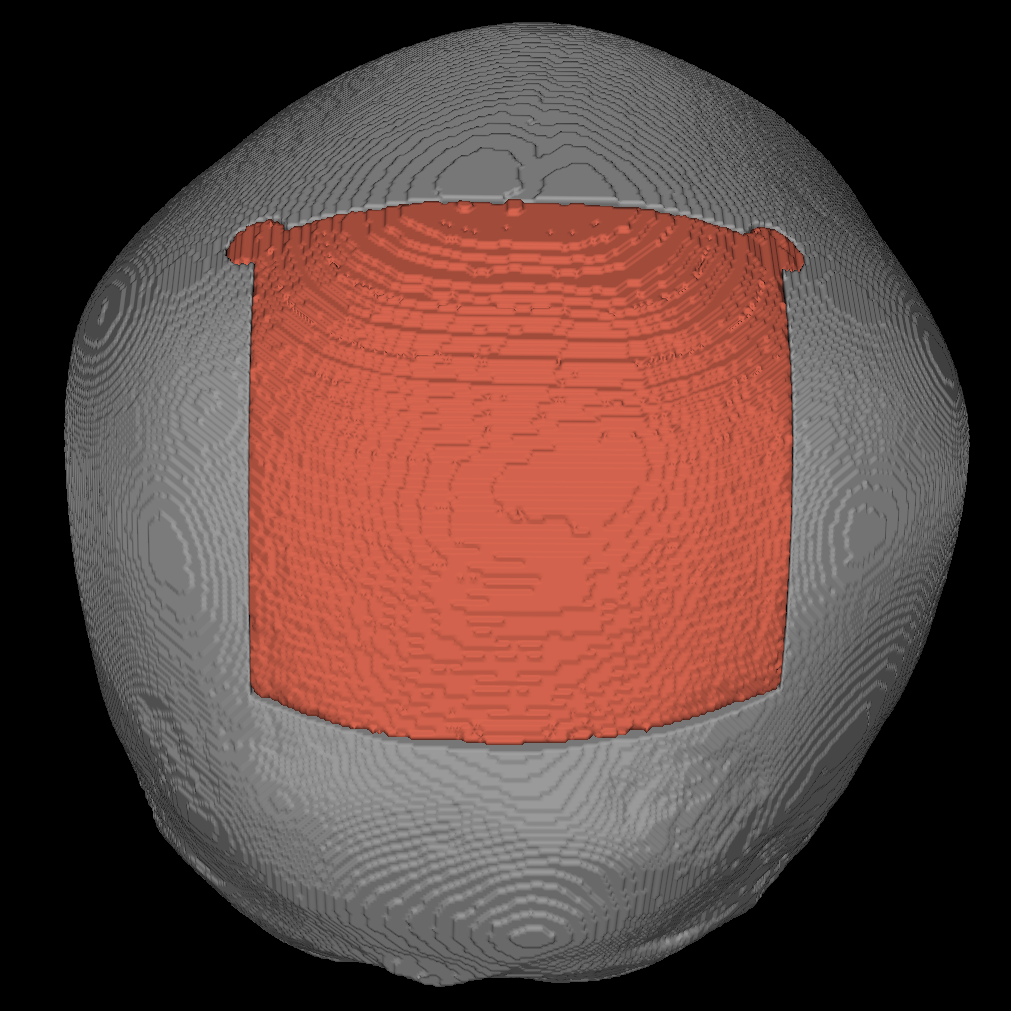}};
    	\node[] at (8, -2)  {\includegraphics[height=0.16\textwidth]{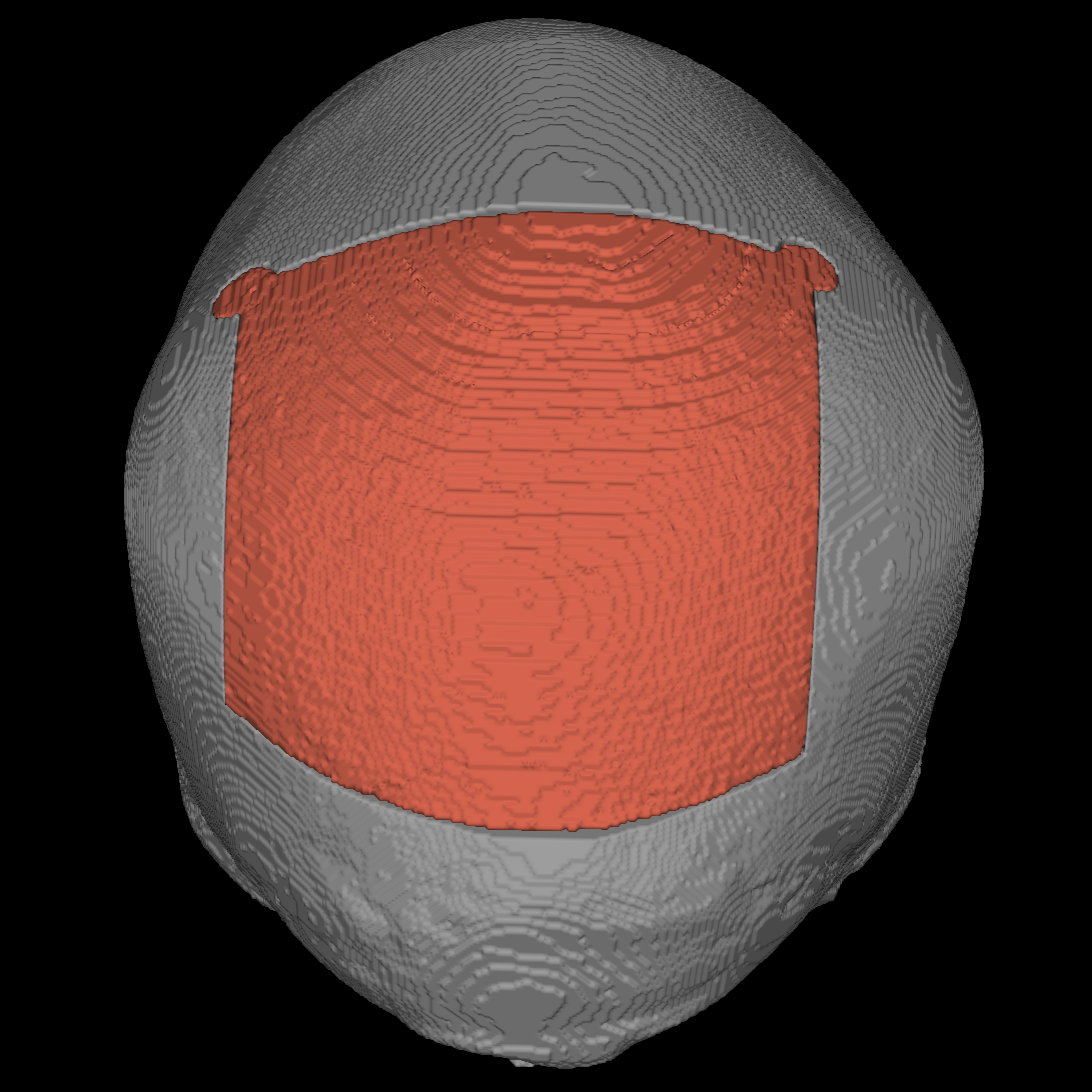}};
	\end{tikzpicture}}
\caption{Additional results of our method for the SkullFix dataset.}
\label{SkullFix}
\end{figure}
\section{Additional Results Ensembling Method}
\begin{figure}[h]
\centering
\resizebox{\textwidth}{!}{
	\begin{tikzpicture}
	    \node[] at (0, 0)  {\adjincludegraphics[width=0.16\textwidth, trim={.15\width, .1\width, .15\width, .4\width}, clip]{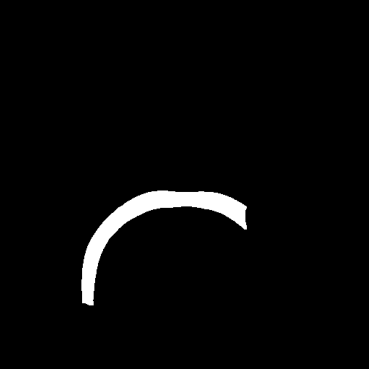}};
    	\node[] at (0, 1)  {\scriptsize Implant 1};
    	\node[] at (2, 0)  {\adjincludegraphics[width=0.16\textwidth, trim={.15\width, .1\width, .15\width, .4\width}, clip]{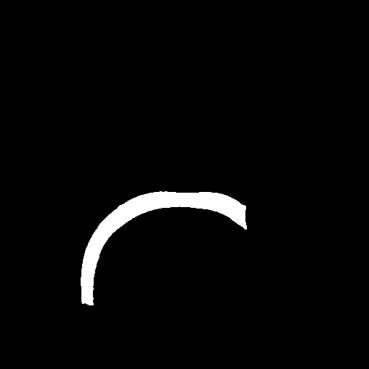}};
    	\node[] at (2, 1)  {\scriptsize Implant 2};
    	\node[] at (3.5, 0){$...$};
    	\node[] at (5, 0)  {\adjincludegraphics[width=0.16\textwidth, trim={.15\width, .1\width, .15\width, .4\width}, clip]{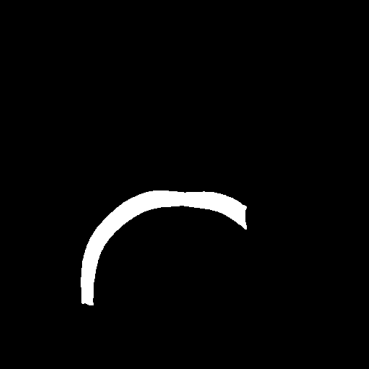}};
    	\node[] at (5, 1)  {\scriptsize Implant 5};
    	\node[] at (7, 0)  {\adjincludegraphics[width=0.16\textwidth, trim={.15\width, .1\width, .15\width, .4\width}, clip]{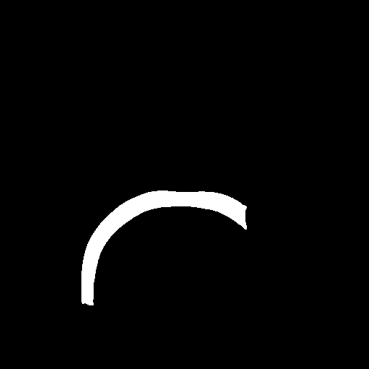}};
    	\node[] at (7, 1)  {\scriptsize Mean};
    	\node[] at (9, 0)  {\adjincludegraphics[width=0.16\textwidth, trim={.15\width, .1\width, .15\width, .4\width}, clip]{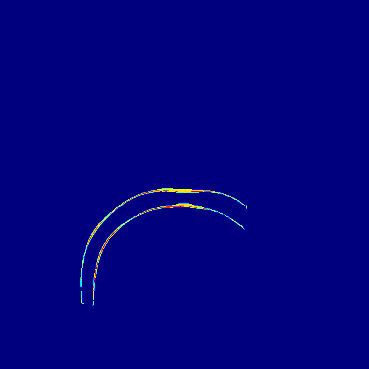}};
    	\node[] at (9, 1)  {\scriptsize Variance Map};
	\end{tikzpicture}}
\caption{Different implants, mean implant and variance map for a single skull defect.}
\label{Ensembling}
\end{figure}
\section{Network Architecture of the Point Cloud Diffusion Model}
\begin{table}[h]
\caption{Architecture of the proposed point cloud diffusion model. The input point cloud is subsequently passed through SA1-4, FP1-4 and a MLP. Time embedding is concatenated to the point features in front of every SA \& FP block.}
\begin{center}
\resizebox{.45\textwidth}{!}{
    \begin{tabular}{|l|c|c|c|c|}
    \hline
        \multicolumn{5}{|c|}{Input size: $30720\times 3$}\\\hline
        \multicolumn{5}{|c|}{Input time embedding size: $64$}\\\hline\hline
        \multicolumn{5}{|c|}{\textbf{Time embedding}}\\\hline
        \multicolumn{5}{|c|}{Sinusoidal embedding dimension: 64}\\
        \multicolumn{5}{|c|}{MLP$(64,64)$}\\
        \multicolumn{5}{|c|}{LeakyReLU$(0.1)$}\\
        \multicolumn{5}{|c|}{MLP$(64,64)$}\\\hline\hline
        \multicolumn{5}{|c|}{\textbf{Set Abstraction (SA) Layers}}\\\hline
                                & SA1     & SA2     & SA3   & SA4\\\hline
        Number of PVConv blocks & $2$     & $3$     & $3$   & $0$ \\
        Input channels          & $3$     & $32$    & $64$  & - \\
        Output channels         & $32$    & $64$    & $128$ & - \\
        Voxelization resolution & $32$    & $16$    & $8$   & - \\
        Use attention           & False   & True    & False & - \\\hline
        Number of centers       & $10240$ & $2560$  & $640$ & $160$ \\
        Radius                  & $0.1$   & $0.2$   & $0.4$ & $0.8$ \\
        Number of neighbors     & $128$   & $128$   & $128$ & $128$ \\\hline\hline
        \multicolumn{5}{|c|}{\textbf{Feature Propagation (FP) Layers}}\\\hline
                                & FP1     & FP2     & FP3   & FP4 \\\hline
        Number of PVConv blocks & $2$     & $3$     & $3$   & $0$ \\
        Input channels          & $128$   & $256$   & $256$ & $128$ \\
        Output channels & $256$ & $256$   & $128$   & $64$ \\
        Voxelization resolution & $8$     & $8$     & $16$  & $32$ \\
        Use attention           & False   & True    & False & False \\
        \hline\hline
        \multicolumn{5}{|c|}{MLP$(64,3)$}\\\hline
        \multicolumn{5}{|c|}{Output size: $30720\times 3$}\\\hline
    \end{tabular}}
    \label{tab:}
    \end{center}
\end{table}